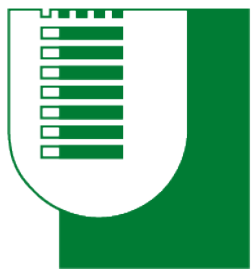

TOR VERGATA
UNIVERSITÀ DEGLI STUDI DI ROMA

Dottorato di ricerca in Fisica - XVI Ciclo

# Energetic Neutral Atom Imaging of Planetary Environments

Alessandro Mura

*Relatore:*

*Dott. Stefano Orsini*

# Contents

# Included papers

*Paper* **I**

A. Milillo, S. Orsini, and A. Mura. Empirical model of H$^+$ fluxes in the equatorial inner magnetosphere, *Conference Proceedings Vol 75,* "Sun-Earth Connection and Space Weather", SIF, Bologna, 2001

*Paper* **II**

Milillo, A., S. Orsini, D. C. Delcourt, A. Mura, S. Massetti, E. De Angelis, and Y. Ebihara, Empirical model of proton fluxes in the equatorial inner magnetosphere: 2. Properties and applications, *J. Geophys. Res.*, 108 (A5), **1165**, 2003

*Paper* **III**

A. Mura, A. Milillo, S. Orsini, E. Kallio, and S. Barabash Energetic neutral atoms at Mars: 2. Imaging of the solar wind-Phobos interaction, *J. Geophys. Res.*, 107(A10), **1278**, 2002

*Paper* **IV**

S. Massetti, S. Orsini, A. Milillo, A. Mura, E. De Angelis, H. Lammer, and P. Wurz. Mapping of the cusp plasma precipitation on the surface of Mercury. *Icarus,* 166, p. 229-237, 2003

*Paper* **V**

A. Mura, S. Orsini, A. Milillo, D. Delcourt, and S. Massetti. Dayside H$^+$ circulation at Mercury and neutral particle emission. To be submitted to *J. Geophys. Res*.

# 1. Introduction

## 1.1 Purpose and structure of this thesis

The aim of this work is to investigate the applications of the neutral atom imaging to the environments of the Earth, Mars and Mercury. This innovative technique permits the study of energetic plasma by means of analysing the result of the interaction of this plasma with a neutral thermal population or with a surface. The main advantage, when compared to the direct ion detection, is that it is possible to have an instantaneous survey of the whole magnetosphere of a planet. An example could help. Before the first ENA data, most of the knowledge about the Earth magnetospheric plasma came from *in situ* measurements of ions, electrons and electromagnetic fields. Those measurements, of course, could not represent any real instantaneous situation, but only an averaged picture of it, since the temporal and spatial variation cannot be easily be distinguished. Some short time scale phenomena, such as substorms, have been found difficult to comprehend without a global and continuous imaging. Even if some information about the plasma may be extracted from other sources, such as UV imaging [like aurorae, e.g. *Horwitz*, 1987], some populations (for example, the ring current) remained invisible.

Furthermore, neutral atom imaging gives information not only about the energetic plasma, but also about the thermal neutral population (in the case of charge-exchange) or about the surface composition (in the case of sputtering). Conversely, it is necessary to set up some dedicated unfolding techniques to recover the 3D plasma distributions from the 2D ENA images. In chapter **2** I will discuss the details of ENA and LENA imaging and how to extract information by using this technique.

Chapters **3**, **4** and **5** are focused on the environments of the Earth, Mars and Mercury; each chapter is followed by a section dedicated to ENA instrumental concepts. In some way, this structure reflects the chronology of my thesis work. In fact, as far as it concerns the Earth and Mars, some papers have been already accepted and published. Here I present a summary of my work, and most of the information is also contained in the included publications. The studies about Mercury, conversely, are in progress. For this reason I include here not only a submitted paper about Mercury, but also other unpublished results of my research in this last year.

*The Earth*. In chapter **3** I will present an empirical model of $H^+$ equatorial distribution in the inner magnetosphere. I will use this model to reconstruct the global magnetospheric configuration during the geomagnetic storm of July 1991. The study could be a preliminary tool for ENA deconvolution techniques; moreover, it investigates the charge-exchange process, as a main loss of the magnetospheric plasma. In this frame, a new concept for a MH-ENA sensor is presented in section **3.4**, as a first example of neutral atom space instrumentation.

*Mars and Mercury*. While the Earth is, in principle, our main scientific objects, it is worth noting that Mars and Mercury represent some complementary "subsets" of the Earth. Mars has a faint atmosphere, but no large-scale magnetic field; Mercury has a magnetic field but only a tenuous exosphere. The complexity of the Earth magnetospheric environment, in addition to the ENA unfolding difficulties, makes it interesting to study those "laboratory planets".

Even if no ENA/LENA actual measurements have been done either for Mars or for Mercury until now, the scientific community is already focusing on these planets. As far as it regards Mars, ESA mission Mars Express (MEX) has reached Mars in December 2003. The ASPERA-3 instrument, on board MEX, is about to collect the first ENA data about Mars; it is hence extremely important to focus our attention on the unfolding methods. Moreover, ENA imaging may also be applied to Phobos, a natural satellite of Mars. In this case, surface properties can be studied, and this will give information about the origin and history of this satellite. In section **4.2** I will present a new model for the exosphere of Phobos, which may also be applied to many other outgassing natural satellites. In sections **4.3** and **4.4** I will introduce an investigation technique for Phobos' outgassing rate, and in sections **4.5** and **4.6** I will briefly resume main ASPERA-3 facts as well as my participation to the instrument development. Last but not least, I will show the first ENA data from ASPERA-3, collected while it was travelling to Mars.

Concerning Mercury, both ESA (with Bepi Colombo) and NASA (with Messenger) space agencies are planning to explore this inner planet of the solar system. In this case, the lack of available data is proportional to the effort in studying this planet. The chance to have a Neutral Analyser (SERENA: Search for Exospheric Refilling and Emitted Neutral Abundances) on board Bepi Colombo has focused the attention of most of the ENA community, and of SERENA team too, on the study of this planet. In fact, before proposing such an instrument, it is crucial to know as many details as possible about the expected ENA/LENA signal. It is also important to point out what information may be obtained, which instrument sensitivity is necessary and which satellite attitude is optimal. My effort here consists in the feasibility study of this kind of remote-sensing applied to Mercury's environment. More particularly, a new model of magnetospheric ion and neutral circulation, which is one of the major goals of this thesis, has been developed and presented here.

Furthermore, ENA/LENA imaging study merges here with the instrument design activity. In section **5.10** I will report designs for some proposed ENA/LENA sensors. Incidentally, just some days before the presentation of this thesis (December 2003), BC mission was definitely approved and SERENA experiment is selected to be part of the Italian payload, which will be proposed to ESA by the Italian Space Agency.

Finally, in chapter **6** I will summarize my thesis work, I will stress the main scientific results, and I will propose future developments.

This thesis includes, as a fundamental part, four published papers and a submitted one. However, I intend this thesis to be readable also without those papers. For this reason, some parts and main results presented in the papers are resumed, expanded or simply reported here, and some figures have been shared.

# 2. The Energetic Neutral Atoms

## 2.1 Energetic and Low Energy Neutral Atoms: A brief history

The existence of Energetic Neutral Atoms (ENAs), arising from charge-exchange of ionospheric plasma with the local exospheric gas, was discovered in the early 50's. During an aurora, *Meinel* [1951] detected a blue-shifted $H_\alpha$ line, indicating that some precipitating Hydrogen ions were being neutralized. Some years later, *Moritz* [1972] reported the detection of protons of about two hundred keV at equatorial low altitudes. He suggested that they could have come from the outer radiation belt, turned into ENA by exchanging their charge with some neutral thermal population, then reached low altitudes without being affected by the magnetic field, and finally been re-ionised just before detection. In the 70's, the ion detectors on board IMP 7, IMP 8 and ISEE-1, while the satellites were orbiting out of the magnetospheric environment, detected some energetic particle flux coming from the Earth. The existence of ENA was again invoked, since no ion flux is expected in these regions. Since ENA fly almost unperturbed after their emission, more or less like photons, the scientific community started talking of "ENA imaging". Roelof [*Roelof et al.,* 1985; *Roelof and Williams,* 1988] reconstructed the first ENA image of the geomagnetosphere (Figure **1.1**) by using an ion detector on ISEE-1 and thus opening the way to an ENA remote-sensing of both planetary magnetosphere and neutral population.

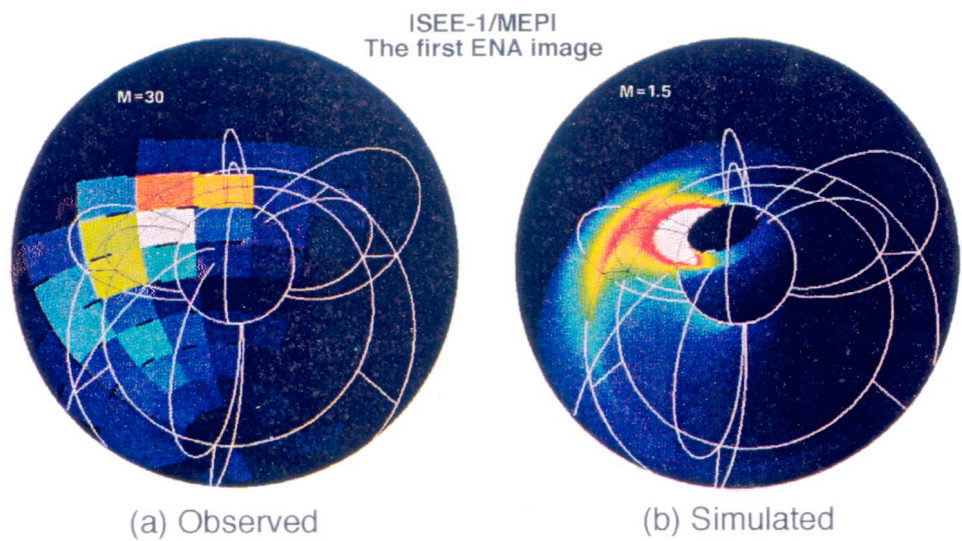

**Figure 1.1 Panel a)** First ENA image reconstructed by Roelof and Williams from ISEE-1/MEPI data. Sun is to the right; the smaller circle represents Earth, with a terminator line crossing it. Other lines are examples of magnetic field lines for L values of 3 and 5. **Panel b)** Same image as **a)**, simulated using a ring current model. From *Roelof,* [1987].

Nowadays, after several other ENA detections by means of ion sensors, new instruments specifically designed for ENA imaging of the Earth magnetosphere are collecting useful data (such as IMAGE/M-H-LENA and CASSINI/INCA) or are about to be launched (such as those on board TWINS and DOUBLE STAR). Moreover, the potential of ENA imaging has become so clear that ENA sensors are also on board extra-terrestrial space exploration missions such as Mars Express (ASPERA-3) or Venus Express (ASPERA-4). A new concept ENA sensor has also been proposed for the future ESA exploration mission to Mercury, Bepi Colombo (BC).

Neutral atom imaging may be applied not only to charge-exchange, but also to all processes producing directional neutral atoms. For example, if plasma interacts directly with a planetary surface, sputtering processes may occur and neutral atoms may be extracted. The detection of those neutrals may give information about both plasma and surface properties. Such a technique can be applied, for example, to the exploration of Mercury's environment and a dedicated instrument package (SERENA) has been proposed on board BC.

While ENAs coming from charge-exchange have energies between some hundreds of eV to some hundreds of keV, the typical energy of the sputtered neutrals is, generally, between fractions of eV to some hundreds of eV. For this reason, and also following a consolidated nomenclature, I will call ENAs only the formers, and I will introduce the term LENAs (low energy neutral atoms) for the latters. In the following, I will discuss those processes separately, but it is worth noting that this separation is made only to better explain the different features of the processes themselves. In fact, a neutral sensor is able to discriminate between emission processes only if the energy ranges are completely separated, and, in principle, this is not always true.

## 2.1 Charge exchange process: imaging plasma-exosphere interaction

When an energetic ion collides with a neutral and non-energetic target particle, it may charge-exchange and be neutralized, becoming ENA. An electron and a small amount of kinetic energy are exchanged between the neutral and the ion. Some examples of this kind of process, active on the Earth and involving energetic ring current ions and cold atmospheric neutrals, are the following:

$$H^+ + H \rightarrow H + H^+$$

$$O^+ + H \rightarrow O + H^+$$

$$H^+ + O \rightarrow H + O^+$$

$$O^+ + O \rightarrow O + O^+$$

In principle, the energy defect of the process is equal to the difference of the two atomic ionisation potentials [*Hasted*, 1964]. Charge-exchange is a resonance process; symmetrical when the species of the ion and the neutral are the same, and accidental otherwise [*Stebbings et al.*, 1964; *Hasted*, 1964]. While the target is scattered at an approximately perpendicular angle with respect to the projectile path, the newly created ENA retains approximately both the energy and the direction of the colliding energetic ion. No longer affected by magnetic or electric fields, the ENA travels along ballistic trajectories (straight lines at these energies), and can be detected outside the interaction region. Figure **2.1** shows a scheme of this process.

Let us consider a vantage point $S_0$ and a unit vector $\hat{v}$. The differential ENA flux of direction $\hat{v}$ observed at $S_0$ is generated by the interaction of the energetic ion fluxes with one or more ambient neutral populations in a generic point $S$ along the line ($S_0$, $\hat{v}$). At $S$, the infinitesimal ion flux $dJ_+$ directed toward $S_0$ is:

$$dJ_+ = n_+(l)\, v\, f(l, \mathbf{v})\, d\mathbf{v}, \qquad (2.1)$$

where $n_+(l)$ is the ion density, $v$ is the velocity, $f(l, \mathbf{v})$ is the ion velocity distribution function and $l$ is the distance $\overline{S_0 S}$. If $f$ is expressed by means of the kinetic energy $E$, the direction $\hat{v}$ and the infinitesimal solid angle $d\Omega$, as usual, the above formula could be changed by using the following one:

$$f(l, \mathbf{v})\, d\mathbf{v} = f'(l, E, \hat{v})\, dE\, d\Omega. \qquad (2.2)$$

At point $S$ the mean free path of ions of energy $E$ relative to charge exchange process is:

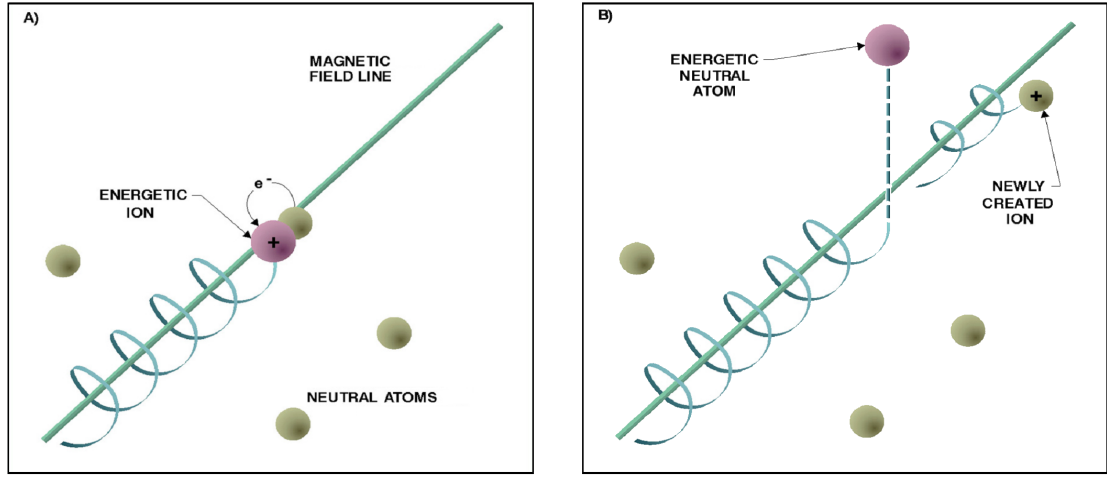

**Figure 2.1** Scheme of the charge-exchange process. Energetic ion hits a neutral target (panel **A**) and gets an electron from it. After collision, it may proceed with a straight-line trajectory (panel **B**).

$$\lambda_+(l,E) = \frac{1}{\sum_k \sigma_k(E) n_k(l)}, \qquad (2.3)$$

where $n_k$ is the density of the $k^{th}$ neutral species at point $S$, $\sigma_k$ is the charge-exchange cross-section between ions and the $k^{th}$ neutral species. We can now estimate the ENA infinitesimal flux $dJ_E(E, \hat{v})$:

$$dJ_{ENA}(E, \hat{v}) = \int_{l=0}^{l=\infty} dJ_{H^+} \frac{1}{\lambda_+(l,E)} dr = \int_{l=0}^{l=\infty} \sqrt{\frac{2E}{m_+}} f'(l, E, \hat{v}) \frac{1}{\lambda_+(l,E)} n_+(l) \, dl \, dE \, d\Omega; \qquad (2.4)$$

and the ENA differential flux integrated along a certain line of sight $\Phi_{ENA}(E, \hat{v})$ :

$$\Phi_{ENA}(E, \hat{v}) = \frac{dJ_{ENA}(E, \hat{v})}{dE \, d\Omega} = \int_{l=0}^{l=\infty} \sqrt{\frac{2E}{m_{H^+}}} f'(l, E, \hat{v}) \frac{1}{\lambda_{ION}(l,E)} n_{H^+}(l) \, dl \; . \qquad (2.5)$$

Charge exchange (CE) cross sections are obviously function of energy (see figure **2.2**) and have significant values only between 1 keV and some hundreds of keV. Typical value for the CE cross section is $10^{-16}$ cm$^{-2}$; even if this value is not very high, in magnetospheric physics cases the integration path may be long enough to cause relevant ENA fluxes at $S_0$.

The ENA flux moving towards $S_0$ may be removed by other processes, such as photon ionisation or stripping [*C:son Brandt*, 1999]. In this case, it is easy to include a destruction function by means of the mean free path $\lambda_{ENA}$ in equation (**2.5**) [*Orsini et al.*, 1994]:

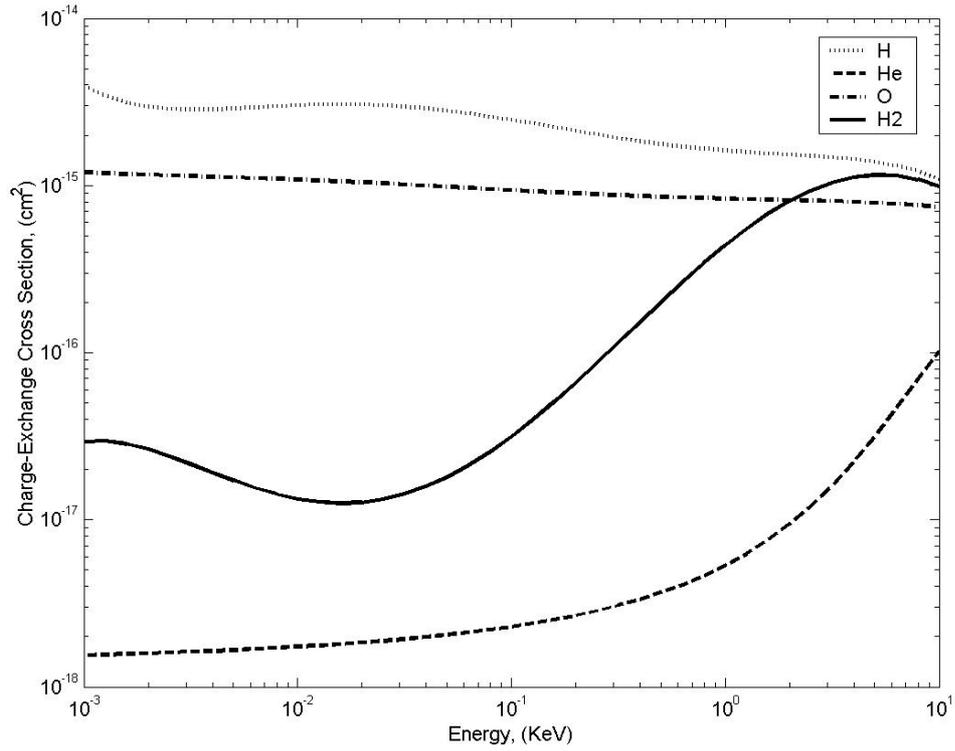

**Figure 2.2** Charge exchange cross sections for H+ with some typical neutral exospheric species. Data from *Barnett et al.* [1977], *Stebbings et al.* [1977], and *Barnett* [1990].

$$\Phi_{ENA}\left(S_0,E,\hat{v}\right) = \int_{l=0}^{l=\infty}\left[\sqrt{\frac{2E}{m_{H^+}}}f'\left(l,E,\hat{v}\right)\frac{1}{\lambda_{ION}\left(l,E\right)}n_{H^+}\left(l\right) - \frac{\Phi_{ENA}\left(S,E,\hat{v}\right)}{\lambda_{ENA}\left(S,E,\hat{v}\right)}\right]dl \quad \textbf{(2.6)}$$

In all cases discussed here, the medium can be considered "optically thin" to ENA, i.e. the mean free path is long enough, and ENA can transport information out of the generation region, thus allowing remote-sensing of the interaction process [*Roelof et al.*, 1985; *Roelof*, 1987; *Henderson et al.*, 1997; *Orsini and Milillo*, 1999; *Milillo and Orsini*, 2001].

At the time CE creates new ENA, it removes ion from plasma flux. The ion destruction rate may be expressed as:

$$\frac{d\Phi_+}{dt} = \frac{\Phi_+}{\lambda_+}v. \qquad \textbf{(2.7)}$$

The integration in equation **2.5** is, almost always, solved numerically over a large number of directions to obtain a simulated ENA image. Moreover, as I will discuss better later in this section, while trying to unfold some ENA data a large number of simulated ENA image may be needed. In many cases, the plasma and exospheric properties are not defined by analytical functions, but by discrete grids. Hence, the integration step $\Delta s$ and the angular resolution $\Delta\alpha$ and $\Delta\beta$ must be small enough to have a very large number of integration points inside the same cell. In the end, this procedure may take a lot of time.

Here I introduce a different kind of ENA simulation, which may be called "weighted-grid integration". It can be applied only if the plasma is defined over a 3D grid, and this happens, for example, if plasma distribution has been simulated numerically. The principle is to consider each plasma cell in the space as if it were a light source, and to treat ENA flux as light. Hence, to obtain the light as seen at vantage point $S_0$, I will sum over all plasma cells, giving them a weight that is inversely proportional to the square of the distance between the cell (point $S_{ijk}$) and $S_0$. The ENA production rate $\dfrac{d\Phi_{ENA}}{ds}$ for any point of the grid may be calculated as:

$$\frac{d\Phi_{ENA}}{ds} = \frac{|v|\,n_H}{\lambda_+},\tag{2.8}$$

and the total flux escaping from a cell located in $S_{ijk}$ and detected in $S_0$ is:

$$\Phi_{ENA}\left(S_0, S_{ijk}\right) = \frac{d\Phi_{ENA}\left(S_{ijk}\right)}{ds}\,\Delta V\,f_\alpha(\hat{v})\,\frac{1}{\left|S_0 S\right|^2},\tag{2.9}$$

where $\Delta V$ is the volume of the $ijk^{th}$ cell; $f_\alpha(\hat{v})$ is the 3D angular distribution function, $\hat{v}$ is the direction from $S$ to $S_0$.

This procedure may be faster than the previous one, especially if the plasma and neutral properties are defined over a discrete grid. In the following, for historical reasons, I will use the former for Mars' ENA imaging, and the latter for Mercury's case.

*ENA deconvolution.* I have already remarked that neutral atoms are good messengers of information. However, this information should be unfolded, because a 2D image, intrinsically, cannot have the amount of information stored in a 3D volume. The first and simplest kind of unfolding method for ENA data is simulating the corresponding ENA image by using theoretical or empirical models of plasma and neutrals, and then comparing the simulated image with the experimental one. A second step is to introduce some free parameters in the models and tune them until a good fit between real and simulated data is found. In section **5.7** I will show an example of the parameter tuning in the case of ENA generation in Mercury's cusps.

Even if the easiest way to obtain information from ENA imaging is the method described above, it has been demonstrated [*Roelof*, 2003] that ENA imaging can produce interesting results even without models, if some intrinsic symmetry or property of the plasma can add enough additional information to remove the ENA ambiguity. In the case of the Earth, for example, the plasma distribution could not have the degrees of freedom of an arbitrary $C^\circ(\mathbf{R^3} \rightarrow \mathbf{R})$ function, if we want it to satisfy the adiabatic expansion law. This consideration, plus the assumption that the pitch angle (PA) distribution is flat at the equator, is able to remove the "ENA ambiguity". In the case of Mars, additional information may be obtained, for example, by considering that the plasma distribution must have, approximately, cylindrical symmetry around the Mars-Sun axis.

Finally, it must be noted that even in the cases where ENA images don't need any magnetospheric model to be unfolded, they do need some to be understood. In fact, a sequence of ENA images, alone, is able to technically reconstruct a sequence of magnetospheric plasma distribution, but only a model is able to provide them with an interpretation, forecast their future development, and distinguish the physical processes involved.

## 2.2 Sputtering process: imaging a planetary surface

ENA image is as useful as the charge-exchange process is effective in removing ions from the magnetospheric plasma. This is the case of the Earth and Mars; elsewhere, such as at Mercury, plasma may interact directly with a planetary surface, and sputtering may be the dominant process (see sections **5.5** and **5.6**). This latter process is the result of the impinging of a particle on a surface; if the impact energy ($E_i$) of this particle is sufficient, another particle (sputtered particle) may be extracted. The energy distribution of the sputtered particles, $f(E_e)$, can be expressed as:

$$f(E_e) = \begin{cases} k \dfrac{E_e}{(E_e + E_b)^3} \left[ 1 - \left( \dfrac{E_e + E_b}{E_i} \right)^{\frac{1}{2}} \right] & E_e \leq E_i - E_b \\ 0 & E_e > E_i - E_b \end{cases} \quad \textbf{(2.10)}$$

[*Siegmund*, 1969, *Sieveka and Johnson*, 1984], where $E_e$ is the ejection energy and $k$ is a normalization constant. The energy $E_b$ is the surface binding energy and depends on the atomic species. Typical values for $E_b$ are few eV, while $E_i$ may reach values of some keV. In this case, it may be shown that the shape of $f(E_e)$, except for the highest energies, depends mostly upon the binding energy $E_b$, which corresponds approximately to the peak energy of the spectrum. Figure **2.3** shows two examples of this function in the cases of Oxygen and Sodium.

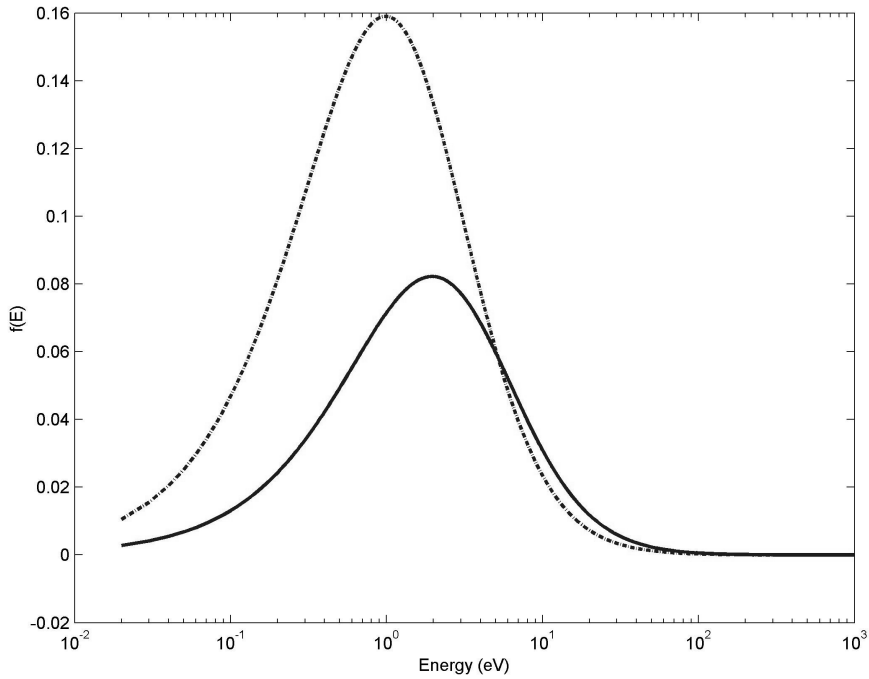

**Figure 2.3** Energy distribution function for sputtered particles, $F(E_e)$, as a function of ejected Na (continuous line) and O particle (dashed line) energy $E_e$ in the case of $E_i = 1$ keV (solar wind protons). Binding energies here are 2 eV for Oxygen [*McGrath et al.*, 1986; *Cheng et al.*, 1987] and 4 eV for Sodium [*Lammer and Bauer*, 1997].

The resulting LENA flux is hence:

$$\frac{d\Phi}{dE} = Y\,R\int_{E\min}^{E\max}\frac{d\Phi_{H+}}{dE_i}f_S\left(E_e,E_i\right)dE_i \ , \qquad (2.11)$$

where $R$ is the fraction of the considered species in the soil composition, and $Y$ is the yield for sputtering production. As for the ENAs coming from charge-exchange, the LENA flux may be simulated once the soil composition and the binding energy (depending on the atomic composition) have been assumed.

The charge of the products arising from this particle bombardment depends both on the composition and the chemical structure of the planetary surface, but generally most of the sputtered particles are neutral. A more oxidized surface produces higher ion content, which is typically between 0.1% and 10% of the total.

The LENA signal coming from the surface of the planet carries information about the precipitating energetic particles (neutrals and ions) and the surface properties. Even if there is no line-of-sight integration, the deconvolution may be difficult as well, and the LENA source is still not univocally determined.  In fact, the low energies of LENAs make their trajectories not rectilinear. In the case of energies close to or below the gravitational escape energy (typically some eV or fraction of eV), the trajectories must be treated taking into account the ballistic laws. In the case of Mercury, a hypothetic LENA sensor on board an orbiting spacecraft may detect particles coming from the other side of the planet. For lower energies, the situation merges with that of an exosphere, and LENA sensor will work as an exospheric neutral detector. In section **5.8** I will show examples of LENA trajectories for medium and low energies.

# 3. The Earth: where all has started.

This chapter is dedicated to my first work as a PhD student: storm-time monitoring of the Earth equatorial magnetosphere. The complexity of the terrestrial magnetosphere makes it extremely difficult to develop time-evolving theoretical models of plasma circulation. Here I will present an original approach, which uses an empirical model of the 90° pitch angle proton distribution in the equatorial plane [*Milillo et al.* 2001]. This model depicts a "virtual magnetospheric configuration", statistically including all possible processes [*Paper* **I**, *Paper* **II**], and in the following is also referred to as *reference model* (section **3.1**).

In section **3.2** I will demonstrate that it is possible to reconstruct different magnetospheric configurations, i.e. to obtain *modified* models from the *reference* one, just tuning some of its numerical constant.

Finally, I will show how to use *in situ* data to guide the model parameters. Their modulation during a geomagnetic storm leads to a time-series of different *modified* models, and hence a sequence of pictures of the magnetosphere during the storm. In other terms, the observed energy spectra collected in a single position are able, via the empirical model, to give information about global plasma properties such as ring current and partial ring current development, L-shell compression/stretching and cross-tail potential drop. In section **3.3** this method will be applied to CRRES/MICS/LOMICS data collected during July 1991 storm.

This study does not directly involve ENA imaging; nevertheless, magnetospheric modelling is the basis for ENA simulation and unfolding. Moreover, the method described here can still be referred to as a "magnetospheric imaging", obtained by applying the model to *in situ* data. Last but not least, this model is able to show how important the losses due to charge-exchange are, and, in turn, how useful ENA imaging is to monitor the magnetospheric configuration.

In section **3.4** I will present the instrumental concepts and some preliminary simulations of a medium-high energy neutral analyser (MH-ENA NAOMI) proposed to ESA for the ISS (International Space Station).

## 3.1 Reference model.

The present analysis is based on an empirical model of the 90° pitch angle proton fluxes in the equatorial plane of the Earth [*Milillo et al.,* 2001]. The model is based on the AMPTE/CCE/CHEM experiment data, and it is a superposition of different functions representing different equatorial populations and/or physical processes; a complete description of the model is given in *papers* **I** and **II**.

The model has been developed using a filtered data set, which represents a statistical configuration (AE < 100 nT) averaged during solar minimum. All large-scale spatial and temporal processes coexist in the model, which does not depict any specific and instantaneous configuration. However, in section **3.2** I will show that the model is able to reconstruct *any* magnetospheric configuration, provided that its numerical constants are opportunely tuned.

The goal of this study is to reconstruct, via the model, the different configurations of the geomagnetosphere during a storm. Hence, among different segments/populations of the model, I am focusing on the ones that describe the injection and the diffusion of protons in the equatorial magnetosphere. These functions, expressed in terms of $H^+$ flux ($\Phi_{H^+}$) versus *L-Shell* (*LS*), are summed to obtain the total $H^+$ differential flux:

$$\frac{\Phi_{H^+}(LS)}{dE\,d\Omega} = \left[ A_{G2} \exp\left( -\frac{(LS - P_{G2})^2}{2W_{G2}^2} \right) + CO \right] \exp\left( -\frac{I_S^2}{(LS - P_S)^2} \right) + A_{G3} \exp\left( -\frac{(LS - P_{G3})^2}{2W_{G3}^2} \right) \quad \textbf{(3.1)}$$

where *A, P* and *W* simply stand for amplitude, peak positions and widths of the *G2* and *G3* gaussian functions. The first addendum in equation **3.1** describes the injected population, in the following simply referred to as *CO+G2*. The second addendum represents the diffused population, or simply *G3* in the following (this nomenclature has been adopted for consistency with the model description in *papers* **I** and **II**).

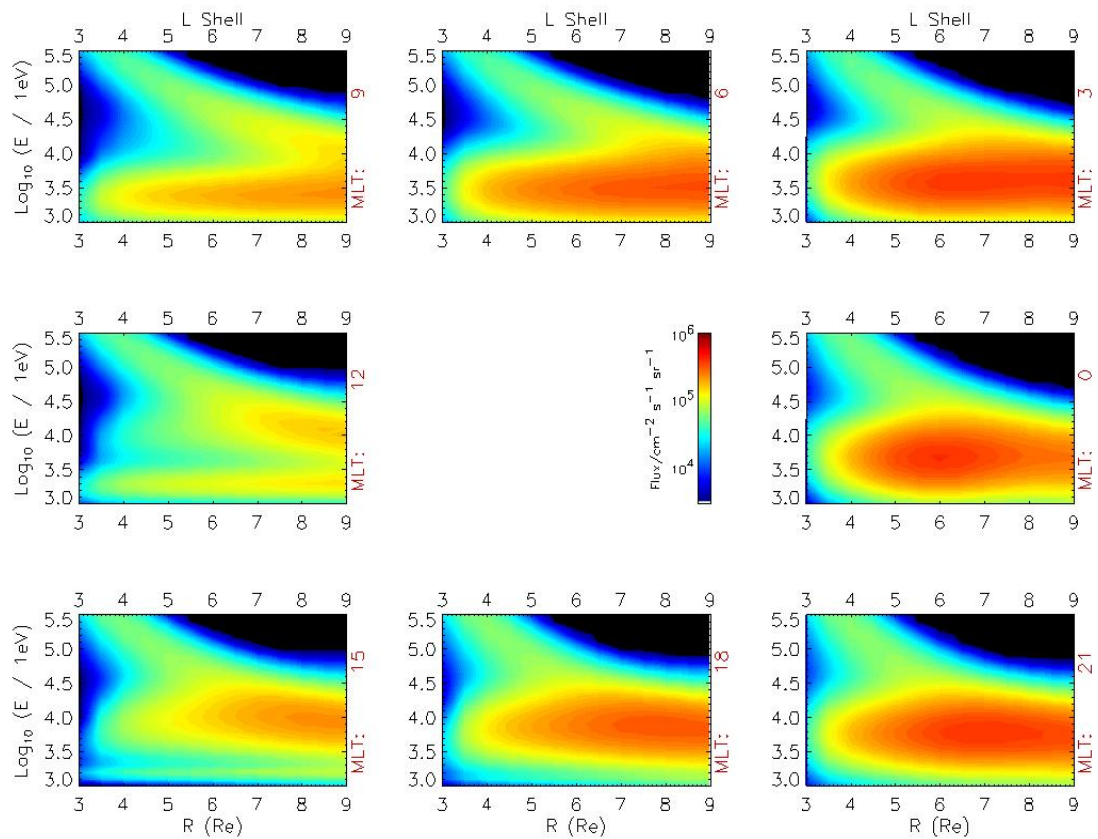

**Figure 3.1** Colour-coded picture of the modelled 90° pitch angle differential proton fluxes as a function of *LS* and *E* at different *MLT*

The $\Phi_{H^+}$ versus energy (*E*) and Magnetic Local Time (*MLT*) dependency is not explicit in equation **3.1,** but it is obtained by other gaussian and trigonometric functions for a total of 37 physical coefficients (see *Paper* **II**). The model is represented in figure **3.1;** each panel is a colour-coded intensity map of the $\Phi_{H^+}$ in the *E* vs. *LS* plane, and different panels refer to different *MLT*.

The *CO+G2* (injection-related) population is represented by the intense peak of $\Phi_{H^+}$ at low energy. The energy corresponding to the peak of this population is function of the *MLT*. In fact, this population is more energetic at dusk than at dawn. This effect, also known as *dawn-dusk* asymmetry, will be discussed below.

The *G3* (diffusion-related) population is distinguishable by its high-energy peak at low *LS*; its shape is almost *MLT*-independent and its peak ($P_{G3}$) has a strong *LS* vs. *E* inverse dependency. This effect follows theoretical predictions and has a strong implication in the monitoring of a geomagnetic storm, as it will be better explained in section **3.2**. The *G3* populations cannot be discriminated from the *CO+G2* one at high *LS*, since there the former is generated by the latter.

The model is able to reconstruct some of the magnetospheric quantities. The velocity distribution function *f(v)* can be obtained from $\Phi_{H^+}$ (equation **B.1** in appendix **B**); then, it is possible to obtain the particle density $n_{H^+}$ and the energy density $\varepsilon$ as the momentum of order two and four of the *f(v)* distribution (equations **B.2** and **B.3**).

If the pitch angle (*PA*) distribution is isotropic, the parallel and perpendicular pressures are both equal to $\frac{2}{3}\varepsilon$ (equation **B.4**), and the current density $J_{H^+}$ is derived from pressure *P* applying equation **B.5**.

The electric field can be obtained from the mean kinetic energy $\overline{K} = \varepsilon/n_{H^+}$ if we also hypothesize that: i) protons motion is adiabatic; ii) **E**×**B** = 0; iii) **E** and **B** are time-independent; iv) all the protons are injected in a small region with similar energy.

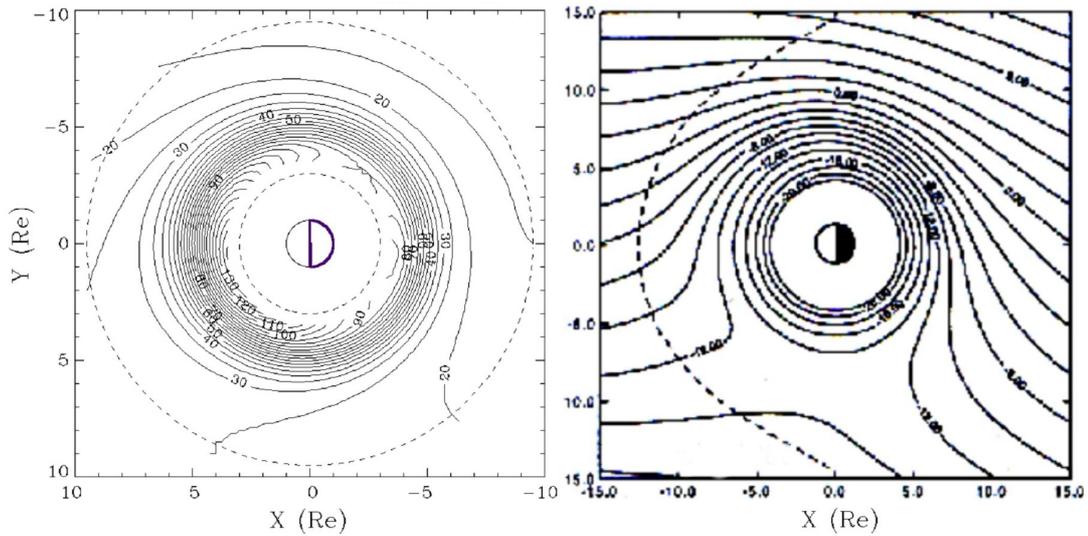

**Figure 3.2** Quiet time electric potential contours, as calculated using the model (left panel) and as predicted by McIlwain [1986].

Under hypotheses i), ii), and iii), and if all protons were injected at one point $r_0$ with *exactly* the same energy $E_0$, they all should follow the same trajectory, ruled by the guiding centre (GC) approximation (see appendix **B**). For the conservation of energy, at any point **r** reached by the protons, the energy spectrum should result in a sharp peak at energy $K(\mathbf{r})$:

$$K(\mathbf{r}) = K_0 + qV(\mathbf{r_0}) - qV(\mathbf{r}) \qquad (3.3)$$

where $V$ is the electric potential and $q$ is the charge. Under the more realistic hypothesis iv), the energy spectrum at point **r** extends to more than one energy $K(\mathbf{r})$, and equation **3.3** should be applied, as an approximation, to the *mean* kinetic energy $\bar{K}$. Under these assumptions, the map of lines of equal mean kinetic energy corresponds to the map of the equipotential lines.

Figure **3.2** shows a comparison between the electric potential estimated by using equation **3.3** and the McIlwain [*McIlwain,* 1986] electric field model associated to quiet conditions. It is evident the good agreement between these models for *LS* > 5; at lower *LS*, however, this model is not able to reconstruct the electric field. In fact, since protons

are supposed to be injected with approximately same energy, they should always have an energy spectrum that depicts a single population. Hence, this approximation is less valid when it is possible to recognize two different populations ($LS < 5$), because the diffused one ($G2$) is driven by non-adiabatic processes.

## 3.2 Modified model.

As mentioned before, it is necessary to tune the model coefficients to make the model fit a given, real magnetospheric configuration. As an example, we may imagine to detect a certain proton energy spectrum, at a given location $P_0$ ($LS_0$, $MLT_0$) and at a given time $t_0$. The situation is described by figure **3.3**; the energy spectrum is a vertical segment at $LS=LS_0$ in the panel corresponding to $MLT_0$. In principle, it is not necessary to modify all the 37 model coefficients to obtain a good agreement between the experimental spectrum and the one predicted by the model. In fact, any spectrum is the sum of two populations (injected and diffused) and it is sufficient to introduce 6 parameters that are able to modulate the peak position, intensity and width of these two populations. For simplicity, these parameters are normalized so that a value of 1 represents the *reference* condition for each parameter. These parameters are summarised in table **3.1**.

The intrinsic correlation between these pure numbers and the model coefficients is discussed in *paper* **II**; however, it is more important to understand how the modification of these parameters reflects different physical configurations of the magnetosphere.

The intensities $B$ and $D$ at some time $t_0$ represent the relative importance of the two different processes versus time, until $t_0$. The modulation of the cross-tail electric field and of the potential drop is related to the *dawn-dusk asymmetry*, which is in turn related to the difference of the energy of the (*CO+G2*) peak (factor *C*). An increase of *C* can be imagined as a translation of the whole (*CO+G2*) population towards higher energy in all panels in figure **3.1**. Since the energy scale is logarithmic, the energy shift in the dusk sector will be higher than the one in the dawn sector. In the end, the difference in mean energy between these sectors will increase.

**Table 3.1**

| Name | Meaning | Population |
|------|---------|------------|
| A | Related to energy of peak position (vs. L-Shell) | G3 Diffused |
| D | Intensity | |
| F | Temperature or width | |
| C | Related to energy peak position (vs. MLT) | CO+G2 Injected |
| B | Intensity | |
| E | Temperature or width | |

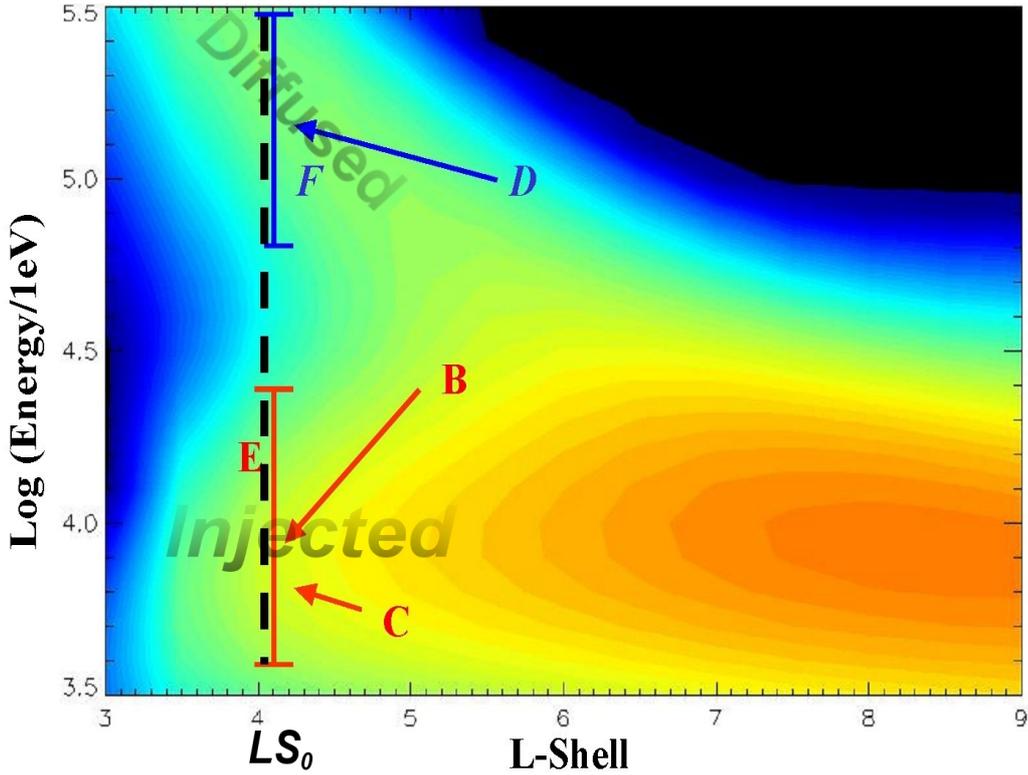

**Figure 3.3** Colour-coded H⁺ differential flux in a *LS* vs. *E* plane, for a generic *MLT*. The dashed vertical line represents an energy spectrum at a single L-Shell (*LS₀*). The coloured segments explain the meanings of the parameters *B* (intensity of *CO+G2*), *C* (related to position of *CO+G2*), *E* (temperature of *CO+G2*), *D* (intensity of *G3*), *F* (temperature of *G3*).

In principle, there is no way to modify the peak position of the diffused population *G3*. This value, actually, follows the theoretical and empirical equation [*Ejiri,* 1978]:

$$\begin{cases} E = \dfrac{1}{LS^3} & PA = 90° \\ E = \dfrac{1}{LS^2} & PA = 0° \end{cases}. \tag{3.4}$$

Hence, an increase (decrease) of the *G3* peak energy can be obtained only if we move our point $P_0$ to lower (higher) *LS*. Alternatively, we may imagine to be at a fixed point $P_0$ and to observe an expanding (compressing) magnetosphere. Hence, I introduce a parameter *A*, which is defined as:

$$A = \frac{LS_1}{LS_0} \qquad\qquad (3.5)$$

where $LS_1$ is the *L-Shell* for which the spectrum is calculated. Hence, *A>1*  *(A<1)* means that we have to move our point outward (inward) in order to find the proper energy spectrum, or that the whole magnetosphere has been compressed (stretched).

In summary, the model is able to reconstruct any given magnetospheric configuration, provided that the six chosen parameters are opportunely tuned.

## 3.3 Imaging the equatorial magnetosphere during a substorm

I have shown that, given an experimental energy spectrum at a certain point $P_0$, it is possible to find a set of parameters for which the model fits the experimental data. Once the *best-fit* parameters have been found, the global image of the magnetosphere can be obtained by using the model (this procedure may be performed iteratively if we have several spectra at different times and different positions). Now, the tasks I want to address myself to are the following:

1) to validate the $H^+$ model with experimental data, by verifying that it is possible to reconstruct any real magnetospheric spectrum;

2) to reconstruct the time-development of the 6 chosen parameters, and discuss it in the frame of the interpretation given in section **3.2**;

3) to use the 6 parameters to obtain the whole equatorial magnetosphere configuration (ring current, electric field, plasma density and spectra).

This method has been applied to proton energy spectra, collected by CRRES MICS/LOMICS experiment along the spacecraft orbit in the midnight-dusk quadrant between L ~4 and ~9, energies between 4 and 400 keV. Figure **3.4** shows the best-fitted parameters, function of time, during July 1991 storm. The coloured lines are the running-window time-averages of the parameters; the running-window covers one orbit (10 hours). In addition, AE, KP and Dst indexes are plotted to help the discussion. In the development of these indexes it is possible to recognize two main storm phases (approximately between DOY 190-191 and between DOY 194-195).

Panel **3.4-B** shows the development of parameter *B*. The good correlation between the

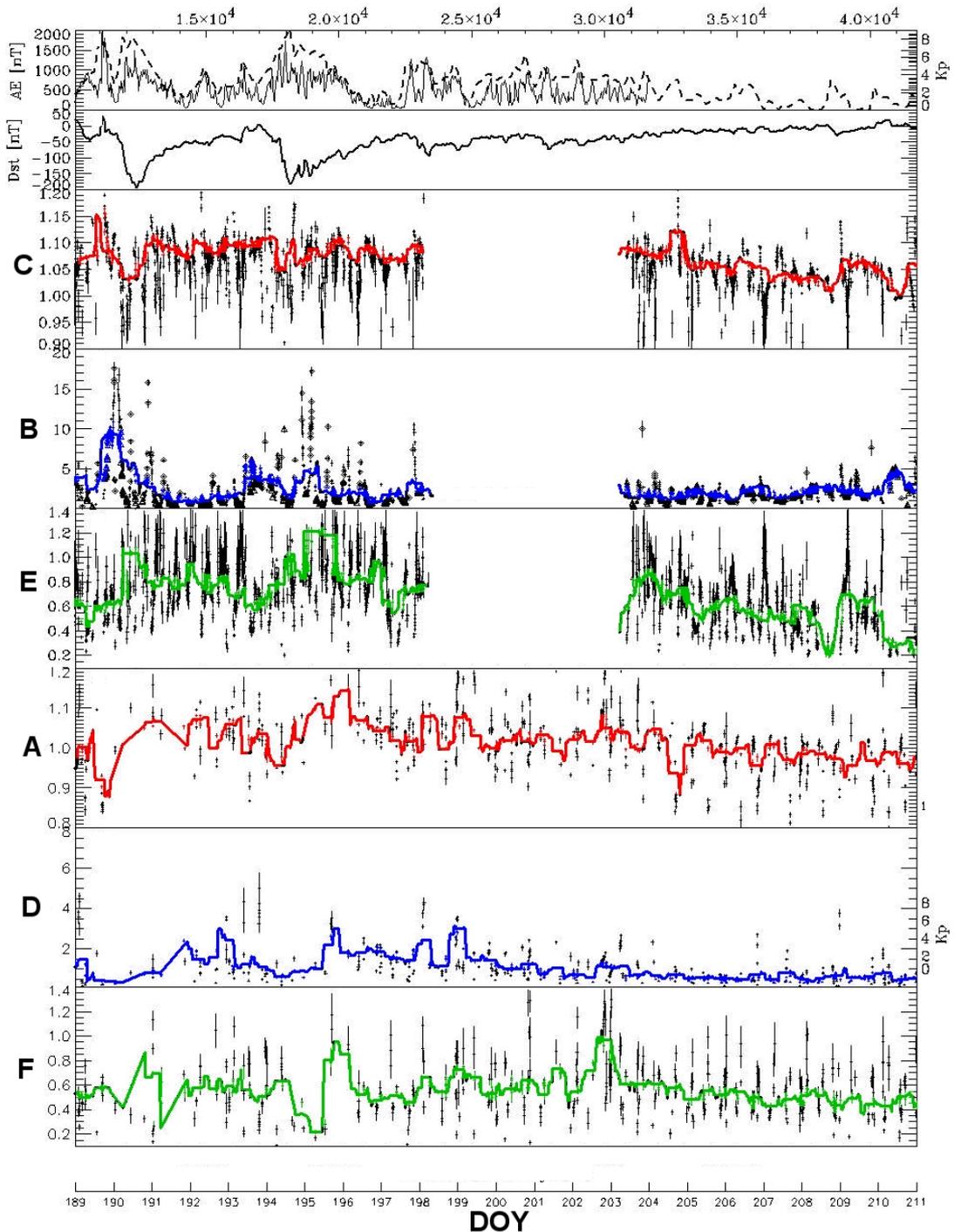

**Figure 3.4** Parameters *C, B, E* (position, intensity and temperature of *CO+G2*) and *A, D, F* (position, intensity and temperature of *G3*) vs. time during July, 1991 storm. Blank interval in first three parameter curves is due to instrumental failure. Dst, AE and *Kp* are shown in the first two panels.

Dst depletions and *B* increases indicates that the model is able to reconstruct the injection and convection of new, fresh particles in the magnetosphere. A similar effect can be recognized in the width of the injected population (parameter *E*). After these processes decrease, loss processes such as charge-exchange rapidly erode the H+ signal at low energy.

Parameter *A* is shown in panel **3.4-A**. During the evolution of a storm, the magnetosphere stretches (*A* <1) before main phase development, then depolarises (*A*>1) just some time after injection of fresh particles [e.g. *Lui*, 2000], and finally slowly gets back to quiet time configuration (*A*=1) during the recovery phase. Actually, two major increases of *A* can be noted just after the two main storm phases, indicating that *A* is a good tracer of the magnetospheric instantaneous geometry.

Parameter *D* is shown in panel **4.3-D**. The intensity of the diffused population is expected to decrease during the main phase of the storm, when this population is removed from the magnetosphere due to the opening of the electric field and of the closed drift paths [*Ebihara and Ejiri*, 1998; *Liemohn et al.*, 1999]. After the main phase, the new protons start to diffuse and *G3* intensity reaches its maximum. As protons get closer to the Earth, however, they reach regions where exospheric densities are higher, and generally several loss processes affect the particle distributions. There, the charge-exchange, as well as other loss processes, starts removing particles, and parameter *D* reacts to that by exhibiting a slow decrease. The reconstructed development follows exactly this theoretical prediction. Moreover, this interpretation explains a little difference between the two analysed recovery phases. In the first one, the *G3* has enough time to reach its maximum, but not enough to decay back to the *reference* model; the second recovery phase is longer, and *D* decreases in an approximately exponential way.

Once the six functions *A*(*t*), *B*(*t*) … *F*(*t*) have been reconstructed, it is possible to use them as an input for the model and obtain all the magnetospheric quantities at any time during the storm. Before attempting such a data-analysis, however, it is necessary to stress two considerations: i) experimental data set has been time-averaged, and since the satellite spends the most of its time in the outer part of the magnetosphere, the spectra collected in this region give a higher contribution to the fitting process; ii) during the injection phase, *CO+G2* vs. *LS* deviate from the model, because electric field shielding decreases and particles penetrate deeper into the inner magnetosphere. For these reasons, during storm time, the inner part of the magnetosphere cannot be properly reconstructed until model improvements are done.

In figure **3.5** I show the global reconstruction of mean kinetic energy (electric potential, panels **A, B, C**), energy density (panels **D, E, F**) and perpendicular current density (panels **G, H, I**) at three different stages of the storm.

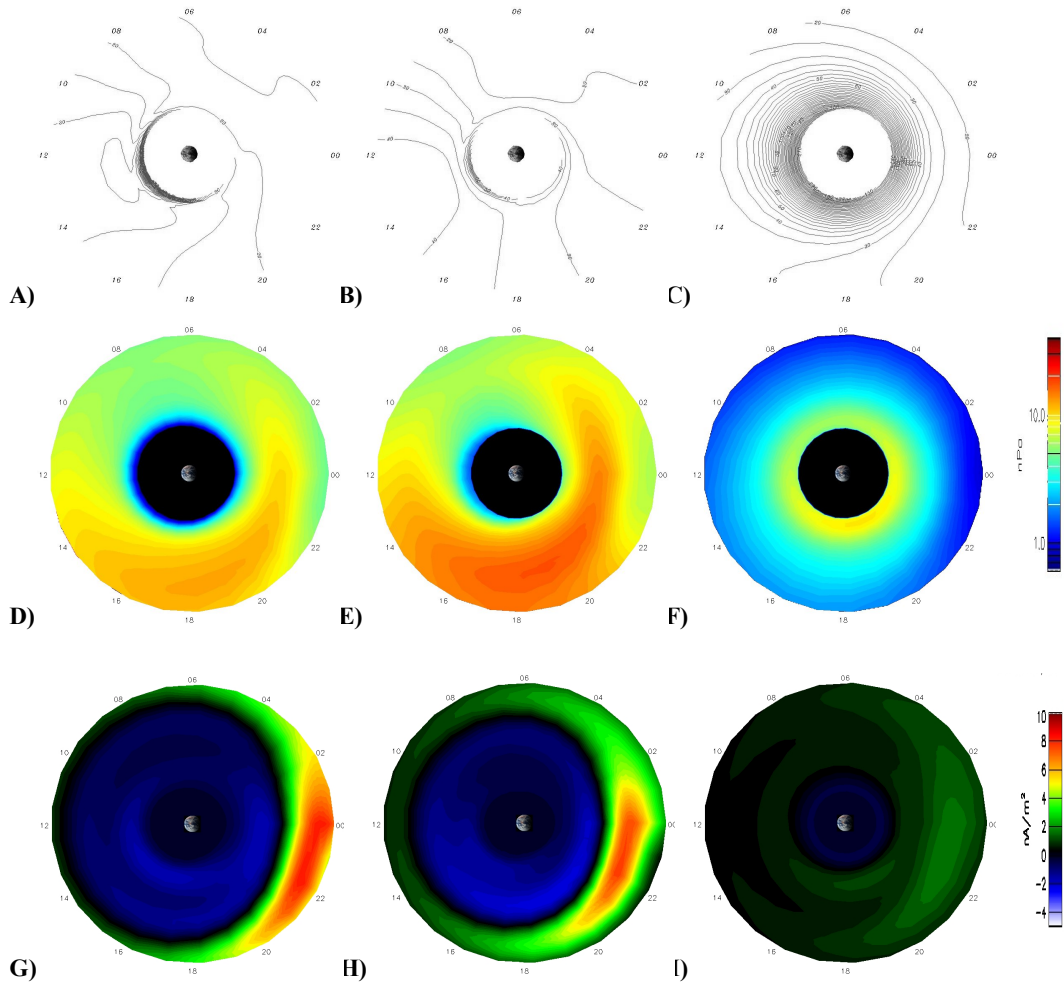

**Figure 3.5** Global reconstruction of electric potential (panels **A, B, C**), energy density (panels **D, E, F**) and perpendicular current density (panels **G, H, I**) at three different stages of the storm. Panels **A, D, G**: initial phase of the storm; panels **B, E, H**: main phase; panels **C, F, I**: recovery phase.

Panels **A, D, G** refer to the initial phase of the storm (*sudden commencement*), just before the *main phase*. The cross-tail electric field penetrates into the inner part of the magnetosphere; the energy density intensifies at dusk because protons flush in that region and escape through the afternoon magnetopause [*Ebihara and Ejiri*, 2000]; at the same time the tail current intensifies and is directed towards dusk.

During the *main phase* (panels **B, E, H**) the electric field is still *dawn-dusk*; the energy density is even higher because particles are being populating the inner-dusk regions; the current density shows the depolarisation of the magnetosphere, as it is localized at lower *LS*.

During the *recovery phase* (panels **C, F, I**) the co-rotating electric field shields the inner regions, the energy density becomes symmetric and the ring current closes around the Earth.

In conclusion, this approach provides a unique tool to understand the dynamics of the inner magnetosphere, by taking into account that any given situation is driven by previous effects which are recorded and must be considered for a correct interpretation. I've already stressed that ENA images are in principle able to depict such real conditions, and a comparison with this model could be able to give the dynamical time profile that has led to the configuration photographed by ENA images (see also final comment in section **2.1**).

## 3.4 A new design for a Medium-High Energy Neutral Detector

In this chapter I have discussed main magnetospheric features, and I have introduced an investigation method that could be considered as alternative or complementary to ENA remote-sensing.

Before leaving this chapter, I will briefly discuss about ENA instrumentation applicable to the Earth's magnetosphere. At present, several neutral particle detectors are part of the payload of spacecrafts actually orbiting around the Earth (IMAGE) or travelling to other planets (Cassini). Different detection techniques have been applied, depending on the energy of the detected particles. A summary of ENA flying detectors can be found, for example, in [*C:son Brandt* 1999] and [*Wurz,* 2000]. Here, as an example, I will introduce a new design of a medium-high energy neutral atom (MH-ENA) detector to be possibly located onboard the International Space Station (ISS): NAOMI (Neutral Atom Observer and Magnetospheric Imager).

The scheme of this sensor is shown in figure **3.6.** The instrument has energy, mass and direction resolutions, and is suited for ENA particles in the 1 keV-300 keV energy range.

The detector consists of a pinhole-focusing ENA camera. First, high-voltage deflection plates remove the ion component of the entering flux and collimate the ENA flux. Then, the collimated neutrals pass through a thin carbon foil (CF). There, secondary electrons are emitted and the neutrals are ionised with a certain efficiency, depending on their energy.

The carbon foil is placed just above a double grid, which is charged at some electric potential $\Delta V$. The electric field that is generated between grids causes the electrons to be accelerated upward; they are detected by a start-MCP (micro channel plate). At the same time, the positive-ionised fraction of the incoming neutrals is post-accelerated and deflected downward; depending on their energy, these ions impact onto a stop-MPC or onto a solid-state detector (SSD).

For each event, it is possible to reconstruct the direction of the incoming neutral by analysing the position of the electron signal on the start-MCP. Moreover, the detector is able to resolve the time-of-flight (ToF) between CF and stop-MCP/SSD.

The particles with energy less than or equal to $q\Delta V$ (some keV) are deflected and impact onto the stop-MCP. The impact position depends on the particle energy before the post-acceleration ($E$); the particle energy after post-acceleration is simply $E+q\Delta V$. This information, in addition to ToF, provides energy and mass resolution. The particles with energy well above $q\Delta V$, as well as the non-ionised particles, travel more or less along straight lines and are collected onto the SSD. In this case, it is possible to roughly determine the particle energy in the ToF chamber by means of the pulse-height analysis on the SSD, and hence the mass of the particle. However, the initial particle energy (before CF) is not known, since is it not possible to determine the charge of the particle, and whether it has been post-accelerated or not.

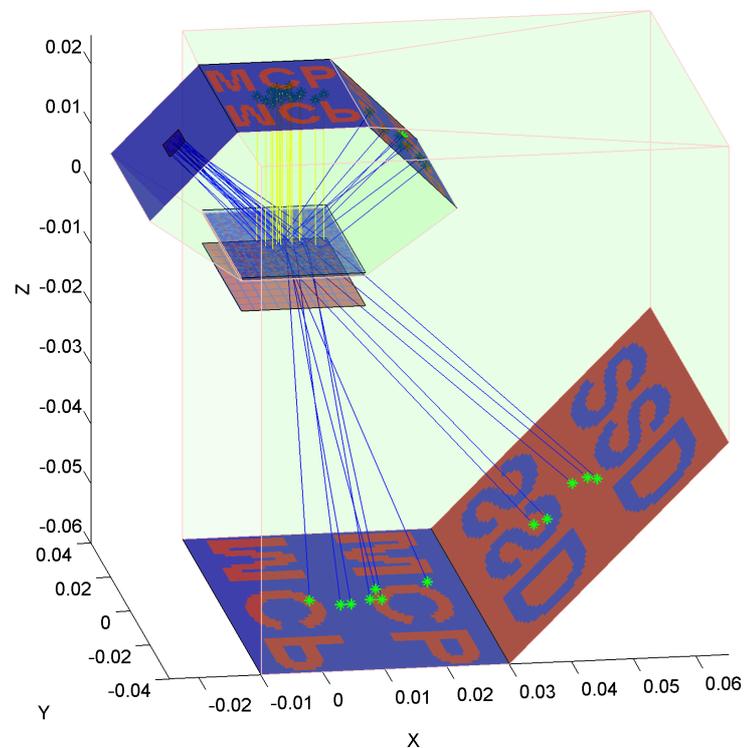

**Figure 3.6** NAOMI scheme and simulated trajectories. Red and blue surfaces are respectively stop-SSDs and start-MCPs. Grid and carbon foil are in the middle of the figure; blue lines are protons' trajectories; yellow lines are electrons' trajectories.

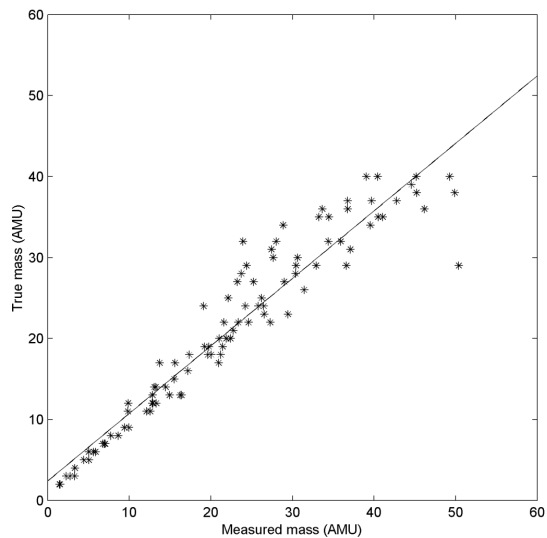

**Figure 3.7** Scatter plot of true mass versus measured mass in a simulated detector. The error on the mass is calculated taking into account only the ToF.

As an example, figure **3.7** shows a scatter plot of measured vs. true mass in a simulation of the instrument, for particles of about 1 keV, and with $\Delta V$=20 kV. The error on the mass resolution has been computed taking into account only that on the ToF.

# 4. Mars and Phobos: waiting for Mars Express

This part of the thesis is dedicated to Mars/Phobos environment in the frame of the forthcoming ESA mission Mars Express (MEX). More particularly, I will present here an application of ENA imaging at Mars: the investigation of the Oxygen population generated from natural satellite Phobos. In fact, the existence and the magnitude of this population could give precious information about the evolution of this moon [*Fanale and Salvail*, 1989]. This study has been already published, and the details can be found in *paper* **III**; here I will summarise the principal topics and the results.

First, I will introduce some empirical models for proton plasma and neutral population at Mars, and a dedicated model for Phobos-originated Oxygen population. These neutral atoms are supposed to outgas from Phobos and fill a torus all along its orbit, but the existence of such a torus is still debated. So far, only indirect evidences of it have been found [*Paper* **III**]. The ENA imaging technique would be a useful investigation tool, since in next few months new data from ASPERA-3 sensor on board MEX will be available. While waiting for these data, the feasibility of such an investigation is discussed here.

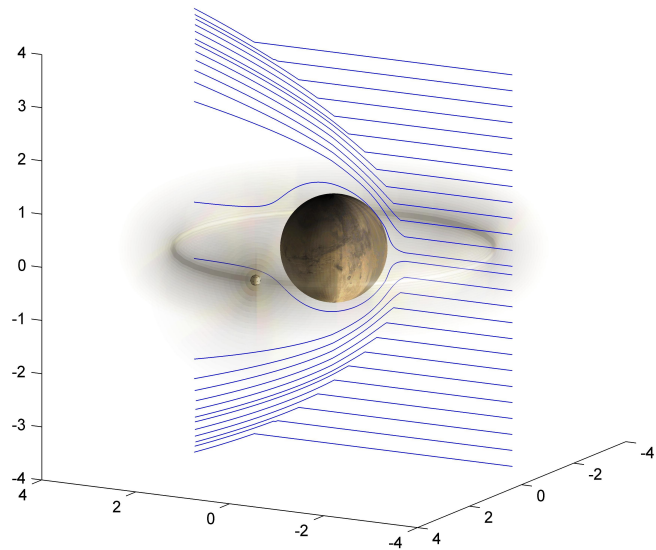

**Figure 4.1** Interaction of solar wind with Mars. Blue lines represent the proton trajectories as calculated by Kallio [1997]. The ring around the planet is a reconstruction of the Phobos Oxygen torus as it has been simulated in *paper* **III**. Dimensions of Phobos are not on scale.

Last but not least, I will present here some fundamental concepts of ASPERA-3 instrument as well as the first ENA data from ASPERA-3.

## 4.1 Mars environment

*Introduction.* The environment of Mars, the last of the terrestrial solar system planets, could have been for many aspects similar to the Earth's one, because of comparable planetary dimensions and distances from the Sun. The main difference is probably the absence of a global magnetic field at Mars: this single lack leads to an enormous amount of consequences in the present and in the past of Mars' environment.

Since the solar wind interacts with Mars mainly through direct impact with the ambient atmospheric/exospheric gas [*Kallio et al.,* 1997], charge-exchange is expected to be very effective. ENA imaging is hence a potentially useful tool to understand the geometry, the physics, and the dynamics of the plasma around Mars.

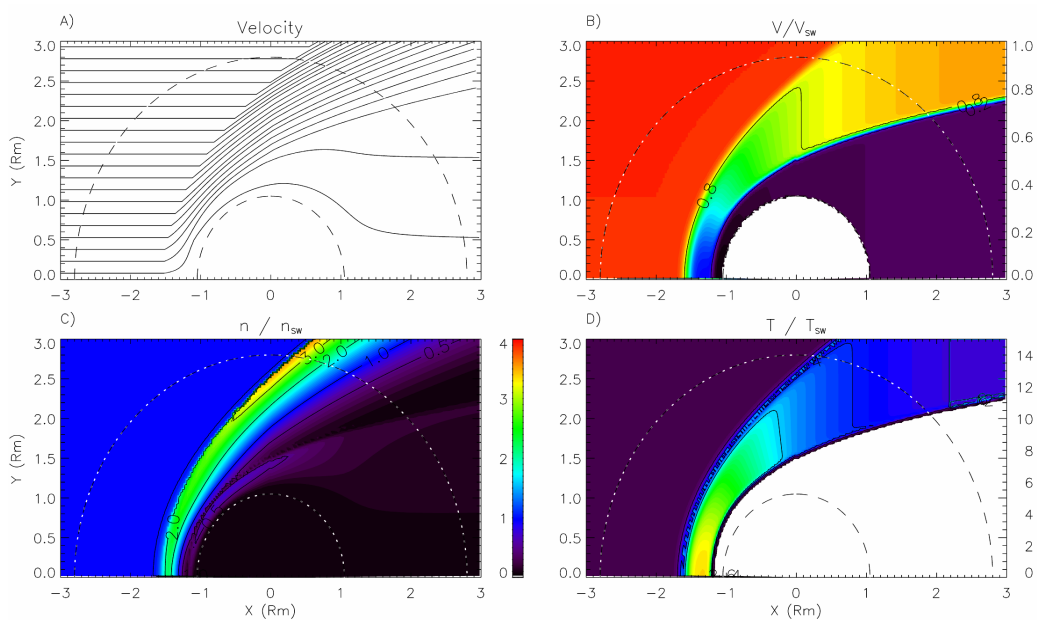

**Figure 4.2** Empirical model of the proton flow, based on ASPERA three-dimensional proton velocity measurements [*Kallio,* 1996]. Panel A) and B): proton streamlines and velocity, panel B) proton density; panel D) proton temperature. An axial symmetry around Sun-Mars axis has been postulated. The inner circumference shows the position of the planetary obstacle boundary; the outer one shows the position of Phobos' orbit.

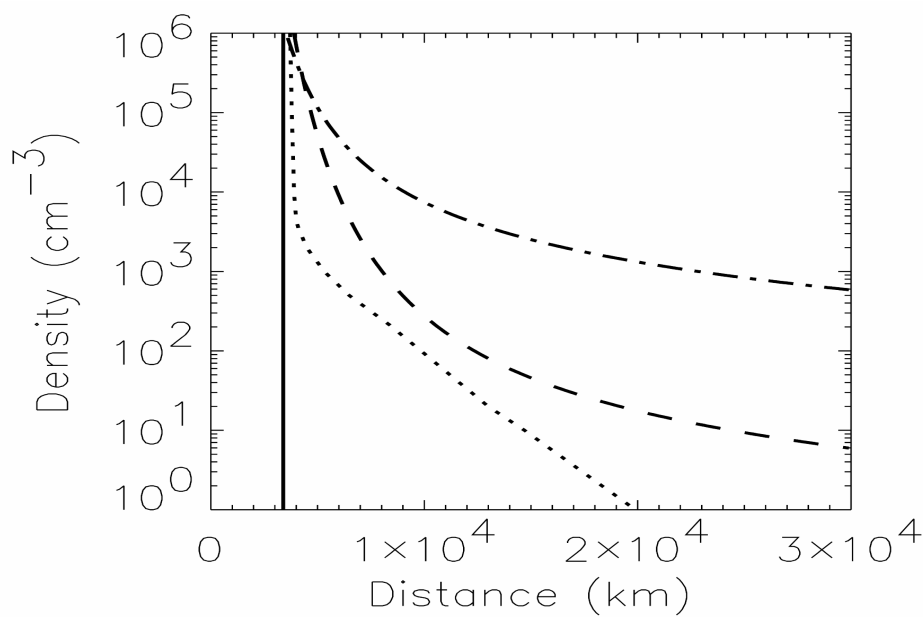

**Figure 4.3** Neutral particle density profiles in the Martian exosphere versus altitude: Oxygen (dotted), Hydrogen (dash-dotted), and molecular Hydrogen (dashed), from *Krasnopolsky and Gladstone*, 1996. The solid line indicates the Mars surface.

*Mars "Magnetosphere" and neutral population.* A schematic of Mars "magnetosphere" is shown in figure **4.1**. Some examples of proton trajectories (blue lines), as well as Phobos' torus, are shown. It could be useful to define specific regions of interaction [e.g. *Vaisberg et al.*, 1990]. Going from the outside towards the planet, we find:

1) a bow shock, where the solar wind starts to be deflected;

2) a "magnetosheath", a region between the bow shock and the magnetopause, where the solar wind particles are effectively deflected and mixed to planetary particles;

3) a "magnetopause", where the plasma speed decreases;

4) a "magnetosphere", the region inside the magnetopause;

5) an obstacle boundary, no plasma or field penetrates it.

The term "Magnetosphere", as well as "Magnetopause" and "Magnetosheath", doesn't seem to be appropriate in Mars context, since there is no magnetic field. Nevertheless, this nomenclature is widely used for consistence with the Earth topology, and for simplicity.

The plasma model used in this study [*Kallio*, 1996] considers the solar wind origin plasma as composed by protons only. Figure **4.2** shows maps of the streamlines, the velocity, the density, and the temperature of the proton flow around Mars. This empirical model, based on PHOBOS-2/ASPERA data (1989), is able to reconstruct all regions mentioned above. More particularly, if we refer to panel **b)**, the red region is the unperturbed solar wind; the blue/green/orange region is the magnetosheath; the dark blue region is the magnetosphere. Bow shock and magnetopause are the boundaries between these regions; obstacle boundary corresponds more or less to planetary surface. According to this model, half of Phobos' orbit is inside the bow shock, but its presence does not modify the model. In fact, due to its very small dimensions, Phobos is not able to deflect the proton flux on large scales.

Our simplified model of neutral population considers Mars' atmosphere and exosphere as composed only of Oxygen, Hydrogen and molecular Hydrogen. In fact, other neutral species are present (such as Helium, for example), but they have a lower charge-exchange cross-section. The neutral profiles for those populations are approximated by exponential functions of the distance from Mars and are shown in figure **4.3;** more details can be found in *paper* **III**.

## 4.2 A model for Phobos' torus

Phobos, the inner of the natural satellites of Mars, is an irregular object with dimensions of 13.5 x 10.8 x 9.4 km and a mass of $1.08 \times 10^{16}$ Kg. Its orbit is located on the ecliptic plane and is approximately circular. The orbital radius ($R_{MP}$) is 9500 km ($\cong 2.7\ R_M$) and the revolution period is about 7 hours.

The suggestive hypothesis that a dust/gas torus may surround Phobos' orbit has been suggested since 1971 [*Soter*, 1971]. Lately, after PHOBOS-2 observation of Mars in 1989, its presence has been invoked to explain some strange effects on IMF and electron density measurements [*Ip and Banaszkiewicz*, 1990; *Horanyi et al.*, 1990; *Dubinin et al.*, 1990; 1991].

More particularly, one of the hypotheses is that Phobos may contain water [e.g.: *Fanale and Salvail*, 1989], and that surface outgassing of Oxygen may occur. The escaping particles should have enough energy to escape the attraction of Phobos, but not that of Mars. They are supposed to travel in ballistic orbits around Mars, thus leading to the formation of the neutral component of the above-mentioned torus. In any case, the presence of such a population is still an open question: if this Oxygen population is really present, which outgassing rate is responsible for it? A possible value, which could explain the PHOBOS-2 disturbances, is $10^{23}$ s$^{-1}$ [*Sauer et al.,* 1995; *Paper* **III**]; so far, no direct observation of this population has been successful. The other hypothesis (a dust ring), however, seems to be less probable since a direct observation of a possible dust ring around Mars [*Duxbury and Ocampo*, 1988] didn't evidence any significant signals.

It is worth noting that Phobos' orbit is partially embedded both in the Martian magnetosheath (in the dusk and dawn sectors) and in the magnetosphere (in the night side). Hence, the outgassing material around the orbit may interact not only with the undisturbed solar wind, but also with the plasma flowing around Mars [e.g.: *Barabash and Lundin*, 1994]. In my analysis, I will focus on the charge-exchange interaction between the Oxygen component of this torus and the energetic protons, and I will discuss on the feasibility of a detection of this population via ENA imaging.

Theoretical studies [*Krimskii*, 1992] suggest that the torus cross section is approximately circular, and that the mean O density $<n_o>$ inside is $\sim 10^4$ cm$^{-3}$. Nevertheless, this value should necessary be higher close to Phobos. In this case, values up to $n_o = 10^6$ cm$^{-3}$ could be found [*Paper* **III**]. In fact, under the hypothesis of a free expansion from Phobos, the oxygen density near the satellite is:

$$n_o = \frac{dN}{dV} = \frac{dN}{4\pi r_P^2 \, dr_P} = \frac{1}{4\pi r_P^2} \frac{dN}{dt} \frac{dt}{dr_P} = \frac{Q}{4\pi r_P^2 \, v_n}; \qquad \textbf{(4.1)}$$

where $r_P$ is the distance from Phobos and $v_n = dr_P / dt$ is the mean particle velocity while escaping from Phobos. This latter value can be obtained by using $v_n = (2 \, k \, T_P /m_O)^{1/2} = 500 \, m \, s^{-1}$, where $T_P = 250$ K is the surface temperature of Phobos, and $m_O$ is the oxygen mass [*Ip and Banaszkiewicz*, 1990; *Krymskii et al*, 1992].

Some preliminary ENA simulations were done using this value, and the result was promising: ENA imaging should be able to investigate the outgassing rate from Phobos. Then I've decided to improve the accuracy on the estimation of the O density function inside the torus, by means of an *ad hoc* numerical, Monte-Carlo model. More than $10^7$ test particles trajectories have been tracked, taking into account gravitational and Coriolis force in the non-inertial frame of Phobos. The test-particle trajectories have been reconstructed with a special mix of analytic and numerical treatment, since full-analytical trajectory solution in a non-inertial frame is not easy (more details can be found in *paper* **III**; used equations are in appendix **B**). The model includes a random source for Oxygen particles, and all possible loss processes.

The result, in terms of neutral distribution $n_o$, is shown in figure **4.4**. Here I adopt cylindrical coordinates: $r$ is the distance from Mars, $\phi$ is the angle from Phobos direction, $\Delta r$=$r$- $R_{MP}$. The two panels refer to two distinct bi-dimensional sections: [$\phi$, $\Delta r$, $z$ = 0] (Panel A), and [$z$, $\Delta r$, $\phi$ = 0] (Panel B). The O density $n_o$ in a small region surrounding Phobos reaches values up to $2 \cdot 10^5$ cm$^{-3}$. In the remainder of the torus $n_o$ is almost $\phi$-independent, with a peak of $\approx 3 \cdot 10^4$ cm$^{-3}$ located at $\Delta r$=0.

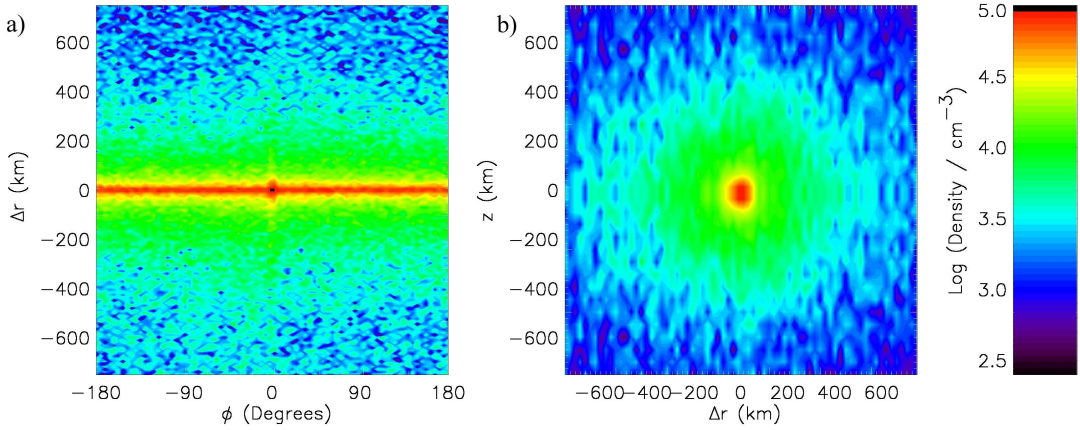

**Figure 4.4** Simulated neutral Oxygen density near the orbit of Phobos for different bi-dimensional sections: $\phi$ versus $\Delta r$, $z$=0 (Panel A); $\Delta r$ versus $z$, $\phi$=0 (Panel B); $Q = 10^{23}$ s$^{-1}$. Phobos position is at $\Delta r$=0, $\phi$=0°, $z$=0; its velocity is directed toward the increasing $\phi$.

An important parameter for the Oxygen distribution simulation is lifetime $\tau$. There are three important causes of O-atom losses in Phobos' torus: charge-exchange, photo-ionisation and electron impact ionisation, and all of them cause relevant losses. A zero-order estimation, using solar wind plasma parameters, leads to a charge-exchange loss frequency $f_{CE} = 10^{-7}$ s$^{-1}$ [*Paper* **III**]. This value is comparable to other loss frequencies. In fact, at Mars' orbit and during solar minimum, the photoionisation loss frequency is $f_\gamma = 10^{-7}$ s$^{-1}$, and the electron impact ionisation frequency is $f_e = 2 \cdot 10^{-7}$ s$^{-1}$ [*Zhang et al.,* 1993]. The total loss frequency $f$ is the sum of all these frequencies ($f = 4 \cdot 10^{-7}$ s$^{-1}$), and I have hence considered an oxygen gas lifetime $\tau = f^{-1} = 2 \ 10^{6}$ s.

## 4.3 Mars ENA imaging

If we want to study the outgassing rate of Phobos by means of ENA imaging, the ENA flux coming from Mars environment should be considered as "background", even if it is, actually, the major objects of ASPERA-3 investigation. The estimation of such a background is a preliminary step for this investigation.

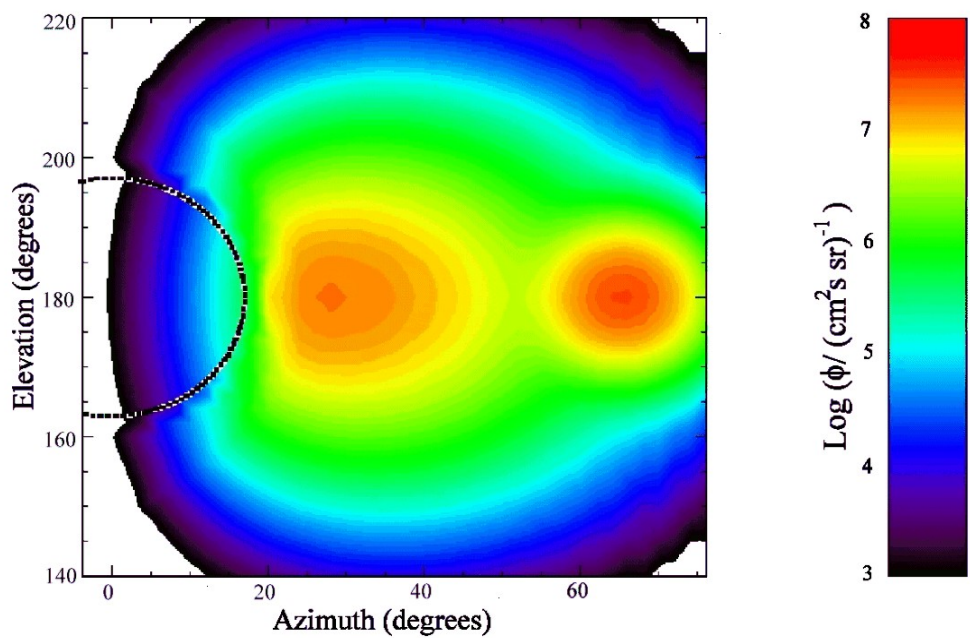

**Figure 4.5** Pseudo-color map of the ENA differential flux, function of the elevation and azimuth angle, if only Mars exosphere was present. The dashed line represents Mars' obstacle boundary. Azimuth angle is measured on the equatorial plane, beginning from Mars' centre; elevation angle is measured perpendicular to the equatorial plane, Mars' centre is at 180°.

The ASPERA-3 package on board MEX includes a Neutral Particle Imager (NPI), with good angular resolution but no energy resolution, and a Neutral Particle Detector (NPD), with low angular resolution (more details in section **4.5**). Hence, for some realistic vantage point configurations, I have simulated some ENA images and the related ENA energy spectra.

An example of Mars' ENA image is shown in figure **4.5**. The vantage point, in this case, is at (-1.3 $R_M$, 2.9 $R_M$, 0) in Mars-Solar-Ecliptic (MSE) coordinates, and it is schematised as a grey point in figure **4.6**. The dotted line represents Mars' obstacle boundary. The ENA differential flux has been calculated by using line of sight integration (equation **2.5**), and then it has been integrated over all energies (from 10 eV to 2 keV). Two peaks can be recognized in the image: a narrow and circular one centred on the sunward direction, and a more extended one closer to the planet. The first is related the ENA flux generated outside the bow shock by the unperturbed solar wind; the second is related to the ENA flux generated in the magnetosheath, by the thermalised solar wind. Figure **4.7 (**solid line**)** shows the ENA spectra relative to Figure **4.5**. The spectrum related to Mars' ENA flux (solid line) is the envelope of the fluxes from all regions. The low energy flux (below 400 eV) is generated very close to Mars' surface (see figure **4.2**); the high energy flux (0.4-2 keV) results both from unperturbed solar wind outside the bow shock, and from deflected solar wind in the magnetosheath.

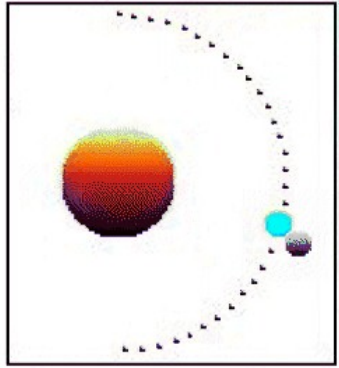

**Figure 4.6** Scheme of the relative Mars-Phobos-ASPERA position, as used for ENA images simulation of figures **4.5, 4.8, 4.9.**

## 4.4 A simple deconvolution: Phobos' outgassing rate

The hydrogen ENA flux generated at Mars can be considered as the sum of two: the one coming from Mars' exosphere and the one coming from Phobos' torus. The first is almost always detectable; the latter, due to the small dimensions of Phobos' torus (see figure **4.4**), is not. In fact, the integration path is, in most cases, too small to produce a noticeable ENA flux. The most favourable configuration for this latter signal occurs when the vantage point is located on Mars' equatorial plane and close to Phobos. During MEX mission, three useful Phobos' close approaches are planned [*Barabash, private communication*].

Figure **4.8** shows the ENA flux generated by Phobos' neutral population, with the exclusion of the background signal reported in figure **4.5**, assuming hypothetically that $Q = 10^{23}$ s$^{-1}$ (see previous comments in section **4.2**).

The ENA image is of course concentrated on the equatorial plane. Here, the peak intensity is higher than the "background" in the same location. However, if we look at the ENA spectrum (figure **4.7**, dashed line), we note that it is well below the background at any energy. Hence, such a measurement is possible only via NPI. I have also simulated ENA images and spectra for other different Mars-Phobos-vantage point configurations, but if the vantage point is placed outside the equatorial plane the ENA flux from Phobos is not detectable. This latter fact suggests a way to make the measurement of $Q$.

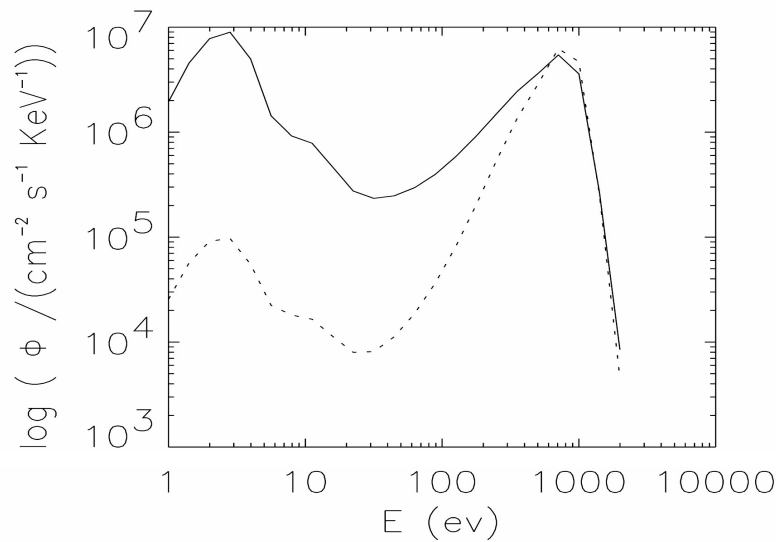

**Figure 4.7** Spectra of the ENA fluxes coming from Mars (solid line) and Phobos (dotted line), integrated over all directions in the sky. Vantage point and Phobos position are the same as in figure **4.5.**

The polar orbit of MEX crosses the equatorial plane at approximately right angles. While approaching Phobos, its period is roughly the same as Phobos' one (7 hours). Even if the torus virtually extends to infinity, the region where O density is noticeable is very thin, i.e. less than 200 km. Hence, ASPERA-3 will be at a "good" vantage point position only for one or two minutes. During this time, the ENA image detected from NPI should be similar to the superposition of figures **4.5** and **4.8**. The result is shown in figure **4.9**. Just before (or just after) this crossing, the ENA image should be the one of figure **4.5**. The difference between those images is the ENA flux coming from Phobos' torus. Since this flux is simply proportional to the O neutral density $n_o$, a rough estimation of $Q$ is possible.

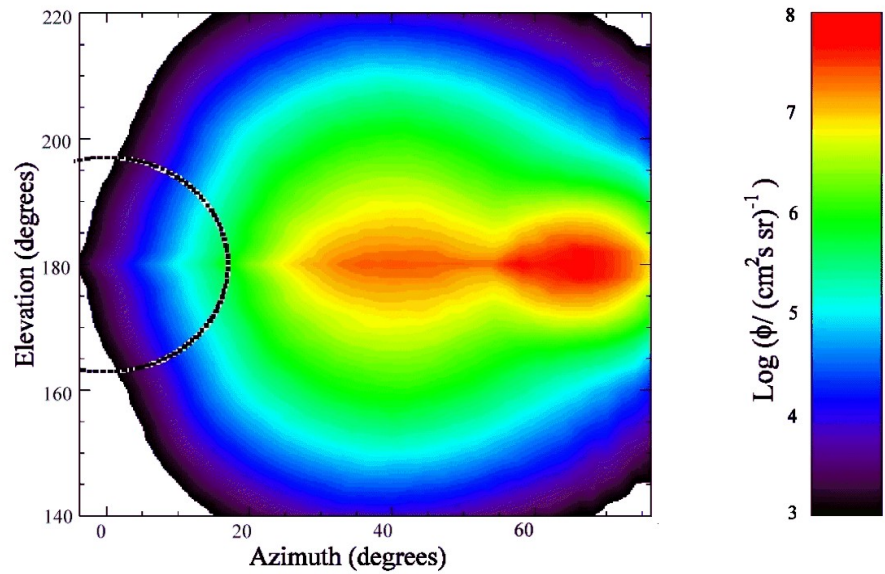

**Figure 4.8** ENA image simulation: same as **4.5**, if only Phobos torus-halo was present.

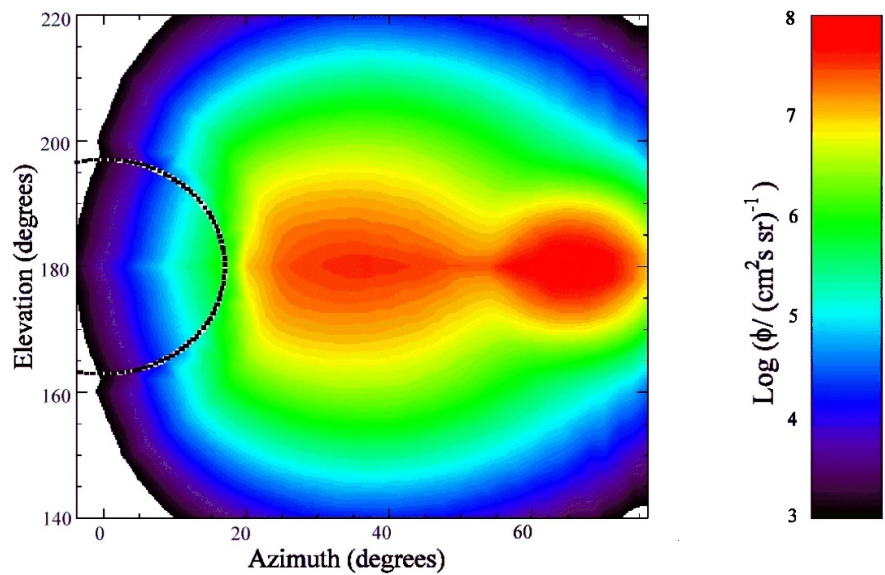

**Figure 4.9** ENA image simulation: overall ENA image resulting from the superposition of ENA images in figures **4.5** and **4.8.**

The very short time scale should guarantee that the external conditions do not change. In any case, this can be verified by comparing the ENA image just before with the one just after the encounter. According to my simulation, the angular resolution of NPI (4.6° × 11.5°) should be good enough to resolve the thin signal in the equatorial plane.

In conclusion, the presence of a neutral population originated from Phobos induces an extra ENA signal that should be detectable. In the considered configuration, this allows the estimation of the outgassing rate $Q$, but only if $Q$ is equal or greater than $10^{23}$ s$^{-1}$. Since I have used the most favourable conditions (vantage point/Phobos configuration), it is not realistic to imagine the detection of a smaller rate. Anyhow, this consideration does not affect the feasibility of this measure. In fact, disturbances on PHOBOS-2 data can be explained by outgassing only if $Q \geq 10^{23}$ s$^{-1}$. If this last hypothesis is right, NPI sensor will detect some extra signal/disturbance during MEX-Phobos approach, and $Q$ will be estimated in detail.

## 4.5 ASPERA-3 Instruments

During my thesis, I actively participated to the development of the ASPERA-3 experiment (Analyser of Space Plasmas and EneRgetic Atoms). More particularly, I was responsible for the development of both data display and real-time analysis software (DDU, Data Display Unit). The scientific instrumentation (figure **4.10**), part of the MEX payload, includes both ion and electron sensors, and two different neutral analysers: NPI (Neutral Particle Imager) and NPD (Neutral Particle Detector). Here I

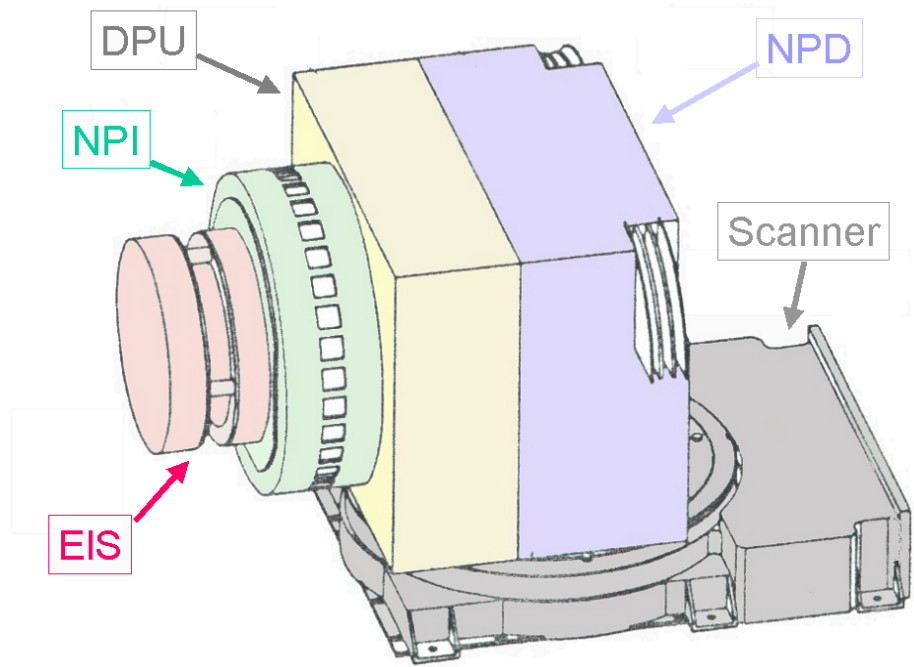

**Figure 4.10** ASPERA-3 Layout.

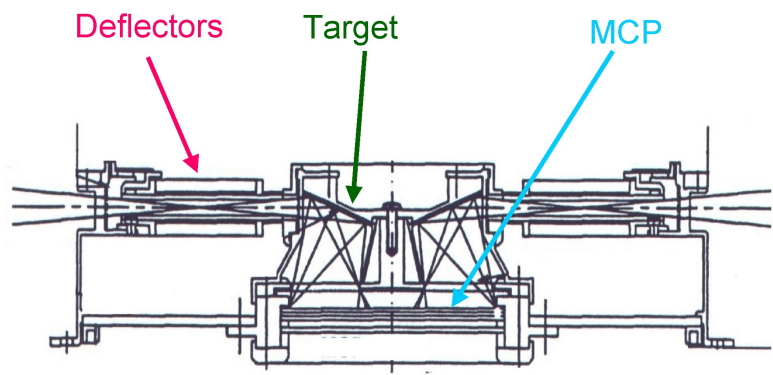

**Figure 4.11** NPI scheme.

will briefly resume some concepts about these two last instruments.

NPI is substantially an imager. It has neither energy nor mass resolution, but good angular resolution. It is placed over a rotating platform, in order to perform a 3D full exploration. It has a cylindrical shape, with symmetry axis at right angles to the rotation axis. Figure **4.11** is a section of the instrument in a plane containing the symmetry axis, showing two of the 32 sectors of the instrument.

The sensor head is provided with electrostatic deflection plates, which remove the incoming ions. The neutrals are allowed to hit a 32-faced target. The sputtered particles produced by the collision, and/or the reflected neutral, can hit the micro channel plate (MCP) below. The MCP, with 32 anodes revealing which sector has been crossed, operates in negative-bias mode, and hence can accept all projectiles except electrons. The UV entering the instrument are suppressed by using a resin-based graphite-dispersion target. A summary of the instrument characteristics is given in table **4.1**.

**Table 4.1**

| Energy Range | 0.1 – 60 KeV |
|---|---|
| Angular Resolution | 4.6° x 11.5° |
| Azimuthal sectors | 32 |
| Geometrical factor | 2.5 $10^2$ cm$^2$ sr |
| Efficiency | ~1% |
| Power | 0.8 W |
| Mass | 0.7 kg |

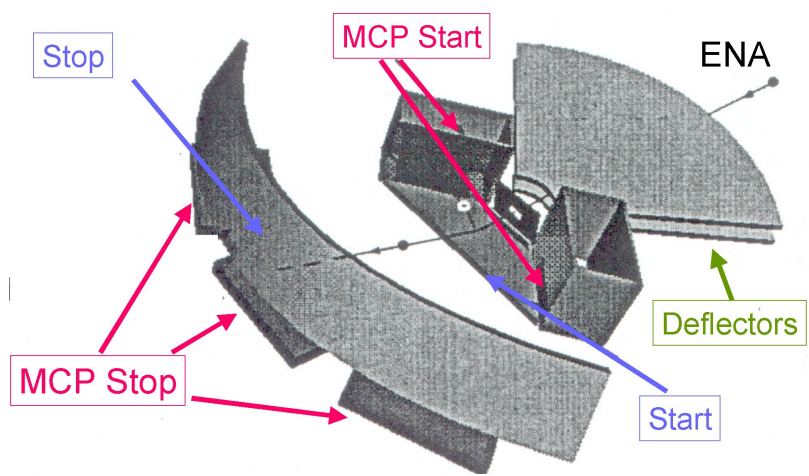

**Figure 4.12** NPD scheme.

NPD is substantially a mass and energy spectrometer, with poor angular resolution (~30°). Two NPD units are included in ASPERA-3, and placed over the rotating platform looking at opposite directions. Detecting the energy or the mass of a neutral implies that this neutral must be perturbed somehow (this consideration has a very important exception: see section **5.10**). In general, this happens either forcing it to cross a very thin carbon foil (as for MH-ENA, section **3.4**), or deflecting it by an ionising surface at very low grazing angles. The ionisation through a carbon foil is not suitable for low energy neutrals; hence, NPD uses the surface-scattering technique. Particles are not necessary ionised; upon their scattering on the surface some electrons are released and the detection of the latters makes the start of a ToF measurement. Figure **4.12** shows a 3D view of the instrument.

After the deflection plates, similar to those of NPI, the neutrals hit a graphite surface.

**Table 4.2**

| Energy Range | 100ev – 10 KeV |
|---|---|
| Energy Resolution | 16 step |
| Azimuthal sectors | 3 |
| Efficiency | ~1%-50% |
| Power | 1.5 W |
| Mass | 1.8 kg |

The result of this collision is the ejections of some electrons, which are collected on two start-MCPs using appropriate electric field. The electrons detection gives the start time. After the collision, the particle (sometimes ionised) is scattered and proceed until it reaches a stop target. The electrons ejected are collected over stop-MCP to give the stop time. The time-of-flight measurement gives the particle velocity, while the number of collected stop-electrons roughly gives the particle energy, and then the masses. Table **4.2** gives a summary of NPD characteristics.

## 4.6 ASPERA-3 First image

During ASPERA-3 development, I have been responsible of the Data Display Unit (DDU), under the direction of Dr. Stefano Orsini. After MEX launch (June 2, 2003) the instrument has been switched-on to perform some on-flight calibrations and tests. I have actively participated to these operations, and DDU s/w has been used extensively. In particular, some data have been collected during a changing in the satellite attitude. Figure **4.13** shows the DDU control panel and the NPI data collected between 18:00 and 21:20 July 9, 2003. The deflection voltage was off, i.e. most of the detected particles were solar wind ions. Figure **4.14** shows, for the same time interval, the Sun's position within the NPI sectors. The movement of the Sun's signal across sectors from 19 to 26 clearly shows that all those sectors are working fine. Conversely, is it not clear why sector 16 is counting particles, since it should be artificially blocked.

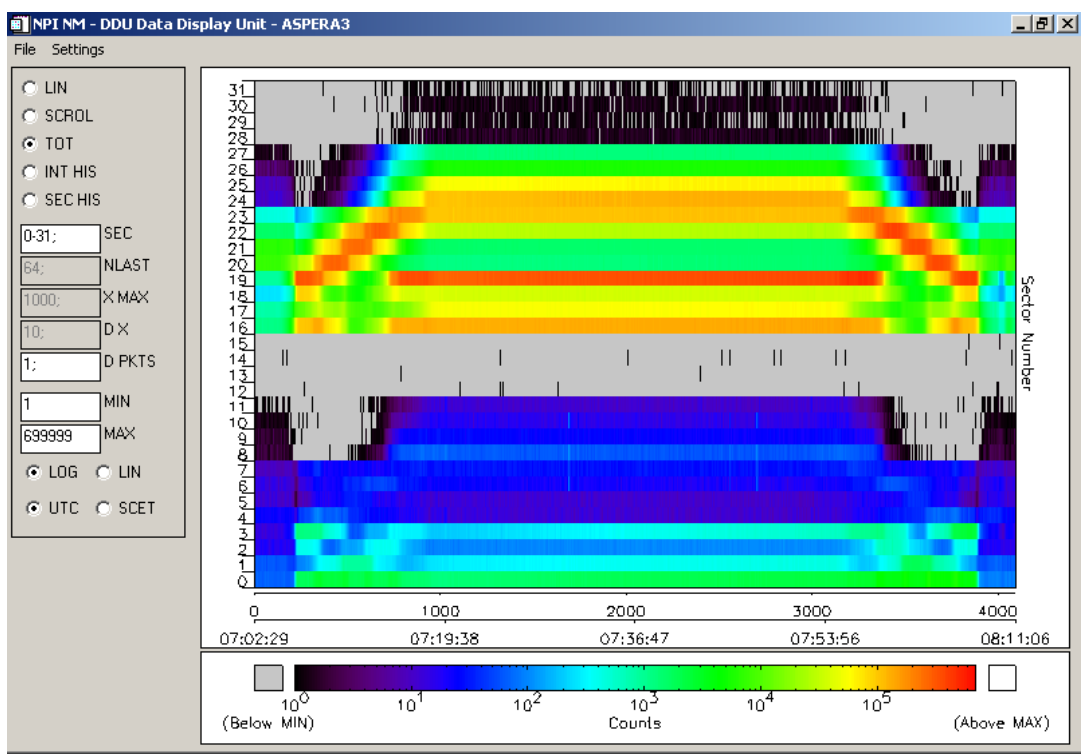

**Figure 4.13** DDU monitor with first ENA data. On *y* axis: sectors. On *x* axis: time. Colours are coded according to count rate.

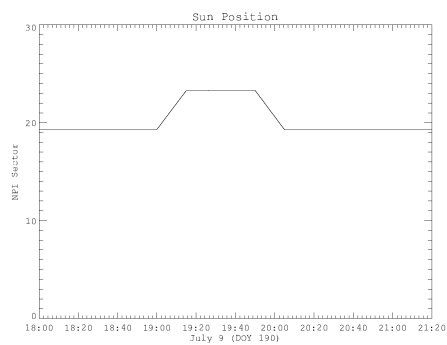

**Figure 4.14** Position of the Sun direction respect to sectors of NPI, function of time, in the same period as in figure **4.12** (from *K. Jonsson,* private communication).

After a six months' flight, Mars Express is going to be inserted into an orbit around Mars. The ASPERA instrument is about to collect and send us the first ENA data of Mars' environment.

# 5. Mercury and the ESA mission Bepi Colombo.

In this part of the thesis, I will discuss about general features of ENA imaging at Mercury. I will first consider the present knowledge of Mercury's environment (**5.1**); then I will introduce the models of Mercury's magnetosphere, exosphere and surface composition adopted for further calculations (**5.2**). In section **5.3** I will present a preliminary simulation of Mercury's equatorial proton circulation, and discuss about the existence of a ring current at Mercury. In sections **5.4** and **5.5**, a general model of proton circulation in all Hermean magnetosphere will be developed, and its results will be applied to ion-sputtering and exospheric profile simulations (**5.6**), and ENA emission simulations (**5.7**). The feasibility of neutral atom imaging will be also discussed (**5.8**), as well as the related instrument requirements (**5.9**). This study includes also a published paper (*Paper* **IV**) and a submitted one (*Paper* **V**).

The importance of this study will been enhanced, however, if we consider that the future Bepi Colombo mission to Mercury will probably include a Neutral Particle Analyser (NPA-SERENA), led by Dr. Orsini and his SERENA group at IFSI. The ESA planning of this mission has induced a big effort by the international community to produce simulations and feasibility studies. This chapter resumes my personal contribution to such an effort. Actually, it may be assumed as one of the main results of my thesis, especially because it provided a major support to the selection of SERENA.

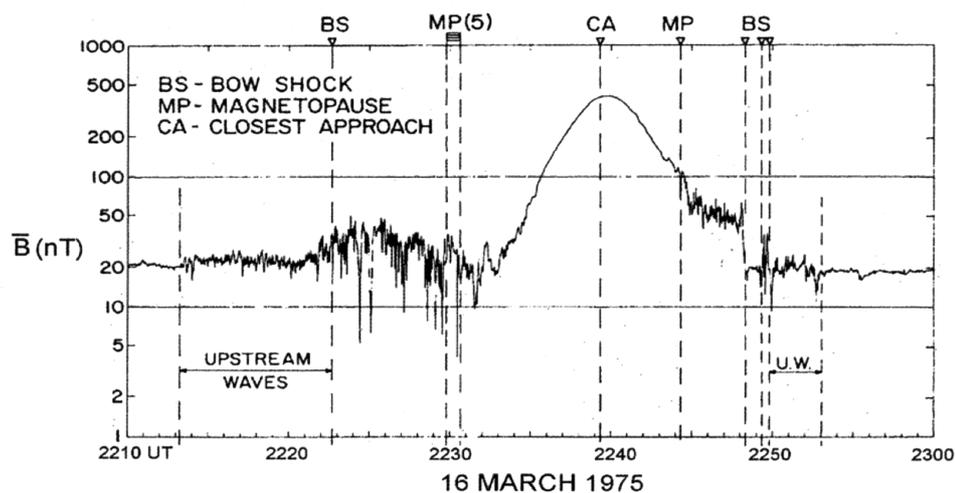

**Figure 5.1** Magnetic field profile inside Mercury's magnetosphere as measured by Mariner 10 in March 1975.

## 5.1 Mercury's Environment

*Prelude.* Why looking at Mercury? The Hermean environment is from almost all angles very different to the Earth's one. Its small distance from the Sun makes surface temperature and daily excursion extremely high; the eccentricity of its orbit causes a high variability in the solar wind conditions at Mercury; the absence of an atmosphere and the weakness of the intrinsic magnetic field result in a peculiar magnetospheric behaviour. These extreme conditions make Mercury a unique case in the solar system: Mercury can be referred to as a "laboratory planet", and all scientific results obtained at Mercury can be applied, even if not transferred, to the Earth.

*Mercury's magnetosphere.* The Hermean magnetosphere was an almost complete mystery until 1974, when the first spacecraft to approach Mercury, Mariner 10, revealed the existence of an intrinsic magnetic field (see figure **5.1**). The magnetic field estimation has large uncertainty, because it is difficulty to separate the internal and external magnetic field components [*Connerney and Ness*, 1988]. Nevertheless, the dipole moment is probably between 284 and 358 nT $R_M^3$ [*Ness et al.*, 1975], pointing approximately 10° from the South Pole. For reference, the Earth's dipole is approximately $3 \times 10^4$ nT $R_E^3$.

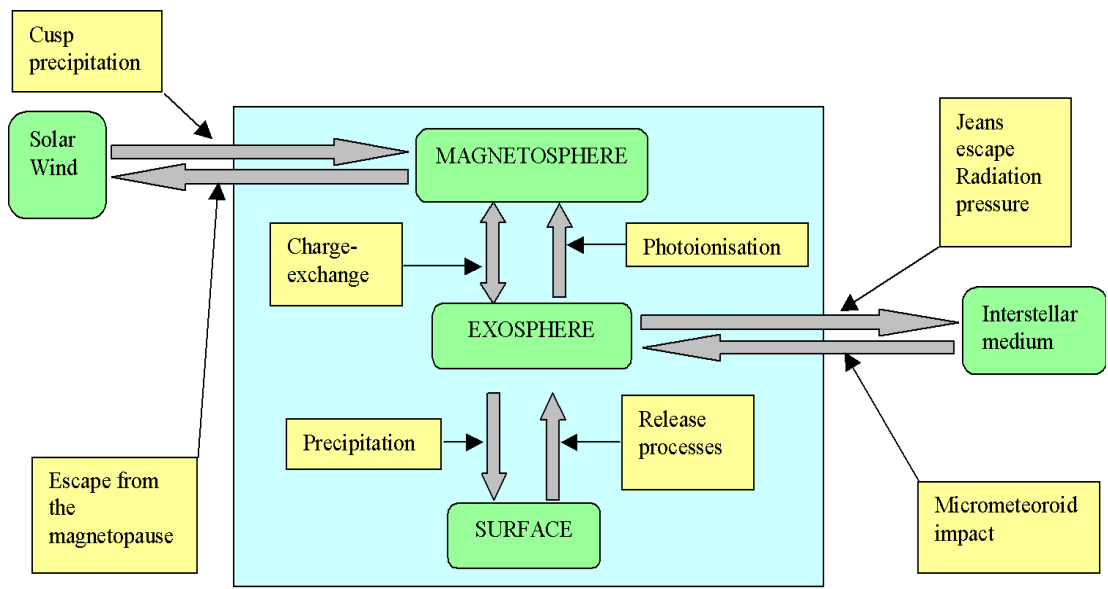

**Figure 5.2** Block diagram showing main exchange processes in Mercury's environment. [*A. Milillo, private communication, 2003*].

Peculiarities in Mercury's magnetosphere arise also from the extreme conditions of the solar wind at Mercury's orbit (0.3÷0.44 AU), which differs substantially from the average conditions present at 1 AU. Parker spiral forms an angle of about 20° with the solar wind radial direction, while it is approximately 45° at the Earth. This implies a change of the relative ratio of the IMF components with respect to the near Earth conditions, and a modified solar wind-magnetosphere relationship. The average solar wind density is about ten times higher than at the Earth, and this value varies considerably due to the high eccentricity of the orbit of the planet, going from 34 $cm^{-3}$ at the aphelion (0.44 AU), to 83 $cm^{-3}$ at the perihelion (0.29 AU). Solar wind speed is approximately 430 km/s, and the dynamic pressure is, on average, 16 nPa.

Applying all above parameters, it has been estimated that the sub-solar point, where the internal and external pressures balance together, is at the distance of 1.5÷1.7 $R_M$ from Mercury's centre [*Siscoe and Cristofer*, 1975; *Goldstein et al.*, 1981], while at the Earth this point is located at 11 $R_E$. This latter consideration may be applied while trying to simulate the Hermean magnetosphere with a scaled model of the Earth's one (see section **5.2**): with respect to magnetospheric dimensions, Mercury appears to be seven times bigger than the Earth. This scale factor has many important implications, for example, in the development of a ring current around the planet. In fact, at the Earth, this current develops at ~ 5 $R_E$, which corresponds to ~ .7 $R_M$ at Mercury, i.e. below the planetary surface. Hence, the injecting ions from the tail fall down to the planet before a complete revolution. This feature will be better discussed in section **5.3.**

Last but not least, a smaller spatial scale implies a smaller time scale at Mercury. Magnetospheric processes are ten times shorter than at the Earth (few minutes instead of approximately 1 hour) [*Siscoe et al.*, 1975], and they are highly conditioned by the solar wind parameters [*Luhmann et al.*, 1998].

*Plasma sources.* The origin of a hermean magnetospheric ion can be essentially either interplanetary (solar wind) or planetary (exospheric thermal ions). Both these sources are located in the dayside.

Solar wind plasma can enter the magnetosphere when magnetic reconnection occurs between IMF and Mercury's intrinsic **B** field [*Luhmann et al.,* 1998; *Killen et al.,* 2001]. Massetti *et al.* [*Paper* **IV**] have estimated that about one half of the magnetosheath ions that lie on reconnected field lines actually crosses the magnetopause, while the remaining part is reflected by the boundary. The ions entering the magnetosphere may either reach the planet surface, causing ion sputtering; or be diffused toward closed field lines and start circulating; or exchange their charge with the thermal exospheric atoms (see section **2.1**), thus producing a H-ENA signal in the keV range. The small inclination of Parker's spiral causes the contribution of the IMF $B_y$ component to be less relevant than at the Earth, and the magnetic reconnection at the dayside magnetopause is essentially driven by the IMF $B_z$ component. Moreover, the increasing weight of the IMF $B_x$ component might play a role in the way Mercury's magnetosphere links with the solar wind [e.g. *Sarantos et al.,* 2001, *Killen et al.*, 2001, *Kallio and Janhunen*, 2003]. The three panels of figure **5,** *Paper* **IV,** show position, extension of the open field lines area, and energy distribution mapped on the northern dayside surface of Mercury in response to different solar wind conditions.

The ions of planetary origin at Mercury may come both from the photo-ionised exospheric gas and from the ionised component of ion-sputtering product. The ion-sputtering is the dominant source in the area where solar wind precipitates. Those ions have generally an initial energy up to some eV, depending both on the process involved in the generation of the exosphere and on the photo-ionisation efficiency. Once extracted, they are trapped by the magnetic field and start to circulate within the magnetosphere.

*Mercury's Exosphere.* The UV spectroscopic observations on Mariner-10 and ground based observations of the gaseous envelope of Mercury established the presence of H, He, O, Na, K and Ca [*Killen and Ip*, 1999; *Morgan and Killen,* 1997; *Bida et al*., 2000]. Moreover, other volatile constituents such as $H_2$, $N_2$, $O_2$, $H_2$ and $CO_2$ are expected [*Wurz and Lammer,* 2003]. All those species, together, do not establish an atmosphere, having too small a column density. The gaseous envelope of Mercury is hence uncollisional, and gas atoms interact mainly with the surface and with magnetospheric plasma. A schematic summary of the refill and loss processes is shown in figure **5.2**.

The main exospheric sources are due to release processes of atoms from the surface. The released particle energy depends on the process involved. Photon stimulated desorption (PSD), thermal evaporation and micrometeoroid impact produce low energy (few eV) particles. Ion-sputtered neutrals reach higher energies (see equation **2.10**). Among the emitted atoms, some are gravitationally bound, and fall back onto the surface; the others (especially the sputtered ones) have enough energy to escape from the exosphere. The major exospheric losses are photo-ionisation, especially in the dayside hemisphere, recombination with the soil and Jeans escape.

Another process involving exospheric neutrals is charge-exchange. This process, however, is numerically less important; it produces high-energy neutrals that escape from the exosphere very fast, and replaces them with thermal ions.

*Final considerations.* Given previous argumentations, Mercury's magnetosphere has sometimes been referred to as a "pocket magnetosphere". The occurring processes are simpler than at the Earth. The whole Mercury environment/system, from the point of view of a space scientist, may be schematised as magnetosphere + exosphere + surface. The lack of knowledge about Mercury, hence, is due mostly to the lack of data, more than to the complexity of the related physics. If we look at the future, we expect to have much more data after the NASA and ESA missions Messenger and Bepi Colombo.

## 5.2 Adopted models

*Magnetic field model.* Since most of the magnetic field features are uncertain, with the exception of the mean dipole moment (see section **5.1**), in this study I have tried to reconstruct the magnetic field around Mercury using two different approaches. First, I started with a pure dipole (M=300 nT $R_M^3$), limited by a parabolic magnetopause in the dayside hemisphere (an ellipsoid: subsolar point at $x$=1.7 $R_M$).

This model has been used in the preliminary study of the equatorial circulation (see section **5.3**), but since it cannot reconstruct most of the magnetospheric features (such as open field lines, for example) it has been discarded in the following of the study (sections **5.4-5.7**). Subsequently, in order to model the dayside proton circulation (section **5.4**), I have modified a T96 Tsyganenko model, according to the study of *Massetti et al.* [*Tsyganenko*, 1996; *Paper* **IV**]. The use of such a complex model introduces the difficulty of having several parameters to guess. Its main features are: i) defined realistic magnetopause; ii) large-scale Region 1 and 2 Birkeland current systems; iii) IMF penetration across the magnetospheric boundary [*Tsyganenko*, 1996]. It accepts both IMF $B_y$ and $B_z$ as independent input parameters.

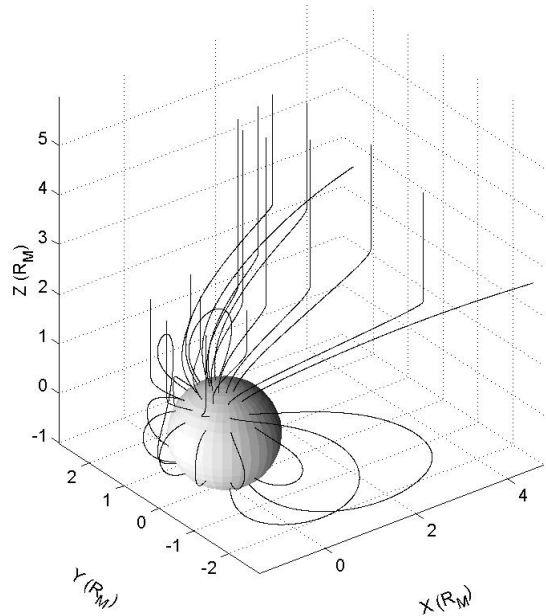

**Figure 5.3** Magnetic field as simulated by the model in the *reference* conditions (see section **5.2**).

First of all, to take into account the differences between both the intensities of the magnetic fields and the radii of Mercury and the Earth, I have divided every space constants in the T96 model by a factor 6.9, like in previous analyses based on the Mariner 10 data [e.g., *Siscoe et al.* 1975, *Luhmann et al.* 1998]. Then, I have removed from the model the contribution of the ring current, because the presence of this structure in the magnetosphere of Mercury is uncertain; a more detailed discussion of this feature will be reported in section **5.3**. I have also assumed a 50% contribution of Birkeland's currents in the T96 model, which is a halfway position between the Earth's case and a null contribution. In fact, due to the expected lack of a conducting ionosphere on Mercury, there is still a wide debate regarding the existence of field-aligned currents [*Slavin et al.*, 1997].

Finally, it must be noted that: i) this model does not describe the field depression caused by the diamagnetic effect of the plasma engulfing the magnetospheric cusps; ii) it is not possible to introduce a non-zero *x* component of IMF magnetic field; iii) the model is stationary, and hence it cannot take into account particle acceleration due to magnetic field line drag.

As a *reference* configuration, here I assume $\mathbf{B_{IMF}}$= **(**0, 0, -20) nT, which can be considered as reasonable values [*Paper* **IV**], if we discard the $B_x$ component. Other configurations of the magnetosphere have been considered by changing some of the IMF components ($B_y = \pm 5$ nT and $B_z = -10$ nT), in order to study the plasma circulation dependence on the above-mentioned parameters. Figure **5.3** shows the magnetic field lines in the *reference* conditions.

*Electric field model.* At very early stages of this study (section **5.3**), I have approximated the electric field with a simple stationary and uniform one, in the dawn to dusk direction (*y* axis), and with different absolute values in the range 0.1-10 mV/m, which seems to be a reasonable range for |**E**| in the hermean magnetosphere (see below). Then, this field has been modified by supposing the surface of Mercury to be, alternatively, both conductive and non-conductive, as there is still some controversy about this feature. This model of electric field has been used in the preliminary study of the equatorial circulation (see section **5.3**), but since it was not consistent with the magnetic field one it has been discarded in the following of the study (sections **5.4-5.7**).

It is possible to develop a model for the electric field taking into account the magnetic field one. In fact, with the assumption:

$$\mathbf{E} \times \mathbf{B} = 0, \tag{5.1}$$

$V(\mathbf{r})$ in an arbitrary point of space can be calculated following the **B** field line as far as it reaches any region where $V(\mathbf{r})$ is well known, and **E** is calculated from *V*. Following the study of *Delcourt et al*, [2003] here I have assumed, as a zero order hypothesis, that the potential at the surface of Mercury is similar to the Volland [*Volland,* 1978] potential at the ionosphere of the Earth. Using appropriate solar wind parameters, and taking into account the smaller size of Mercury's magnetospheric cavity with respect to the Earth's one, we can estimate a cross-polar cap potential drop (*PD*) between 10 and 100 kV [*Delcourt et al.,* 2003]. Willing to test the particle trajectories under different conditions, here I use *PD* = 10 kV as "*reference*" condition [*Ogilvie et al.* 1977], *PD* = 100 kV as "*high PD*" condition and the less realistic value of 1 KV as "*extremely low PD*" condition. This last value intentionally reduces the influence of **E**, and can be also used to evaluate the importance of **E** in the reconstruction of the proton circulation.

The mean electric field in the central magnetotail is, in these three cases, equal to 1, 10 and 0.1 mV/m respectively.

*Exosphere model.* As mentioned before, the exospheric components may come from several sources. My model includes only those species that have a significant CE cross-section in the case of proton projectiles, i.e. H, $H_2$, He and O. According to *Wurz and Lammer* [2003], H, $H_2$ and He may come from thermal emission as well as from micrometeorite impacts; their density vertical profiles are shown in figure **5.4.** It is worth noting that $H_2$ density has been depleted as indicated by authors. Oxygen is supposed to be produced substantially from ion-sputtering [*Wurz and Lammer,* 2003]. In this latter case, the process is included in the dayside proton circulation model (section **5.4**). Hence, I have estimated the O vertical profile, and compared the result with experimental estimations. The vertical profile for O is discussed in section **5.6** and shown in figure **5.4**. Exceptionally, also ion-sputtered Sodium has been considered as an exospheric constituent, and its vertical profile has been simulated (section **5.6**). Even if it has no implications with CE, due to its low cross-section, this element may be considered as an important tracer of Mercury exospheric behaviour [*Killen et al.*, 1990] since it can be observed from the Earth.

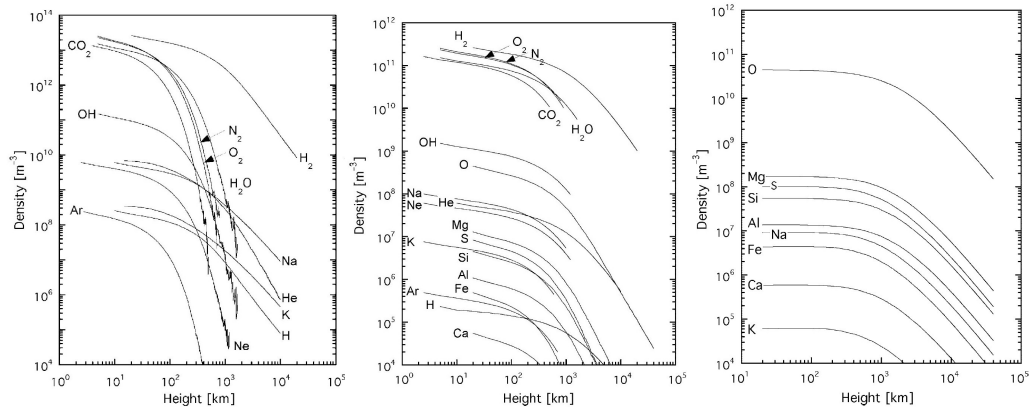

**Figure 5.4** Exospheric density vertical profiles at Mercury (dayside) for different components. From *Wurz and Lammer* [2003].

*Surface Model.* The surface model includes only Oxygen and Sodium. Even if there are some discrepancies among authors on the binding energies of these two species, theoretical computations suggest that the binding energies of Oxygen and Sodium are respectively 2 eV [*McGrath et al.*, 1986; *Cheng et al.*, 1987] and 3-4 eV [*Lammer and Bauer,* 1997], and these values have been used in the following of my study. Other authors [*Weins et al.*, 1997], after having measured that the empirical distribution of Na particles sputtered from a $Na_2SO_4$ surface (pressed powder sample) peaks at 1.5 km/s, have deduced lower values (.54 eV) for the related binding energy. If those empirical measures were found correct, and considering that the lower is $E_b$ the less is the function $F$ at energies well above $E_b$, the sputtering flux calculated in the following should be reduced of a factor 10 approximately. In section **5.6** I will discuss the implications of the sputtering process for the exospheric abundance of Na and O.

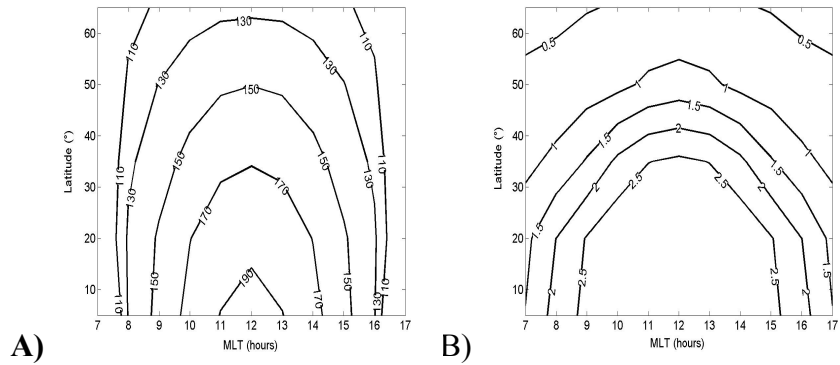

**Figure 5.5** Panel **A):** Lines of equal H$^+$ density at the magnetopause, in the MLT versus Latitude plane. Panel **B):** same as **A**), with lines of equal H$^+$ mean energy at the magnetopause.

*Proton source.* In my simulation I start tracking the protons as they cross the magnetopause (MP), entering the magnetosphere from the magnetosheath. In this latter region the protons of solar wind origin are decelerated, compressed and heated. The H$^+$ density exhibits a bulk in the subsolar region; as the protons move tailward along the flanks of the magnetosphere, their velocity grows and their temperature and density decrease.

In agreement with the T96 model, I have approximated the dayside magnetopause with an ellipsoid, and all test particle trajectories start from this surface. According to this model, the distance between the surface of Mercury and the magnetopause at the subsolar point is approximately 0.5 $R_M$.

Proton mean energy and density functions over this surface have been derived from the work of *Spreiter et al.* [1966], and adapted to the case of Mercury [*Paper* **IV**]. Figure **5.5** shows contours of H$^+$ density and mean energy on the MP surface, in a MLT vs. geographic latitude reference system.

It should be noted that the mean proton energy peaks near the subsolar point, whereas the proton group velocity increases while they move away from it, as they are dragged together with the frozen-in IMF. Here we consider the energy of those protons in the reference system of the magnetic field lines, since the magnetic field is steady in our model. Hence, the energy we use as initial condition for our test particle reaches its maximum near the subsolar point, where the proton acceleration due to magnetic reconnection is higher [*Paper* **IV**].

It may be useful to estimate the total H$^+$ flux impacting the MP of Mercury, as predicted by the model, in order to have a reference value when discussing the proton precipitation on the planetary surface. Integrating the data shown in figure **5.5** over all the MP, a total flux between 2 and $3\times10^{27}$ s$^{-1}$ is found. It is worth noting that this value, as well as the distribution of protons over MP and the shape of the MP itself, depend on the external conditions, such as the SW velocity and density. However, in this model I use a fixed source. A change in the unperturbed SW density reflects linearly upon the estimated magnetospheric H$^+$ density, and can be included by applying a simple proportion; the change of the MP shape has been neglected for simplicity.

## 5.3 Equatorial proton circulation at Mercury: a partial ring current?

The main purpose of this study is modelling the proton precipitation in the cusp regions, and the subsequent dayside circulation. Before doing this, one question must be answered: is the tail sunward convection capable to feed the equatorial plasma circulation? Most of the authors [O*rsini et al*., 2001; *Lukyanov et al.,* 2001] suggest that this process, if present, cannot turn-on a steady ring current or even a partial ring current. In fact, as mentioned before, the dimension of the planet, if compared to its magnetosphere, is too big to let room enough for a ring current. To better validate this hypothesis I have run an extremely simple model, based on the guiding centre approximation and on the continuity equation. The simulation considers only the protons circulating in the night-side equatorial plane ($r,\varphi$) of Mercury and with a 90° pitch angle. To demonstrate the absence of a stable ring current there is no need of a whole 3-dimensional simulation. In fact, the particles that have 90° PA and populate the equator experience, in general, fewer losses with respect to those particles that have different PA. These latter particles, actually, get closer to the planet at their mirroring points, where CE is more efficient and planetary impacts are more probable.

The guiding centre (GC) approximation supposes that the proton trajectories come from the superimposition of two motions: a circular motion around the guiding centre, due to Lorenz force, and the drift of the guiding centre itself. Hence, on the average, a particle moves only if its GC does, and this may happen either if there is an electric field ("**E**×**B** drift") or if the magnetic field is non-homogeneous ("**B**-gradient drift" and "**B**-curvature drift"). In principle, this approximation is valid if the GC drift velocity may be neglected if compared to the velocity of the particle itself. As I will discuss in detail later, this assumption is not valid in the whole space around Mercury. However, in the equatorial plane this assumption is generally reasonable: a preliminary estimation of the error indicates that the discrepancy between the GC path and the real path is less than 10% of the path length. For the purpose of this preliminary study, this approximation is hence acceptable.

Since protons are treated as a fluid, we have the continuity law:

$$\frac{dn_{H+}}{dt} + \nabla\left(n_{H_+}\mathbf{u}\right) = F - S\,,\tag{5.2}$$

where $n_{H+}$ and $\mathbf{u}$ are respectively the proton density and velocity, while $F$ and $S$ are two functions for the proton sources and sinks. Here, this equation must be applied to a 3 dimensional space: $r$ and $\varphi$ are the polar coordinates of the GC of the protons in the equatorial plane; $K$ is the kinetic energy of the protons themselves. The first two components ($u_r$ and $u_\varphi$) of the velocity $\mathbf{u}$ are given by the $\mathbf{E}\times\mathbf{B}$ plus the $\mathbf{B}$-gradient drift, using electric and magnetic models described in section **5.2**. The third component of the velocity ($u_k$) can be estimated considering that, as a zero order hypothesis, only the electric field is able to change the kinetic energy of the particle. Hence, we have:

$$u_k = \frac{dK}{dt} = \frac{dK}{dx}\frac{dx}{dt} + \frac{dK}{dy}\frac{dy}{dt} = E_x u_x + E_y u_y\,.\tag{5.3}$$

In our case, the function $F$ for the sources is assumed to be zero everywhere except at instant $t=0$, when an arbitrary source has been considered. In fact, since we want to demonstrate the impossibility of a stable circulation in the equatorial regions of Mercury, it is sufficient to suppose some a-priori extended distribution in the tail. If this source leads to a rapid disappearing of protons, then all possible sources do the same. The $n_{H+}(0)$ function used here is a simple gaussian function in the three dimensions $r,\ \varphi,\ K$.

The function $S$ for the sinks is calculated taking into account charge exchange and collisions with the planet. Charge-exchange losses have been estimated using neutral density of H, He and O given by *Wurz and Lammer* [2003], and relative cross-sections (figure **5.4**). As far as it concerns the planetary collisions, I have adopted an *ad hoc* function *dp* representing the probability for a proton to hit the surface after each integration step *dt*. In fact, GC method approximates the trajectory of a particle with a drifting circle or radius equal to Larmor radius ($R_L$). Hence, a particle close to the planetary surface may have part of this circle inside the planet. Function *dp* is proportional to the particle velocity *v* and inversely proportional to $R_L$:

$$dp = \begin{cases} 0 & r > \left(R_M + R_L\right) \\ \dfrac{v\,dt}{2\,\pi\,R_L} & R_M < r < \left(R_M + R_L\right) \\ 1 & r < R_M \end{cases} \tag{5.4}$$

Equation **5.2** has been solved numerically. To have a good accuracy, the integration step *dt* and the grid sizes ($\Delta r$, $\Delta \varphi$, $\Delta K$) must satisfy the relation:

$$dt << \min\left(\frac{v_r}{\Delta r}; \frac{v_\varphi}{\Delta \varphi}; \frac{v_K}{\Delta K}\right) \tag{5.5}$$

in every point of the grid. A good compromise between machine-time consumption and accuracy has been found using a grid of 50×50×50 bins and *dt*=10⁻² s.

Figure **5.6** shows an example of the evolution of the H⁺ population, in the case of Tsyganenko modified **B** model and with a "conductive" planetary surface. The mean **E** field in this case is 1 mV/m. Protons move sunward under the **E×B** drift and then clockwise around the planet under the **B**-grad drift, depending on their energy. In fact, this latter drift is energy-dependent, and after a while the higher energy protons will

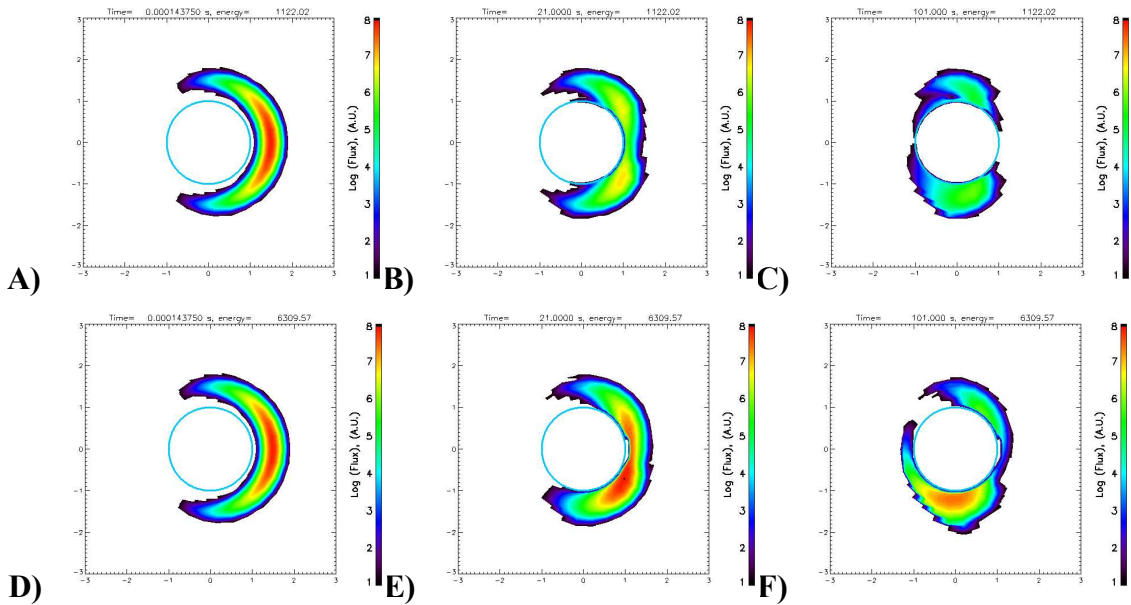

**Figure 5.6** Equatorial H⁺ flux as simulated using the continuity equation and guiding centre approximation and an arbitrary source. Panels **A, B, C**: H⁺ Flux @ 1 keV, respectively after 0, 20, 100 s. from the injection; panels **D, E, F**: H⁺ Flux @ 5 keV, after 0, 20, 100 s from injection. Colour is coded according to H⁺ flux using an arbitrary scale.

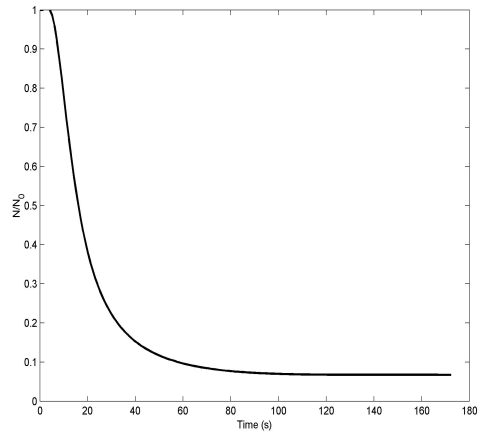

**Figure 5.7** Relative abundance of protons in the equatorial plane of Mercury, as a function of time elapsed since injection, respect to proton abundance at $t$=0. The $H^+$ distribution at is arbitrary.

concentrate in the dusk side. This feature is similar to what happens at the Earth, and has been already discussed. It is worth noting that, for all other possible external configurations, the time evolutions of protons are very similar to this one.

I have run similar simulations with different external configurations, starting with the "dipole **B** + uniform **E**" configuration described in section **5.2** and then using the "Tsyganenko **B** + Volland **E** " one. In all cases, no ring current or partial ring current has ever been able to form. The CE and the planetary collisions rapidly remove the protons from the magnetosphere. In figure **5.7** I show the relative intensity of the proton flux (@ 90° PA) versus the elapsed time, in the same case as figure **5.6**.

As a matter of fact, even if some short-time proton circulation is possible in the night side, it will hardly extend to the dayside. A stable circulation of protons around Mercury is extremely difficult, and a complete and stable ring current is not able to form. This latter consideration is important also while reconstructing the **B** field by means of a Tsyganenko model (see section **5.2**).

## 5.4 Dayside proton circulation at Mercury: model description

**Simulation of H⁺.** To obtain a full spatial, pitch angle and energy distribution of the protons in the magnetosphere of Mercury I have used a Monte-Carlo (MC) single-particle model. I have simulated the trajectories of a large number of protons (hereafter called "test particles") by solving the full equation of motion (including electromagnetic and gravitational forces). In fact, Larmor's radius of protons in the energy range I am analysing (approx. 0.1÷10 keV) is generally not small when compared with the dimension of Mercury magnetosphere. Hence, non-adiabatic effects may become of crucial importance for most of the ion magnetospheric transport [*Delcourt et al.*, 2003], and the use of an adiabatic treatment based on the guiding centre approximation will lead to non-negligible errors when reconstructing H$^+$ paths in full 3D space around Mercury.

Each test particle trajectory begins at the bow shock of Mercury (see section **2.4**) and finishes if either:

1) the test particle hits the surface of the planet; or

2) the test particle reaches a non-connected IMF field line; or

3) the test particle gets too far from Mercury in the nightside ($r > 10 R_m$).

The H$^+$ distribution function in space, energy ($E$) and pitch angle ($\alpha$) has been reconstructed over a 5-dimensional grid, and it is represented by a matrix $N_{ijklm}$. The grid extends from 1 to 2 Mercury radii ($R_M$), 0 to 24 $h$ (MLT), –90° to 90° (latitude), 0° to 180° (pitch angle) and 100 eV to 10 keV (energy); the number of bins for each dimension is respectively 8×48×24×4×4.

In this simulation, each test particle represents a certain number of real protons. If we assume that all those protons were "born" (i.e. they enter the simulation region) from the same point $Q$ on the MP within an infinitesimal surface $\Delta S_Q$, then they all have the same trajectory. We can also imagine that those protons were "born" from $\Delta S_Q$ with a fixed frequency $1/\Delta T$. If the H$^+$ flux at $Q$ is $\Phi_Q$, the time interval between protons is:

$$\Delta T = \frac{1}{\Phi_Q \, \Delta S_Q} \qquad\qquad (5.6)$$

Since they are equally spaced in time along the trajectory (forming a sort of "train"), the total number of protons $N$ within an arbitrary cell *ijklm* is simply $T/\Delta T$, where $T$ is the time a proton stays within that cell. This time is the same for both protons and the associated test particle:

$$T = N \, \Delta T = n \, dt \qquad\qquad (5.7)$$

where $dt$ is the time integration step for the test particle and $n$ is the number of times that, during its trajectory, the associated test particle is inside that cell. It follows that:

$$N = \frac{n\,dt}{\Delta T} = n\left(dt\,\Phi_Q\,\Delta S_Q\right) \tag{5.8}$$

Hence, after every integration step $dt$, I just have to increase the value of the bin corresponding to its position ($N_{ijklm}$) of a value $w_{H+}$:

$$w_{H+} = dt\,\Phi_Q\,\Delta S_Q\,f_{CE}(t)\,. \tag{5.9}$$

The function $f_{CE}$ takes into account the losses due to charge-exchange with the neutral exosphere of Mercury. This function has been calculated solving numerically the following differential equation:

$$\begin{cases} f_{CE}(0) = 1 \\ \dfrac{df_{CE}}{dt} = |\mathbf{v}|\sum_i \sigma_i\,n_i \end{cases}, \tag{5.10}$$

where $\sigma_i$ and $n_i$ are the charge-exchange cross section and the local number density of the $i^{\text{th}}$ neutral species (see figures **2.2** and **5.4**). The 5-D number density ($\frac{dn_{H+}}{dE\,d\alpha}$) of $H^+$ in any point of the grid is obtained by dividing the bin value $N_{ijklm}$ by its 5D-volume, while the proton flux $\Phi_H$ is simply $n_{H+}$ times $v(E)$. The surface $\Delta S_Q$ may be estimated just dividing the total surface where the generation of test particles occurs, by the number of test particles that have been run.

In principle, to perform an accurate numerical integration of the motion, it should be necessary to calculate the electric field for a very large number of points along each trajectory. This procedure, in general, is very expensive in terms of machine-time, because at each point it is necessary to trace the magnetic field line up to the planetary surface (see section **5.2**). To permit the simulation of a large number of test particles, I have preliminarily estimated the electric potential $V$ on the vertexes of a three-dimensional cubic grid. The potential in any other point can be calculated via 3-linear interpolation using the eight nearest vertexes. The bin size is 500 km, which leads to an error on the potential, which has been empirically found to be less than 5 ‰. This error may be neglected if compared to other uncertainties in the model.

**Simulation of neutral particles.** In this model, neutral particle emission is related to ion circulation through two processes: ion-sputtering and charge exchange. In fact, such processes are included in the model as proton losses, and both of them may produce neutral particles. In section **6** and **7** I will present simulated neutral signal coming from both processes.

*Sputtering.* This process has been discussed in section **2.2**. Here I solved the integral in equation **2.11** in a numerical way. Every time a test-particle hits the surface, it is converted to a test-neutral or a test-ion, with random initial direction. The energy is chosen randomly according to energy distribution in equation **2.10** (see figure **2.2**); the species of the sputtered particle is chosen randomly according to different relative abundances $R_k$; the weight $w$ in equation **2.11** is multiplied by a factor $Y_k$. This new test-particle may, of course, hit the surface again: the procedure exposed here is then performed again, as far as its weight is greater than one and its energy is able to produce sputtering ($E_i > E_b$).

In this study I have concentrated my attention to Oxygen and Sodium only. In fact, both O and Na have been observed in the exosphere of Mercury [*Broadfoot at al.,* 1976; *Potter and Morgan,* 1985, 1997]. The overall fraction of Oxygen ($R_O$), if we consider all molecules containing O, may be up to 50% [*Wurz and Lammer*, 2003]. The sodium content of the surface ($R_{Na}$) has been estimated to be within 0 and 1.4 % [*Goettel*, 1988]; here I have chosen a halfway value, i.e. 0.5%. Yields for those species strongly depend on the molecular structure of the surface; here I have adopted a rough value of 0.1 for both O and Na, since the uncertainty on the surface abundance probably overcomes the error on the yield. It is worth noting that a bad estimation of either $Y$ or $R$ does not affect the estimated vertical profile of the neutral emitted population, but it can be considered as a scale factor for the whole density distribution function of neutrals. More particularly, an overestimation of the O sputtering flux may come from considering the same yield for all surface constituents containing Oxygen.

The surface binding energy, conversely, has a deep impact on the energy spectrum of the particles emitted towards the exosphere. For any $E_i \gg E_b$, in fact, it may be shown that $f_S (E_e, E_i)$ peaks approximately at $E_b/2$. Surface binding energies are, generally, of the order of some eV; here I have adopted, respectively, 3.5 eV and 2 eV for O and Na. Some recent studies [e.g. *Weins et al.*, 1997] indicate lower values (~.5 eV) for the Na binding energy. Considering that the lower is $E_b$ the less is the function $F$ at energies well above $E_b$, if this estimation is found correct, the Na flux from ion-sputtering calculated in the following should be reduced of a factor of 10 approximately.

*Charge-Exchange.* This process has been discussed in section **2.1**. The ENA flux and images have been obtained by using equation **2.9**; cross-sections for $H^+$ colliding various neutral species (H, $H_2$, He and O) are shown in figure **2.2**; estimated exospheric densities (from *Wurz and Lammer* [2003]) are shown in figure **5.4**; all processes have been considered. It is worth noting that the $H_2$ density has been depleted as suggested by authors. The depletion coefficient here is 0.01, in order to be conservative with the estimation of ENA fluxes.

## 5.5 Dayside proton circulation at Mercury: results

Here I present the simulation results in terms of $H^+$ circulation. Six different sets of boundary conditions have been used, and about $3 \times 10^5$ test-particles have been tracked for each set. A particular set of boundary conditions, called "*reference*" conditions, is: $\mathbf{B_{IMF}}$=(0, 0, 20) nT, $PD = 10$  kV. Then I have changed PD (1 kV and 100 V), $B_y$ (± 5 nT) and $B_z$ (-10 nT). To reduce the amount of runs, I have not explored all combinations of these values, but just all variations of a single parameter from the "*reference*" one. This must be taken into account in general, since variations of potential drop and $\mathbf{B_{IMF}}$ often occur simultaneously.

To obtain the $H^+$ number density $n_{H^+}$, the differential number density ( $\frac{dn_{H^+}}{dE \, d\alpha}$ ) has been integrated over all pitch angles and over two different energy ranges: 0.1-1 keV and 1-10 keV. Particles outside these two ranges have been discarded.

To help the discussion, I present an overall summary of results (table **5.1**), sections of $H^+$ density (figure **5.8**) and maps of $H^+$ flux on the surface (figure **5.9**). Table **5.1** shows three parameters that synthesize some features of the $H^+$ flux impacting the surface, i.e. the total flux ($F_T$), the mean MLT ( $\overline{MLT_S}$ ) and the mean latitude ( $\overline{\varphi_S}$ ). Those parameters have been calculated for different boundary conditions and energy ranges. In particular, the mean position of the $H^+$ spot on the surface corresponds roughly to the bulk of the emission of the sputtered neutrals, which will be monitored by the NPA-SERENA orbiting sensor; the total flux of $H^+$ impacting the surface has important implications on the exospheric refilling.

Panels in figure **5.8** show colour-coded sections of the $H^+$ density distribution. Each panel shows $n_{H^+}$ over the superposition of two different surfaces: the one outside the red circle is the *x-z* plane; the one inside the red circle is a hemisphere just above Mercury's surface, in the dawn side. The Sun is on the right. Different panels refer to different external conditions and energy ranges. It is worth noting that panels in figure **5.8** do not show the protons that cross the magnetopause (thus becoming elements of the simulation) in the south hemisphere. In this way, the protons that are found in the southern hemisphere can be immediately identified as "bouncing protons". To obtain the complete distribution it is sufficient to superimpose this $H^+$ distribution to another one, which is symmetric to the first with respect to the equatorial plane.

Panels in figure **5.9** show polar-stereographic projections of the $H^+$ flux on the northern surface of Mercury, colour-coded according to the $log_{10}$ of the intensity. This flux has been obtained by using the $H^+$ density distribution function (position, energy and PA), in the cells just above the surface. However, a small error will arise from this procedure because those cells have a height of about 300 km, and even if the **B** field lines can be considered more or less parallel over this short scale, some protons may bounce inside this cell. The flux obtained in this way is hence an upper limit for the real flux at the surface. Different panels refer to different external conditions and energy ranges.

**Table 5.1**

| Energy Range (keV) | $B_y$ (nT) | $B_Z$ (nT) | P. Drop (V) | Total Flux ($10^{25}$ s$^{-1}$) | Mean MLT (*hh:mm*) | Mean Lat (°) |
|---|---|---|---|---|---|---|
| *Total:* .1÷10 | -5 | -20 | $10^4$ | 18 | 11:36 | 44 |
| | 0 | -10 | $10^4$ | 16 | 11:40 | 48 |
| | 0 | -20 | $10^3$ | 27 | 10:40 | 37 |
| | 0 | -20 | $10^4$ | 19 | 11:39 | 44 |
| | 0 | -20 | $10^5$ | 5.6 | 11:38 | 63 |
| | 5 | -20 | $10^4$ | 21 | 11:28 | 44 |
| *Low Energy:* .1÷1 | -5 | -20 | $10^4$ | 3.6 | 11:54 | 57 |
| | 0 | -10 | $10^4$ | 3.3 | 12:04 | 60 |
| | 0 | -20 | $10^3$ | 4.8 | 11:53 | 53 |
| | 0 | -20 | $10^4$ | 3.7 | 11:54 | 57 |
| | 0 | -20 | $10^5$ | 0.94 | 11:51 | 67 |
| | 5 | -20 | $10^4$ | 3.8 | 11:50 | 57 |
| *High Energy:* 1÷10 | -5 | -20 | $10^4$ | 16 | 11:28 | 38 |
| | 0 | -10 | $10^4$ | 13 | 11:27 | 42 |
| | 0 | -20 | $10^3$ | 22 | 10:24 | 31 |
| | 0 | -20 | $10^4$ | 16 | 11:32 | 39 |
| | 0 | -20 | $10^5$ | 4.7 | 11:32 | 60 |
| | 5 | -20 | $10^4$ | 17 | 11:19 | 38 |

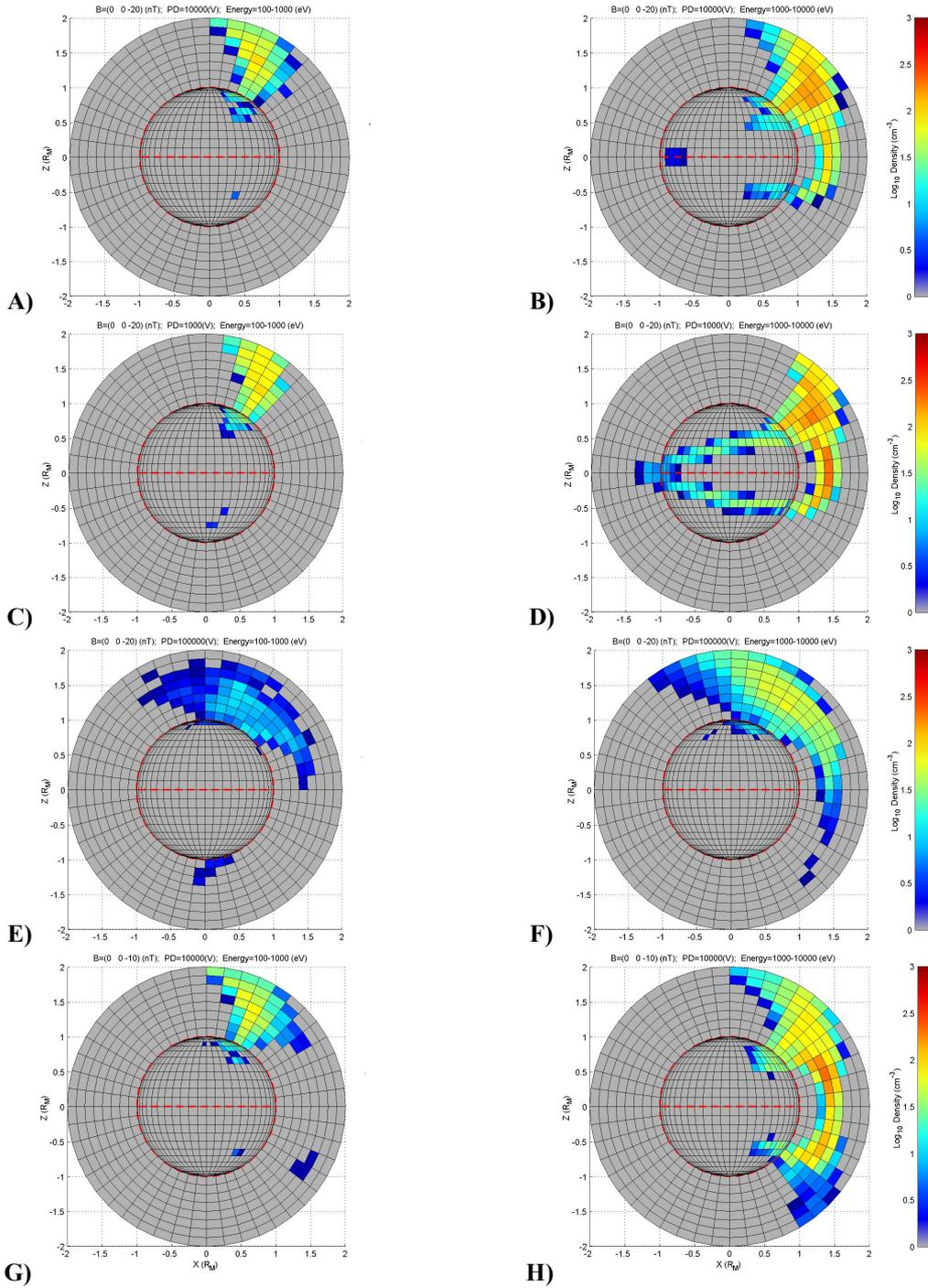

**Figure 5.8** Colour-coded maps of H$^+$ density over two different surfaces: *x-z* plane (outside the red circle) and Mercury surface (inside red circle). Panels **A, C, E, G**: low proton energy (100 eV – 1 keV); Panels **B, D, F, H**: high proton energy (1-10 keV). Panels **A, B**: *reference* condition. Panels **C, D**: same as **A, B**, but low potential drop (*PD*=1 kV). Panels **E, F**: same as **A, B**, but high potential drop (*PD*=100 kV). Panels **G, H**: same as **A, B**, but low IMF (*B$_z$* = -10 nT).

**General features.** In a typical H$^+$ simulated distribution it is possible to recognize three different and peculiar populations, whose relative intensity is a matter of interest of the present study. The first one (α), present in all external conditions, is the population precipitating towards the planet along the magnetic field lines. Some of them are bounced by the increasing magnetic field; the others reach the planetary surface. During this motion, particles are drifted northward by the **E×B** drift and westward by the grad-**B** drift. The former is energy-independent; the latter is more efficient for the highest energies. The proton source is symmetric around noon, and any longitudinal shift in the surface precipitation (evidenced by the parameter $\overline{MLT_S}$) is due to **B**-grad drift.

The second population (β) is made by protons that are bouncing in the dayside, and is concentrated between the subsolar point and the surface. The third population (γ) is made by protons that move westward and eventually hit the surface in the nightside. Both these two last populations are generated from the first, because of the diffusion through field lines. Diffusion occurs if the temporal or spatial scale of the **B** field variations is shorter, respectively, than the Larmor period or radius. In the present study, however, only the second case is of interest, because the **B** field is static.

***Reference* conditions.** Panels **5.8A** and **5.8B** show the precipitating H$^+$ in the "*reference* conditions" for different energy ranges. In both cases the **α** population extends down to the surface, but high energy protons (1-10 keV, panel **5.8B**) impact Mercury at lower latitudes  with respect to low energy ones (100 eV – 1 keV, panel **5.8A**). In fact, the mean energy of the H$^+$ at the magnetopause is higher when close to the sub-solar point (see figure **5.4**). The population close to the Low Latitude Boundary Layer [*Paper* **IV**] is able to diffuse towards lower latitudes and start bouncing, raising the **β** population. Low energy protons are injected at higher latitude and cannot experience any bouncing. No **β** population is, in fact, recognizable in panel **5.8A**.

The H$^+$ flux impacting the surface (see panels **5.9A, 5.9B**) is up to $10^9$ cm$^{-2}$ s$^{-1}$ for low-energy protons and up to $5\times10^9$ cm$^{-2}$ s$^{-1}$ in the case of high-energy protons. The mean flux intensity in the open field line area is approximately $10^8 \div 10^9$ cm$^{-2}$ s$^{-1}$ for both energy ranges, which is in agreement with previous studies [*McGrath at al.* 1986]. The difference of the distribution for the two energy ranges is more evident if we consider the position of the precipitating protons (see table **5.1** and panels **5.9A, 5.9B**), which is (11:32 MLT, 39°) for protons above 1 keV, and (11:54 MLT, 57°) for protons below 1 keV. The $\overline{MLT_S}$ shift between these two populations is due to the energy-dependence of the **B**-grad drift.

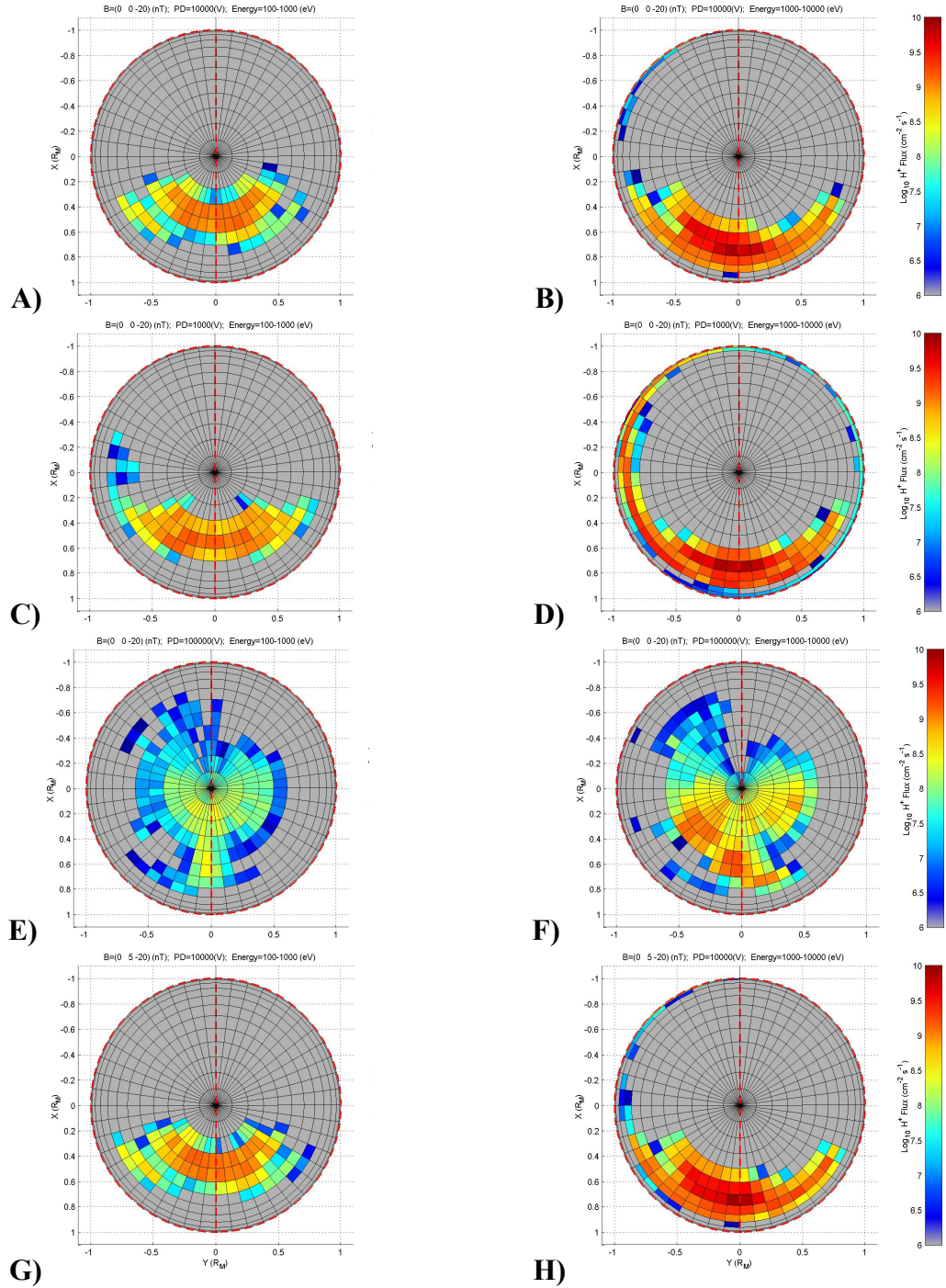

**Figure 5.9** Colour-coded maps of H$^+$ flux over Mercury's surface. Panels **A, C, E, G**: low proton energy (100 eV − 1 keV); Panels **B, D, F, H**: high proton energy (1-10 keV). Panels **A, B**: *reference* condition. Panels **C, D**: same as **A, B**, but low potential drop (*PD*=1 kV). Panels **E, F**: same as **A, B**, but high potential drop (*PD*=100 kV). Panels **G, H**: same as **A, B**, but different IMF (*B$_y$* = -5 nT).

**Electric field variation.** In this analysis I have used two extreme values for the potential drop (1 kV and 100 kV) in order to have two extreme conditions for the electric field. The mean |**E**| in each case is respectively .1 mV/m and 10 mV/m (see section **2.3**). The first, clear effect of the increasing/decreasing **E** is the

increasing/decreasing of the northward shift of the surface footprint due to the **E×B** drift. This effect is energy-independent, and actually is clearly recognizable by comparing figures **5.8C** with **5.8E** (low energy), and **5.8D** with **5.8F** (high energy). The same comparison may be done, with identical results, on figure **5.9** (panels **C, E, D,** and **F**), using the H$^+$ flux at the surface instead of the density.

Another evident effect, originated from the same cause, is the modulation of β and γ population magnitude. With the increasing of the **E** field, in fact, protons are drifted northward, and no β population is able to form. Conversely, if the potential drop is low, the **B**-grad drift becomes predominant for high and low energy protons, and the γ population is drifted more westward, while β population does not change significantly. In this case, high-energy particles may reach the nightside. Panel **5.8D** clearly indicates that γ particles are bouncing during their drift, and they are gradually getting closer to the planet. Hence, a sort of small "partial ring current" is able to form; an important consequence of this drift is the possibility of proton precipitation in the nightside and the subsequent emission of sputtered particles from that region (panel **5.9D**). As far as it concerns low energy protons (panels **5.8C** and **7C**), some of them are drifted westward, but the bulk precipitates directly onto the surface with a small spread.

According to table **5.1**, intense electric fields cause less surface precipitation. The protons are supposed to be deflected northward and to be lost in the nightside.

In order to evaluate the effect of the **E×B** drift on the H$^+$ precipitation in the *reference* conditions, we may assume that $PD = 1$ kV reduces the influence of the electric field, so that it roughly corresponds to a "*zero electric field*" condition. By comparing total precipitating flux as reported in table **5.1**, in the cases of $PD$=10 kV ($19 \times 10^{25}$ s$^{-1}$) and $PD$=1 kV ($26 \times 10^{25}$ s$^{-1}$), we can conclude that at least one third of the protons that could precipitate onto the planetary surface are removed by **E×B** drift.

**Magnetic field variation.** The efficiency of the reconnection process is roughly proportional to the negative *z* component of the **B$_{IMF}$** field. This phenomenon has been studied here by comparing the "*reference*" case with the case of **B$_{IMF}$**=(0, 0, -10) nT. The total precipitating flux, in this latter case, is lower ($4.2 \ 10^{26}$ cm$^{-2}$ instead of $5 \ 10^{26}$ cm$^{-2}$). Moreover, protons impact at higher latitudes on the average (see table **5.1**). There is no other evident modulation on the density distribution shape (see figures **5.8G** and **5.8H**); the β population is slightly more evident, and there is less nightside precipitation.

In conclusion, the influence of $B_z$ parameter is more on the intensity rather than on the shape of the $n_{H+}$ distribution function.

The influence of the $B_y$ component, as one may aspect, is mainly on the footprint longitude, with a westward shift of about 3° for each 10 nT. There is no other evidence of an influence of the $B_y$ component on the general shape of the H$^+$ distribution, probably because IMF drives the input of the plasma more than the circulation after injection (see figures **5.9G** and **5.9H** for example).

It is worth noting that, in terms of influence, the **B$_{IMF}$** variation seems to be less relevant than *PD* also because in the first case I have chosen a more conservative position. Actually, more extreme conditions can be expected at Mercury; the $B_z$ component, for example, may range between –30 and 30 nT, to say nothing of the $B_x$ component absence.

## 5.6 Sputtering at Mercury

In this study I concentrate on the sputtering of Oxygen and Sodium. Both of them have been proved to exist in Mercury's exosphere: the first has been observed by Mariner 10 UVS [*Broadfoot at al.,* 1976]; the second has been detected by ground-based observations [*Potter and Morgan,* 1985]. I will first discuss the simulated vertical profiles for these species, as they are a general matter of interest; then I will concentrate on the feasibility of monitoring the cusp proton precipitation by means of a neutral particle imager (ELENA).

Figure **5.10** shows the estimated mean vertical profile for these species in the dayside exosphere (solid lines). These densities have been calculated considering sputtering as the only source; since other source processes (i.e. photo-stimulated disorption and thermal emission) are possible, such densities may be considered as lower limits. Moreover, I don't take into account the "ambient atoms" [*Killen and Ip,* 1999], i.e. atoms that are not at their first ballistic orbit. Dashed lines represent the vertical profiles over the point of maximum proton precipitation, and can be considered as an upper limit for sputtering-generated exosphere. These values have been obtained using the "*reference*" conditions, and the vertical profile density in other cases can be estimated just applying a simple proportion, using total flux in table **5.1**. Sputtering emission, in fact, is roughly proportional to precipitating flux ($E_i$ does not modify much the emission spectrum, except for very high $E_i$). Hence, densities may be up to 50% higher (for

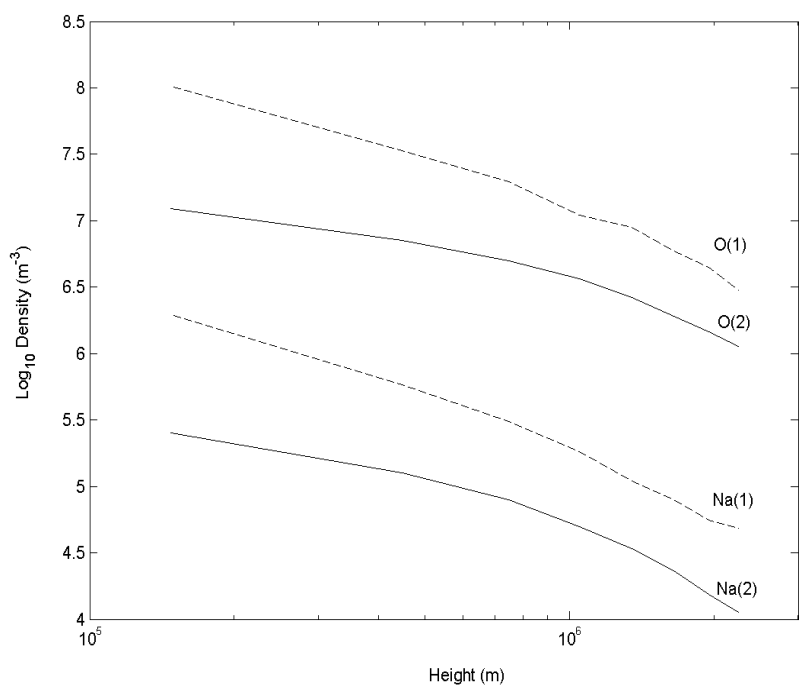

**Figure 5.10** Exospheric density for O and Na, function of altitude from planetary surface, calculated considering sputtering as the only release process (solid lines, **2**). Dashed lines (**1**) still consider only sputtering, but refer to the vertical profile in the region of maximum ion precipitation.

example in the minimum potential drop conditions), or down to 80% lower (in maximum potential drop conditions).

As far as it concerns Oxygen, the estimated total column density is between $5 \times 10^8$ and $5 \times 10^9$ cm$^{-2}$, the surface density is between 20 cm$^{-3}$ and 200 cm$^{-3}$ and the scale-height is between 500 and 1000 km. Once again, it is worth noting that the thermal part of this population is underestimated: the first measure of the Oxygen exospheric content made by Mariner UV camera indicates $7 \times 10^3$ cm$^{-3}$ as an upper limit, even if the scale-eight estimation was doubtful.

As far as it concerns Na, it has been noted [*Killen et al.*, 2001] that its temporal variability was found to be compatible with solar activity, which is in turn related to sputtering. The estimated total column density is up to $10^8$ cm$^{-2}$, whereas the measured value is considerably higher (between 1 and 3 x$10^{11}$ cm$^{-2}$ [*Killen et al.*, 1990]). This may be explained if sputtering were not the first source for exospheric sodium. Another reason, which must be taken into account while discussing the results, is that I have intentionally discarded the ambient atom (i.e. neutral scattering on the surface plus accommodation/release), which have lower energy and reside in the lower part of the exosphere. Their inclusion in the model will be a further step, and will lead to a modelled sodium exosphere more concentrated in the lower altitudes. Scale-height analysis seems to confirm this last hypothesis. For example, simulated scale-height for Na is about 500 km, and a factor 10 greater than that observed (50 km) [*Killen et al.* 1990, *Wurz and Lammer,* 2003].

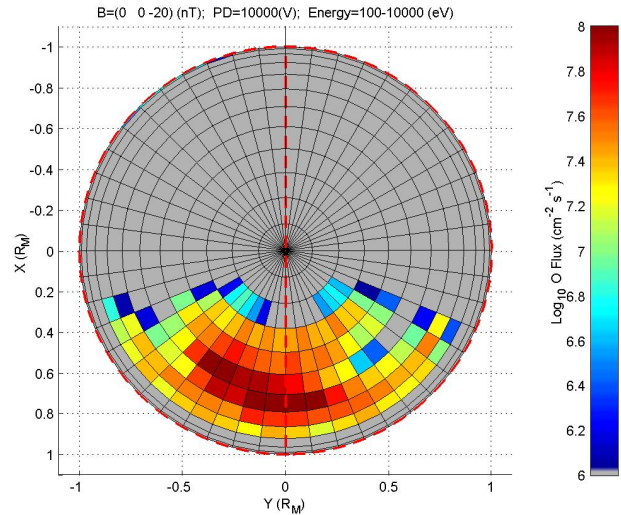

**Figure 5.11** Colour-coded O-ENA flux generated by ion-sputtering on the northern surface of Mercury. Flux is integrated over all energies above escape energy (1.4 eV).

We now want to speculate on the possibility of remote-sensing the plasma-surface interaction. The estimated total proton flux on Mercury surface is between $10^{26}$ s$^{-1}$ and $7 \times 10^{26}$ s$^{-1}$, depending on external conditions. For comparison, we can obtain the amount of protons removed in one second by the charge-exchange process by integrating the charge-exchange loss function (equations **2.3** and **2.7**) in all the 5D space. This value is between $10^{22}$ s$^{-1}$ and $10^{24}$ s$^{-1}$, depending on the boundary conditions and exospheric models. We can conclude that surface impact is the first cause of loss, in terms of magnitude, for proton plasma circulation. Consequently, sputtering process is an important source for the refilling of the exosphere, and since both sputtering and CE result in neutral atoms production, neutral atom imaging of the sputtering process is probably the best way to make remote-sensing of the plasma circulation. The Bepi Colombo/planetary orbiter (BC-MPO) is expected to have a neutral particle analyser on board. The proposed package (SERENA) includes a neutral particle imager (ELENA) with angular (~2°) and energy (~5%) resolution. Such an instrument would be able to detect the high-energy tail of the neutral sputtering emission, approximately above 10 eV. These particles travel more or less along straight lines (escape energy is 1.4 eV for O and 2 eV for Na) and their detection at MPO orbit (height: 1500 km at apocentre, 400 km at pericentre) would give an image of the sputtering release at the surface. Figure **5.11** shows the simulated intensity of sputtered O flux at the surface, integrated over all energies above 10 eV and over all directions, in the "*reference conditions*"; in the same configuration, the Na flux is similar in shape and approximately 1% of the O flux. In both cases, the intensity is high enough to be detected by ELENA; the angular resolution of the instrument results in a surface resolution down to tens of km. It is hence possible to monitor the variations in shape and intensity of the H$^+$ precipitation on the surface by the ELENA sensor on board BC. Since ELENA sensor can discriminate masses, the high surface resolution would permit the observation of differences in the surface local composition. Last but not least, the global analysis of the sputtering erosion of the surface would give unique information about the present and the past of Mercury's surface, revealing if (and how) the solar wind/surface interaction, in the past millions of years, has influenced the present surface composition.

## 5.7 ENA production at Mercury

Even if charge-exchange is not the main loss mechanism for magnetospheric protons, monitoring the product of such an interaction allows having important information about the magnetospheric and exospheric properties [*Barabash*, 2001]. In this frame, I will first discuss the general feasibility of ENA imaging and then its implications for neutral particle sensors on board MMO and MPO.

ENA images have been simulated, for different vantage points, by applying the procedure described in section **2.1** (*weighted cell integral*) to the simulated H$^+$ density shown in figure **5.8**. After a preliminary study, I have focused on two different positions for the vantage point. The first one is in the tail region, outside the equatorial plane but still in the shadow of the planet; as an example, I have chosen the point $P_1$: (-1.8, 0, 0.8) $R_M$. From this point, by looking sunward, it is possible to monitor simultaneously the plasma precipitating into the cusp, and that escaping in the dawn region. The second point $P_2$, which can be located in the dawn hemisphere, approximately on the $y$ axis and with $r < 2$, is a good point to look at the $\gamma$ population that drifts from dayside to nightside. In my study I have located $P_2$ at (0, -2, 0) $R_M$.

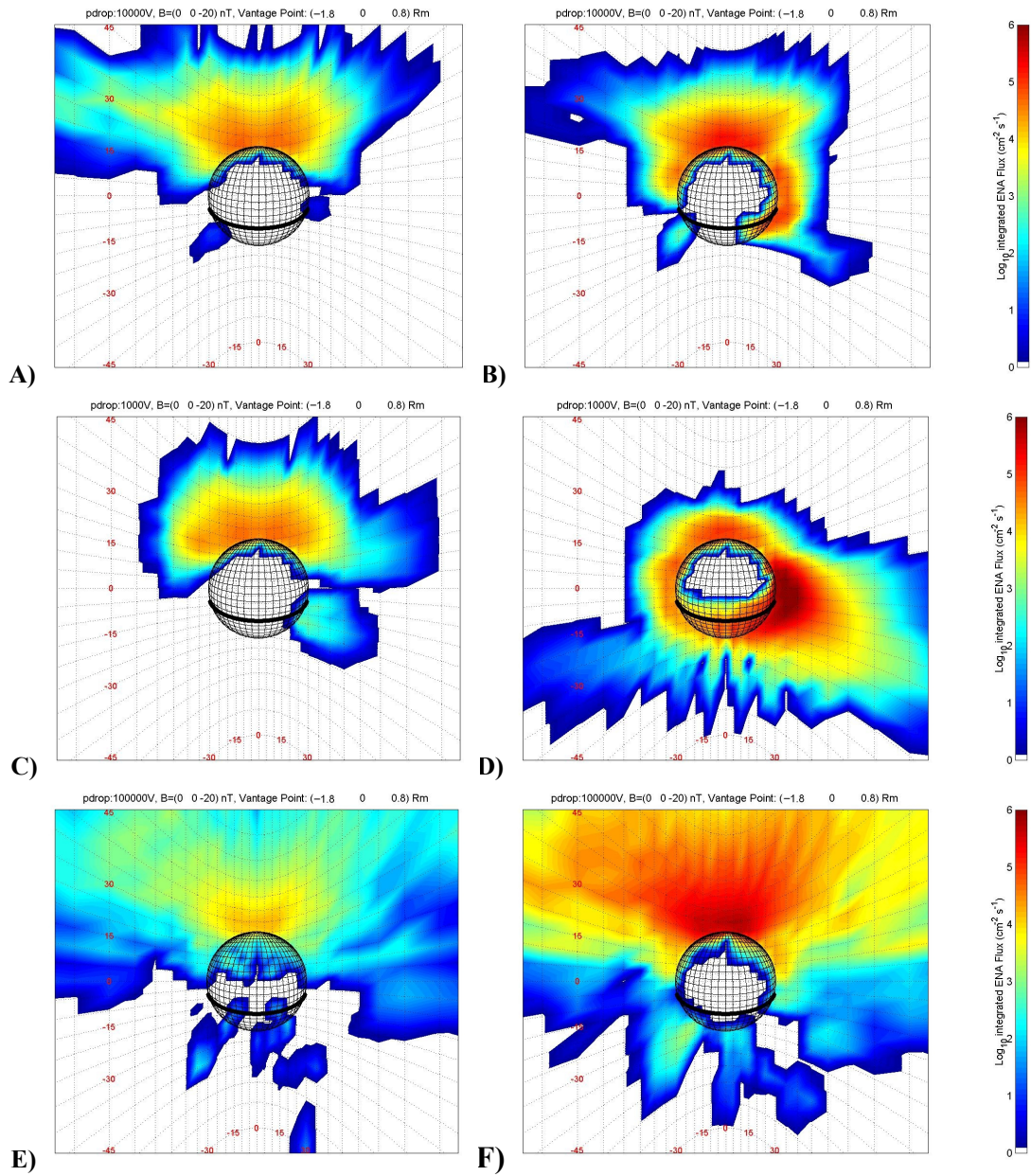

**Figure 5.12** Simulated ENA images, from a vantage point in the nightside ($P_I$=[-1.8, 0, 0.8] $R_M$). Colour is coded according to ENA flux, integrated over two energy ranges: 100 eV - 1 keV (panels **A, C** and **E**) and 1-10 keV (panels **B, D** and **F**). The boundary conditions are: $\mathbf{B_{IMF}}$ = (0, 0, -20) nT; $PD$ = 10 kV (panels **A** and **B**), $PD$ = 1 kV (panels **C** and **D**), $PD$ = 100 kV (panels **E** and **F**).

Figure **5.12** shows simulated ENA images, as seen from $P_I$ in a "fish-eye" projection, for different $PD$ values and integrated over two different energy ranges: 100 eV-1 keV, and 1 keV-10 keV. As for figure **5.8**, protons that enter the simulation area in the southern hemisphere are not shown. Generally, in these images, it is possible to observe ENA coming from two different H$^+$ populations. The first one is **α**, made by protons

that are precipitating into the cusp regions or that are passing above the North Pole before escaping in the nightside, and that get close enough to experience charge-exchange with the hermean exosphere. This process results in an intense signal in the upper part of the ENA image. The second population (γ), which is clearly evident especially for low *PD* and in the higher energy range, is made of protons that are drifting westward and are bouncing during their motion. These protons reach low altitude at their mirroring points, and even if their flux is generally lower than the flux of the precipitating ones, they may produce an intense ENA signal, distinguishable in the right side of ENA images.

The westward proton circulation is generated by **B**-grad drift, the intensity of which is proportional to proton energy. Hence, the second ENA signal is present only at energies above 1 keV; moreover, the **B**-grad drift dominates the **E×B** drift if the *PD* is low. Hence, the second ENA peak is more evident for low values of *PD*.

Similar considerations apply to ENA images simulated from point $P_2$ (Figure **5.13**). In this case the ENA signal has been integrated over all energies. In addition to α and γ populations, which are clearly distinguishable, those protons that are mirroring in the dayside (β) generate a clear ENA signal as well. The ENA signal in the centre of the image comes from the γ population, and hence increases as the *PD* decreases.

Generally, the information carried by ENA images need to be unfolded, applying models of plasma and exosphere. However, the simulations suggest that ENA imaging at Mercury, alone, is able to give some useful information. This happens, for example, if the vantage point is in $P_1$. There, the proton precipitation into the cusp can be directly

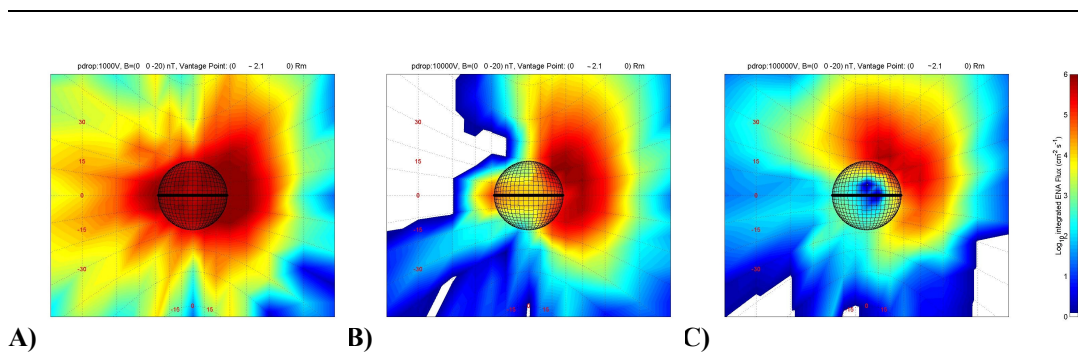

**A)**                                            **B)**                                            **C)**

**Figure 5.13** Simulated ENA images, from a vantage point in the dawn sector ($P_2$=[0, -2.1, 0] $R_M$). Colour is coded according to ENA flux, integrated over all energies (100 eV-10 keV). The boundary conditions are: **B$_{IMF}$**=(0, 0, -20) nT; *PD* = 10 kV (panel **A**), *PD* = 1 kV (panel **B**), *PD* = 100 kV (panel **C**).

monitored: any increase is immediately reflected in an increase of the ENA flux, since no other sources contribute. The ENA sensor on board MMO could address itself to such investigations. Moreover, in the case of a vantage point located in position similar to $P_2$, it is also possible to instantaneously monitor the open field line area on the surface. The chosen position for the vantage point is suitable for the MPO, and hence for NPA-SERENA. The SERENA/ELENA sensor will cover a field of view of about $2° \times 60°$. Providing a proper spacecraft attitude, the detection of the α and β populations in the dayside will be possible by means of ENA imaging. At the same time, it will be possible to detect the γ population that is drifting in the space between the sensor and the surface (through ENA imaging), as well as the γ population that is precipitating onto the surface of Mercury (through the imaging of the directional neutral atoms coming from ion-sputtering).

## 5.8 Neutral atom detection at Mercury. Instrumentation and feasibility

*Instrumentation.* The ENA sensors discussed so far, such as NAOMI (section **3.5**) and NPD (section **4.5**) need to make a two-step measurement for each incoming ENA in order to resolve simultaneously mass and energy. To do this, it is necessary, in principle, to interact with the ENA; such an interaction (with a carbon foil or surface) perturbs the neutral and affects the precision of the measurement.

An innovative technique is presented here: the sensor ELENA (Emitted Low-Energy Neutral Atoms) is a Time-of-Flight Detector, based on the state-of-the art of ultra-sonic oscillating choppers (>100 kHz) and mechanical gratings.

The incoming neutral particles are separated from ions by HV plates. Then the micro-valve choppers let neutrals enter the chamber with a definite timing. Particles are then flown into a TOF chamber, and finally detected by a 2-dimesional stop array (based on MCPs and discrete anode sets). The mechanical layout of the instrument is shown in figure **5.14**: the stop surface is on the top; the particle entrance, with shutters, is on the bottom.

The MCP "sees" a sequence of pulses generated by the oscillation of the random mask with respect to a fixed collimator (see figure **5.15**); the random intervals are multiples of $\Delta t$. We can imagine this "opening/closing" as a sequence of ones and zeros, i.e. an array $s_i$ of length $N$. Each element of array $s_i$ represents a fixed time interval $\Delta t$. Let $F(t)$ be the time of flight distribution of neutral particles; $F_i$ is the discrete representation of it, with $i=1,N$. The signal on the MCP in the interval $j$ ($Z_j$) is hence equal to [*Wilhelmi and Gompf*, 1969]:

$$Z_j = C\,\Delta t \sum_{i=1}^{N} s_{j-i} F_i + U_j \qquad (5.11)$$

where $C$ is a normalisation constant and $U_j$ is introduced to take into account the noise. The array $s_i$ should be a random array, i.e. its autocorrelation function should be zero. Let $S_{jk}$ be a *circulant* matrix, obtained from $s_i$:

$$S_{ij} = s_{j-i} = \begin{pmatrix} s_N & s_{N-1} & \dots & s_1 \\ s_1 & s_N & \dots & s_2 \\ \vdots & \vdots & & \vdots \\ s_{N-1} & s_{N-2} & \dots & s_N \end{pmatrix}; \qquad (5.12)$$

in this case, equation **5.11** can be written in a more compact notation:

$$Z_j = S_{ij} F_i + U_j \qquad (5.13)$$

and, in principle, the ToF distribution $F$ can be obtained from the MCP signal $Z$ by inverting matrix $S$.

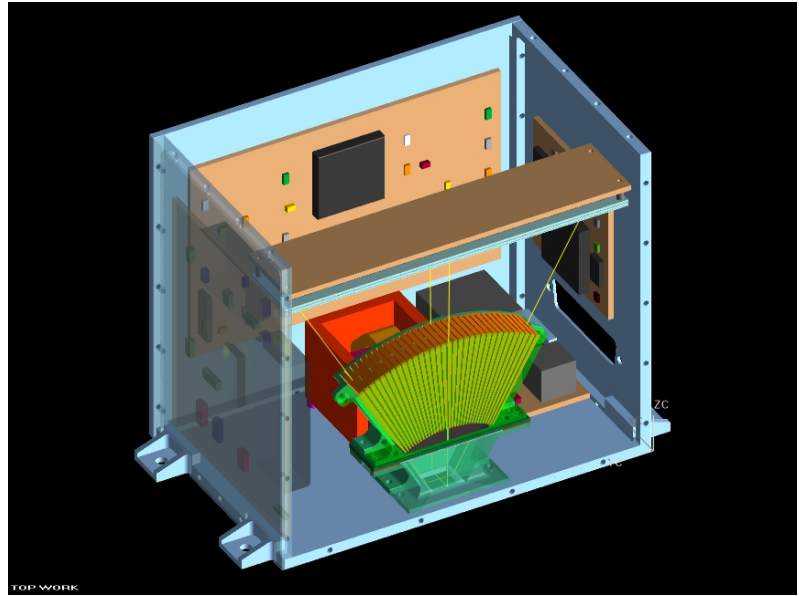

**Figure 5.14** ELENA mechanical layout.

Figure **5.16** shows the transform simulator for ELENA. Red line is the energy spectrum of an input, mono-energetic signal. Blue line is the energy spectrum of the reconstructed signal, as simulated by using a Monte-Carlo model and equation **5.13**. If the signal/noise ratio were too small, an alternative solution is to use non-random

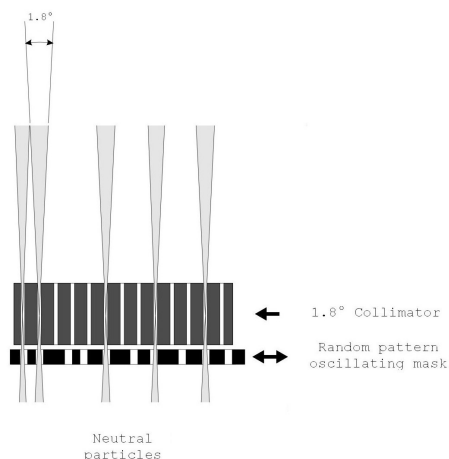

**Figure 5.15** ELENA shuttering concept.

sequences, with $s_1$=1 and $s_{j>1}$=0. We can measure the ToF of each single particle by assuming that it has entered the chamber in the middle of the first "opening". In this case, it is necessary to find a compromise between the ToF resolution ($\Delta t$), the number of ToF channels ($N$-1) and the duty cycle ($N \cdot \Delta t$). The particle energy is measured in the same way as NPD (see section **4.5**); the particle mass is known once having measured its energy and velocity. The main characteristics of ELENA sensor are shown in table **5.2**.

*Feasibility.* One of the main scientific purposes of this study is to investigate the feasibility of neutral atom imaging at Mercury. In doing this, we must take into account that different production processes need different treatment. Neutral atoms coming from charge-exchange have been simulated by using a Monte-Carlo model (in sections **5.4** and **5.5**), and their detection has been discussed**.** Neutral atoms coming from the surface could be generated by using the same model, but in this case the resolution of the numerical grid is probably too rough. Those latter neutrals, in fact, do not travel in straight lines (because of their lower energy), hence a little variation in energy and direction at the surface may completely change their trajectory. Moreover, since this study may also give some advice about suitable ELENA allocation on board BC, several tests must be done in a short time. The study of the trajectory from source to detector, hence, needs a separate, analytical treatment.

**Table 5.2**

| Intrinsic field-of-view (IFOV) | 1.8° x 1.8° |
|---|---|
| Total field-of-view (FOV) | 60° x 1.8° |
| swath width (cross track, 8 pixels) | 102.96 km @t 400 km altitude |
|  | 386.40 km @ 1500 km altitude |
| swath length (along track, 1 pixel) | 12.87 km @ 400 km altitude |
|  | > 48.30 km @ 1500 km altitude |
| dwell time | 5 s |

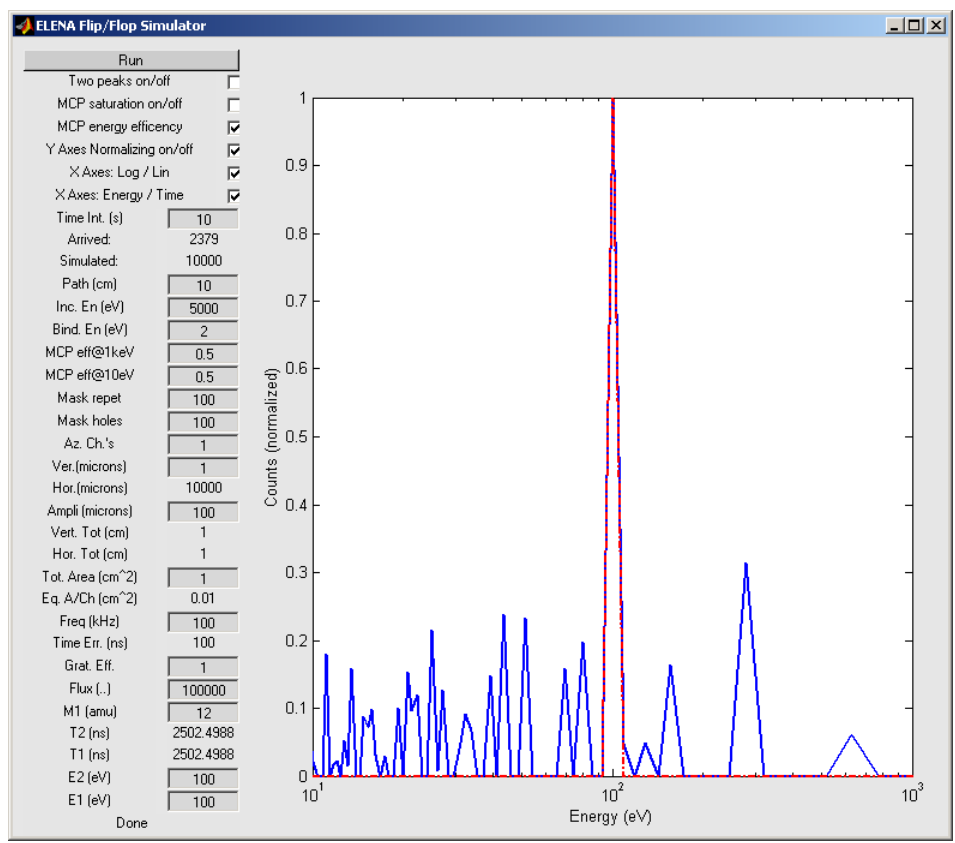

**Figure 5.16** Sample transform simulator for ELENA. Red line is the energy spectrum of an input, mono-energetic signal, simulated using a Monte-Carlo model. Blue line is the energy spectrum of the measured signal, as reconstructed by using equation **5.13**.

To achieve this goal, I have developed a dedicated *tool of analysis*. The philosophy of this tool is in the questions: "for a given instrument allocation, field of view and energy range, which signal is detectable?" and "can a strong surface composition anisotropy be resolved by the neutral flux entering the instrument?". To give the answers, the model just backtracks the particle trajectories from the sensor to the surface. This can be done analytically by solving the gravitational equation of motion. The mathematical procedure is very simple and is not reported here; the used equations are in Appendix **B**.

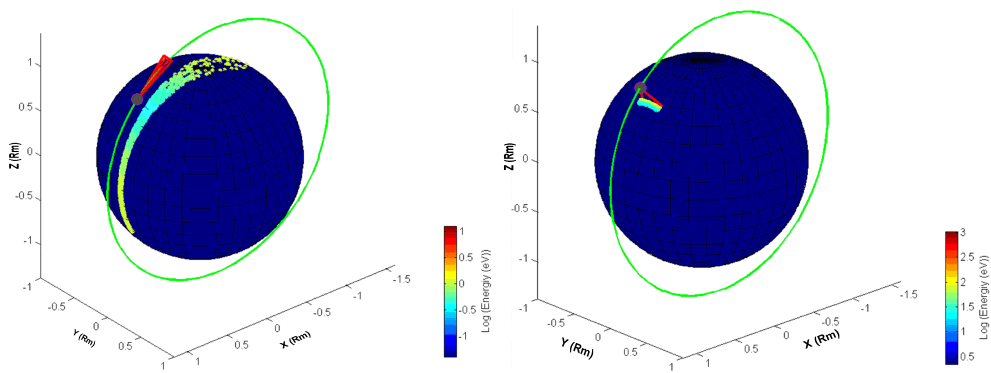

**Figure 5.17** Footprints of LENA trajectories that can be detected from BEPI COLOMBO spacecraft. Coloured points on the surface of Mercury are colour-coded according to the sputtering energy. Green curve is the spacecraft orbit, red lines represent the sensor viewing angle for two different proposed sensors: STROFIO, a low energy exospheric sensor (left panel), and ELENA, a mid energy directional sensor (right panel).

As an example, I show the simulations of the source that may produce a signal in the range 10-1000 eV. The results are shown in figure **5.17**. The simulation shows that neutral mass spectrometer, detecting particles below 50 eV (namely: STROFIO, part of the SERENA package) detects a signal that is generated in a narrow region. If STROFIO is pointing towards RAM (as in figure **5.17A**) this region lies around the surface projection of the satellite's orbit. Within this region, however, it is not possible to determine the location where sputtering has occurred. As far as it concerns ELENA (figure **5.17B**), some uncertainty arises from the energy detection. In fact, an error on the detected energy results in an error on the estimation of the source position. This error is due to the combination of the satellite velocity (2÷3 km/s) and the detected particle velocity, which depends on the energy. Hence, this error can be neglected for energies above hundreds of eV. For low energy particles, however, this additional error should be considered. As an example, a 50% error on the energy of detected sodium @ 20 eV results in a 25 km error on the position of the source, which must be added to the intrinsic error due to IFOV (see table **5.2**); this error decreases to less than 10 km for sodium @ 200 eV and is negligible above 500 eV.

# 6. Summary and Conclusions

Here I'm presenting a summary of the scientific results of this work. Among them, I'm focusing on the three most important ones, i.e. the global reconstruction of the ring current during a geomagnetic storm, the feasibility of the ENA investigation applied to Phobos and a global modelling of the Solar Wind/Mercury interaction.

## 6.1 Earth equatorial proton distribution during storm time

Nowadays, the importance of ENA imaging among the instruments of investigation of the magnetosphere of the Earth has been widely accepted. Actually, several orbiting spacecrafts include ENA sensor as part of their payload. Nevertheless, ENA imaging needs to be used in addition to accurate plasma circulation models to produce the best results.

It is important to understand how plasma modelling and ENA imaging are complementary. As an example, we may consider an arbitrary configuration of the model presented in chapter **3** and a given ENA image of the same plasma circulation. ENA imaging gives 2D information about a 3D volume; the model gives information and interpretation about the $H^+$ distribution over a 2D equatorial plane. Even if ENA imaging alone is able to give an accurate reconstruction of the plasma circulation, only a model is able to focus and distinguish the physical processes occurring, to provide a correct interpretation of the present and, eventually, a forecast of the future.

The main scientific steps/results of this study are:

1) the model is able to reconstruct any instantaneous real magnetospheric configuration;

2) it is possible to monitor the storm-time development of proton distribution by using this model tuned by best-fitting some selected parameters with experimental data,

3) details on partial/global ring current evolution, energetic particle flow patterns and electric potential can be obtained through model analysis.

At present, the model has been tested using CRRES/MICS/LOMICS data collected during July 1991; in the future, the use of more experimental data (if possible, observed by two or more spacecrafts simultaneously) is recommended. Moreover, the future inclusion of a pitch angle analysis in the model will allow to reconstruct the full 3D magnetospheric circulation. In this way, the comparison between ENA images as predicted by the model and as actually observed by, for example, IMAGE/MENA-HENA will be possible.

## 6.2 ENA and Mars-Phobos environment

One of the main open questions about Phobos' exosphere composition and outgassing rate is related to the presence of a gas/dust torus along its orbit. Its presence has been suggested, but never observed, after PHOBOS-2 observations in 1989; a definite clarification of such an intriguing issue would be of fundamental interest to determine the time evolution of Phobos.

In chapter **4**, I have presented a method able to investigate Phobos' exosphere properties. I have simulated the ENA signal/disturbance generated by the interaction between the solar wind and the outgassed Oxygen, as expected according to an original O-torus model. The simulations show that this ENA signal may be discriminated from the background for particular positions of the vantage point. This technique can give relevant information, and determine the major characteristics of neutral outgassing, whose rate has not been directly measured. The importance of such a study is clear, if we consider that the simulated ENA images will be compared with *in situ* measurements. In fact, ESA mission Mars Express (MEX), launched in June 2003, will have at least three close approaches to the natural satellite Phobos. The data collected by the ASPERA-3 instrument on board MEX will probably give the scientific community an ultimate answer to the Phobos-torus dilemma.

More generally, many issues related to the effects of the interaction between solar wind and Martian environment will be addressed by ASPERA-3. In fact, the ENA imaging technique is a powerful tool to understand the fundamental role played by the solar wind for the planetary evolution.

## 6.3 Neutral Atom imaging of Mercury's Environment

In this study I have presented a Monte-Carlo model of the magnetosphere of Mercury, and I have used it as an input for the simulation of neutral particle emission from both the surface (via ion-sputtering) and the exosphere (via charge-exchange).

The main purposes of this study are: an investigation of the dayside $H^+$ circulation, and a feasibility analysis of the neutral particle imaging technique while applied to it.

*Major steps.* The development of the magnetospheric model involves:

1) modifying a T96 magnetic field model;

2) reconstructing the electric potential by using a Volland model of ionospheric potential;

3) reconstructing the $H^+$ distribution on the magnetopause;

4) estimating the error caused by the actual neglecting of the $H^+$ convection from the tail, by using a very simple model of Hermean "ring-current". Incidentally, this last estimation has given another proof of the non-existence of such a current in the equatorial plane of Mercury.

*Model limits.* Among the limits of the model, I have to mention the absence of a $B_x$ component in the modelled $\mathbf{B_{IMF}}$, and the fact that the magnetopause surface has not been modified while tuning the external conditions. In any case, it is important to remember that the high uncertainties about Mercury's environment may cause the highest errors.

The accommodation/release of particles on the surface has not been taken into account; this process will be included in a future study. Here, it causes an underestimation of the exospheric Na and O, but this fact does not affect the estimation of the neutral atom fluxes as detected by ELENA. In fact (see figure **5.11**), this estimation has been done by considering only "directional" particles (i.e. above escape energy), which by definition cannot be "re-accommodated".

As a matter of fact, one of the exospheric species that give a higher contribution to ENA generation in our model is $H_2$. The exospheric density of $H_2$, as given by *Wurz and Lammer* [2003], has been intentionally depleted as suggested by authors. The factor adopted here (0.01) is intentionally low, and leads to a conservative estimation of ENA fluxes.

*Main results.* This model has reconstructed the general circulation of protons in the dayside hemisphere of Mercury; so far, this model does not foresee any intense $H^+$ circulation in the nightside. Three main $H^+$ populations have been identified, and their behaviour under different external conditions has been discussed.

The model is able to estimate the total flux on the surface in different external conditions. This value has important implications in modelling the ion-sputtered-originated exosphere of Mercury (see section **5.6**). If we compare it to the total $H^+$ crossing the MP ($\sim 2 \times 10^{27}$ $s^{-1}$, see section **5.2**), we can conclude that about one tenth of it is able to precipitate onto the planetary surface, depending on the external condition. The remaining part is mostly bounced back; actually, the $\mathbf{E \times B}$ drift prevents a considerable amount of protons from falling down to Mercury's surface. According to this estimation, and taking into account that we have used extreme potential drop conditions, the modulation of the surface precipitation made by the potential drop seems to be comparable to the one made by the IMF magnetic field.

After the $H^+$ circulation has been reconstructed for different external conditions, I have focused my attention on plasma loss processes, more particularly on surface collisions and charge-exchange. Both these processes open the way to a remote-sensing of the plasma circulation via neutral atom imaging; in fact, both of them produce neutral atoms. Moreover, sputtering process resulting from surface collisions is expected to be a powerful tool of investigation, since: i) proton fluxes on the planetary surface may be extremely high; ii) surface composition may be analysed as well as plasma circulation.

In this frame, I have studied the feasibility of neutral atom remote-sensing made by the SERENA/ELENA instrument on board Bepi Colombo/MPO. As a matter of fact, such an instrument should be able to address itself to all the scientific issues discussed in this study.

# Acknowledgements

I would like to express my deep gratitude to Dr. Stefano Orsini and Dr. Anna Milillo, my advisors at the Interplanetary Space Physics Institute. Their support during these last three years has been fundamental for me as a person and as a scientist. I am very happy to look back and see what a long way we have gone together. Thank you.

# Appendix A: Acronyms

| | |
|---|---|
| AE | Auroral Electrojet Index |
| AU | Astronomical Unit ($1.469 \times 10^{11}$ m) |
| CE | Charge exchange |
| CF | Carbon Foil |
| Dst | Disturbance Storm index |
| ENA | Energetic Neutral Atom |
| FOV | Field of View |
| GC | Guiding Centre |
| *Kp* | Planetary K index |
| LENA | Low Energy Neutral Atom |
| *LS* | L-Shell |
| MC | Monte-Carlo |
| MCP | Micro Channel Plate |
| MLT | Magnetic Local Time |
| *PD* | (Cross-tail) Potential Drop |
| $R_E$ | Earth Radius |
| $R_M$ | Mars/Mercury Radius |
| SSD | Solid State Detector |
| SW | Solar Wind |
| TOF | Time Of Flight |

# Appendix B: Symbols, useful equations and laws

**B.1** *Electro-magnetism and plasma properties*

| | |
|---|---|
| $f(v)$ | Velocity distribution function |
| $m$ | Particle mass |
| $\Phi(v)$ | Particle flux |
| $n$ | Particle density |
| $v$ | Particle velocity |
| $\varepsilon$ | Energy density |
| $\mathbf{v}_E$ | (Guiding centre) $\mathbf{E}\times\mathbf{B}$ drift velocity |
| $\mathbf{v}_{\nabla B}$ | (Guiding centre) Grad-$\mathbf{B}$ drift velocity |
| $\mathbf{E}$ | Electric field vector |
| $\mathbf{B}$ | Magnetic field vector |
| $B_r,\ B_\theta,\ B_\phi$ | Magnetic field components in spherical coordinates |
| $M_B$ | Magnetic moment |
| $r_L$ | Larmor radius |
| $P_\perp$ | Perpendicular plasma pressure |
| $P_{//}$ | Parallel plasma pressure |

$$f(v) = \Phi(v)\, m\, v^{-2};\qquad \textbf{(B.1)}$$

$$n_{H+} = 4\Pi\int_0^\infty f(v)v^2\, dv;\qquad \textbf{(B.2)}$$

$$\varepsilon = 2\Pi m\int_0^\infty f(v)v^4\, dv\qquad \textbf{(B.3)}$$

$$P_\perp = P_{//}\,\frac{2}{3}\,\varepsilon\qquad \textbf{(B.4)}$$

$$J_\perp = \frac{\mathbf{B}}{B^2}\left[\nabla P_\perp + \left(P_{//} - P_\perp\right)\frac{\mathbf{B}\times\nabla B}{B^2}\right]\qquad \textbf{(B.5)}$$

$$\mathbf{v}_E = \frac{\mathbf{E}\times\mathbf{B}}{B^2}\qquad \textbf{(B.6)}$$

$$\mathbf{v}_{\nabla B} = \frac{1}{2}v_\perp\, r_L\,\frac{\mathbf{B}\times\nabla\mathbf{B}}{B^2}\qquad \textbf{(B.6)}$$

$$r_L = \frac{mv_\perp}{q\,|\mathbf{B}|}\qquad \textbf{(B.7)}$$

$$\begin{cases} B_r = -\dfrac{2M_B}{r^3}\sin\theta \\[2ex] B_\theta = \dfrac{M_B}{r^3}\cos\theta \\[2ex] B_\phi = 0 \end{cases} \tag{B.8}$$

**B.2** *Ballistic orbits equations*

| | |
|---|---|
| $G$ | Gravitational constant |
| $M$ | Planetary mass |
| $\mathbf{r}$ | particle position |
| $\mathbf{v}$ | particle velocity |
| $a$ | Apocentre |
| $b$ | Pericentre |
| $L$ | Momentum |
| $E$ | Energy |

$$E/m \;=\; \frac{1}{2}\,v^2 - \frac{GM}{r} \tag{B.9}$$

$$L/m \;=\; \mathbf{r} \times \mathbf{v} \tag{B.10}$$

$$\begin{cases} b = \dfrac{\left(\dfrac{L}{m}\right)^2}{GM + \sqrt{(GM)^2 + 2\left(\dfrac{E}{m}\right)\left(\dfrac{L}{m}\right)^2}} \\[5ex] a = \dfrac{\left(\dfrac{L}{m}\right)^2}{GM - \sqrt{(GM)^2 + 2\left(\dfrac{E}{m}\right)\left(\dfrac{L}{m}\right)^2}} \end{cases} \tag{B.11}$$

# Appendix C: Planetary fact sheet

**Table C.1**

|  | Mars | The Earth | Mercury |
|---|---|---|---|
| Mass ($10^{24}$ kg) | 0.64185 | 5.9736 | 0.3302 |
| Volume ($10^{10}$ km$^3$) | 16.318 | 108.321 | 6.083 |
| Equatorial radius (km) | 3397 | 6378.1 | 2439.7 |
| Polar radius (km) | 3375 | 6356.8 | 2439.7 |
| Mean density (kg/m$^3$) | 3933 | 5515 | 5427 |
| Surface gravity (m/s$^2$) | 3.71 | 9.80 | 3.70 |
| Escape velocity (km/s) | 5.03 | 11.19 | 4.3 |
| Bond albedo | 0.250 | 0.306 | 0.119 |
| Solar irradiance (W/m$^2$) | 589.2 | 1367.6 | 9126.6 |
| Black-body temperature (K) | 210.1 | 254.3 | 442.5 |
| Maximum surface Temperature (K) | - | - | 700 |
| Minimum surface Temperature (K) | - | - | 90 |
| Semi-major axis ($10^6$ km) | 227.92 | 149.60 | 57.91 |
| Sidereal orbit period (days) | 686.980 | 365.256 | 87.969 |
| Orbit eccentricity | 0.0935 | 0.0167 | 0.2056 |
| Length of day (hrs) | 24.6597 | 24.0000 | 4222.6 |

**Mars Atmosphere**

Surface pressure: 6.36 mb at mean radius (variable from 4.0 to 8.7 mb depending on season) [6.9 mb to 9 mb (Viking 1 Lander site)]

Surface density: ~0.020 kg/m$^3$

Scale height: 11.1 km

Total mass of atmosphere: ~2.5 x 10$^{16}$ kg

Average temperature: ~210 K (-63 C)

Diurnal temperature range: 184 K to 242 K (-89 to -31 C) (Viking 1 Lander site)

Wind speeds: 2-7 m/s (summer), 5-10 m/s (fall), 17-30 m/s (dust storm) (Viking Lander sites)

Mean molecular weight: 43.34 g/mole

Atmospheric composition (by volume):

> *Major*: Carbon Dioxide ($CO_2$) - 95.32%; Nitrogen ($N_2$) - 2.7% Argon (Ar) - 1.6%; Oxygen ($O_2$) - 0.13%; Carbon Monoxide (CO) - 0.08%

> *Minor* (ppm): Water ($H_2O$) - 210; Nitrogen Oxide (NO) - 100; Neon (Ne) - 2.5; Hydrogen-Deuterium-Oxygen (HDO) - 0.85; Krypton (Kr) - 0.3; Xenon (Xe) - 0.08

**Mercury Atmosphere**

Surface pressure: ~10$^{-15}$ bar

Average temperature: 440 K (167 C) (590-725 K, sunward side)

Total mass of atmosphere: <~1000 kg

Atmospheric composition: 42% Oxygen ($O_2$), 29% Sodium (Na), 22% Hydrogen ($H_2$), 6% Helium (He), 0.5% Potassium (K), possible trace amounts of Argon (Ar), Carbon Dioxide ($CO_2$), Water ($H_2O$), Nitrogen ($N_2$), Xenon (Xe), Krypton (Kr), Neon (Ne)

**[From http://nssdc.gsfc.nasa.gov/planetary/factsheet]**

**Table C.2 . Atmospheric abundances and loss rates at Mercury**

| Species | Surface abundance ($cm^{-3}$) | Total Zenith Column ($cm^{-2}$) | JeansFlux ($cm^{-2}\,s^{-1}$) T=550 | Photo-ionisation lifetime (sec) R=0.386 AU | Photoionisation rate ( $cm^{-2}\,s^{-1}$) R=0.386 AU |
|---|---|---|---|---|---|
| H | 23; 230 [a] | $3\times10^9$ [h] | 2.7 [l] | $2.0\times10^6$ [l] | $1.5\times10^3$ [l] |
| He | $6.0\times10^3$ [a] | $<3\times10^{11}$ [h] | 5.3 [l] | $2.8\times10^6$ [l] | $1.1\times10^5$ [l] |
| O | $4.4\times10^4$ [a] | $<3\times10^{11}$ [h] | $6.3\times10^{-9}$ [l] | $7.4\times10^5$ [l] | $4.0\times10^5$ [l] |
| Na | $1.7\text{-}3.8\times10^4$ [a] | $2\times10^{11}$ [i] | $6.9\times10^{-15}$ [l] | 5500-14000 [m-e] 14000-38000 [m-t] 25000 [n] | $2.1\times10^7$ [m-e] $7.4\times10^6$ [m-t] |
| K | $3.3\times10^2$ [b] | $1\times10^9$ [b] | | $9.0\times10^4$ [l] 6700 [n] | $1.5\times10^5$ [l] |
| Ar | $<6.6\times10^6$ [a] | $<2\times10^{13}$ [h] $1.3\times10^9$ [k] | | $4.8\times10^5$ [l] | $4.2\times10^7$ [l] |
| $^{20}$Ne | $6\times10^3$ day [c] $7\times10^5$ night [c] | $3.7\times10^{10}$ [h] | | | |
| $H_2$ | $<2.6\times10^7$ [a] | $<8.7\times10^{14}$ [h] | | $2.3\times10^6$ [l] | $8.8\times10^8$ [l] |
| $O_2$ | $<2.5\times10^7$ [a] | $<9.6\times10^{13}$ [h] | | $2.6\times10^5$ [l] | $2.7\times10^9$ [l] |
| $N_2$ | $<2.3\times10^7$ [a] | $<1\times10^{10}$ [h] | | $4.1\times10^5$ [l] | $4.5\times10^4$ [l] |
| $CO_2$ | $<1.6\times10^7$ [a] | $<4.5\times10^{13}$ [h] | | $1.9\times10^5$ [l] | $2.0\times10^8$ [l] |
| $H_2O$ | $<1.5\times10^7$ [a] | $<1\times10^{12}$ [e] $<1\times10^{14}$ [h] | | $3.7\times10^5$ [l] | $2.0\times10^8$ [l] |
| OH | $1.4\times10^3$ [d,e] | $>1\times10^{10}$ [d,e] | | $6.2\times10^5$ [l] | $2.7\times10^6$ [l] |
| Mg | $7.5\times10^3$ [d] | $3.9\times10^{10}$ [d] | | | |
| Ca | 387 [d] <239 [f] | $<1.2\times10^9$ [d] $<7.4\times10^8$ [e] $1.1\times10^8$ [j] | | | |
| Fe | 340 [d] | $7.5\times10^8$ [d] | | | |
| Si | $2.7\times10^3$ [d] | $1.2\times10^{10}$ [d] | | | |
| S | $5\times10^3$ [d] $6\times10^5$ [g] | $2.0\times10^{10}$ [d] $2.0\times10^{13}$ [g] | | $1.3\times10^5$ [l] | $1.5\times10^5$ [l] $1.5\times10^8$ [l] |
| Al | 654 [c] | $3.0\times10^9$ [d] | | | |

**a** *Hunten et al*. [1988]: measurements or upper limits

**b** *Potter and Morgan* [1988]

**c** *Hodges* [1974]: model abundance

**d** *Morgan & Killen* [1997]: model abundances

**e** *Killen et al*. [1997]:  model abundances

**f** *Sprague et al.* [1993]: measured upper limit

**g** *Sprague et al*. [1995]: prediction

**h**  *Shemansky* [1988]: Mariner 10 measurements

**i**  *Killen et al*. [1990]: measured abundance

**k** *Killen et al*. [2002]: model abundance

**j** *Bida et al.* [2000]

**l** *Killen and Ip* [1999]

**m** *Huebner et al*. [1992] ionisation rates: experimental (e), theoretical (t) for quiet and active Sun

**n** *Cremonese et al*. [1997]

[*A. Milillo*, private communication]